\documentclass[12pt,twoside]{article}
\usepackage{overcite,array}
\usepackage{epsfig}
  \usepackage{amsthm,amssymb}
\usepackage{amsbsy,amsfonts,amsmath}
\usepackage{rotating}
\makeatletter\input{dvips.def}\makeatother
\makeatletter 
\def\@cite#1{\mbox{$\m@th^{(\hbox{\@ove@rcfont#1})}$}}
\renewcommand{\@biblabel}[1]{{#1}.}

\def\@sect#1#2#3#4#5#6[#7]#8{\ifnum #2>\c@secnumdepth \def\@svsec{}\else
 \refstepcounter{#1}\edef\@svsec{\csname the#1\endcsname.\ }\fi
 \@tempskipa #5\relax \ifdim \@tempskipa>\z@ \begingroup #6\relax
 \@hangfrom{\hskip #3\relax\@svsec}{\interlinepenalty \@M #8\par} \endgroup
 \csname #1mark\endcsname{#7}\addcontentsline
 {toc}{#1}{\ifnum #2>\c@secnumdepth \else
 \protect\numberline{\csname the#1\endcsname}\fi #7}\else
 \def\@svsechd{#6\hskip #3\relax \@svsec #8\csname #1mark\endcsname
 {#7}\addcontentsline {toc}{#1}{\ifnum #2>\c@secnumdepth \else
 \protect\numberline{\csname the#1\endcsname}\fi #7}}\fi \@xsect{#5}}
\def\section{\@startsection {section}{1}{\z@}{-3.5ex plus -1ex minus
 -.2ex}{2.3ex plus .2ex}{\sffamily\bfseries\normalsize}}
\def\subsection{\@startsection{subsection}{2}{\z@}{-3.25ex plus -1ex minus
 -.2ex}{1.5ex plus .2ex}{\sffamily\bfseries\normalsize}}
\def\subsubsection{\@startsection{subsubsection}{3}{\z@}{-3.25ex plus
 -1ex minus -.2ex}{1.5ex plus .2ex}{\sffamily\bfseries\normalsize}}
\long\def\@makecaption#1#2{%
  \vskip\abovecaptionskip
  \sbox\@tempboxa{\sffamily\bfseries{#1. #2}}%
  \ifdim \wd\@tempboxa >\hsize
   \footnotesize #1.\ \ #2\par
  \else
    \hbox to\hsize{\hfil\box\@tempboxa\hfil}%
  \fi
  \vskip\belowcaptionskip}

\def\footnoterule{\kern-3\p@
  \hrule \@width 0.5in \kern 2.6\p@}
\renewcommand{\@makefntext}[1]{%
    \parindent 0em%
    \noindent
    \hb@xt@.5em{\hss\@makefnmark}\hskip.2em#1}
 \makeatother

\font\mysmall=cmr10 at 7pt
\font\mymsbm=msbm10 at 12pt
\font\mymsbms=msbm10 at 9 pt
\def\myb{\boldsymbol}
\def\cases#1{\left\{#1\right.}
\font\mysmallit=cmmi10 at 9pt
\renewcommand{\thetable}{\Roman{table}}
\def\upstrut{\raisebox{1ex}{\vbox to 1em{}}}
\begin{document}
\pagestyle{myheadings}
\makeatletter
\renewcommand{\@evenhead}{\sffamily\bfseries\footnotesize\thepage
\hfil Au-Yang and Perk}
\renewcommand{\@oddhead}{\sffamily\bfseries\footnotesize
$\myb {Q}$-Dependent Susceptibility on Penrose Tiles\hfil\thepage}
\makeatother
\title{\bf Wavevector-Dependent Susceptibility
\break in $\myb Z$-Invariant Pentagrid Ising Model \hfill}
\author{\bf Helen Au-Yang and Jacques~H.H.~Perk%
${}^{1,2}$\hfil}
\date{ }
\maketitle
\thispagestyle{plain}
\setcounter{footnote}{1}
\footnotetext{Department of Physics, Oklahoma State University,
Stillwater, OK 74078-3072, USA.}
\addtocounter{footnote}{1}
\footnotetext{Supported in part by NSF Grant No.\ PHY 01-00041.}
{\leftskip=5em\par\noindent\footnotesize
{\it Received  September 17, 2004}
\par\noindent\hrulefill\vglue 5pt
\par\noindent
We study the ${\bf q}$-dependent susceptibility $\chi({\bf q})$ of a
$Z$-invariant ferromagnetic Ising model on a Penrose tiling, as first
introduced by Korepin using de Bruijn's pentagrid for the rapidity lines.
The pair-correlation function for this model can be calculated exactly
using the quadratic difference equations from our previous papers. Its
Fourier transform $\chi({\bf q})$ is studied using a novel way to
calculate the joint probability for the pentagrid neighborhoods of
the two spins, reducing this calculation to linear programming.
Since the lattice is quasiperiodic, we find that $\chi({\bf q})$ is
aperiodic and has everywhere dense peaks, which are not all visible at
very low or high temperatures. More and more peaks become visible as the
correlation length increases---that is, as the temperature approaches the
critical temperature. 
\par\noindent\hrulefill\vglue 5pt
\par\noindent{\sffamily\bfseries\footnotesize KEY WORDS:} Ising model;
quasiperiodicity; Fibonacci sequence; pentagrid; Penrose tiles;
$Z$-invariance; correlation functions;
${\bf q}$-dependent susceptibility.
\par\leftskip=0pt}
\vglue 0pt

\advance\jot by 7pt
\section{Introduction}\label{sect1}

In an experiment \cite{SBGC} done in 1984, Shechtman and his coworkers
found fivefold symmetry in the diffraction patterns of some rapidly cooled
alloys. As such a symmetry is incompatible with lattice periodicity, it
was concluded that the crystalline structures of these alloys, if any, must
necessarily be quasiperiodic. This theoretical explanation came forward
almost immediately, as Penrose, de Bruijn, and Mackay
\cite{Penrose,Pen0,Pen1,Bruijn1,Mackay1,Mackay2} had already studied
tilings that have fivefold symmetry, well before this experimental
discovery. Quasiperiodic tilings are types of almost periodic
structures that permit sharp peaks in the diffraction patterns, but have
normally forbidden symmetries \cite{Mackay2,LSt0,BGM,CJ,Henley,Baake}.

Already in 1986, Korepin \cite{K1} introduced a $Z$-invariant eight-vertex
model on Penrose tiles. The $Z$-invariant inhomogeneous models are
completely integrable\cite{BaxZI} even on irregular lattices and their
critical exponents are known to be the same as those of homogeneous
systems on regular lattices. Thus, the critical behaviors of these
quasiperiodic $Z$-invariant models\cite{BaxZI,K2,AK1,AK2,BGB,GBS,GB,Choy}
have to be the same,\footnote{Universality of
the critical exponents of ferromagnetic Ising models on quasiperiodic
lattices has been confirmed for non-$Z$-invariant cases also using
real-space renormalization group techniques,\cite{AO1} Monte
Carlo simulations,\cite{AAA,BHJ,ON1,ON2,SJR,TJ,RB} series expansion
methods,\cite{AbeDot,DotAbe,Repet} and the study of Yang--Lee
zeros.\cite{SBG,SB,RGS2}} independent of the lattice
structure.\footnote{Unlike the ${\bf q}$-dependent susceptibility,
thermodynamic quantities like the free energy, the specific heat, and the
bulk susceptibility do not probe the lattice structure, although subtle
lattice effects do show up in corrections to scaling.\cite{APsc}}

In such models, the order parameter is the same \cite{BaxZI,K1} for all
sites and it vanishes towards the critical point. Therefore, the Fourier
transform of the one-point function of a $Z$-invariant Ising model is
the product of this order parameter and the lattice sum
$\sum{\rm e}^{{\rm i}{\bf q}\cdot{\bf r}}$. Experiments that probe the
resulting ``magnetic" Bragg peaks are restricted to the low-temperature
phase and the corresponding theory is essentially the zero-temperature
theory, well-studied in the literature. \cite{Mackay2,LSt,SoSt} The
aforementioned Bragg peaks will broaden, if we allow the underlying
quasicrystalline lattice to become distorted by lattice
vibrations.\cite{LSSBH,Hof} However, we shall not consider this
possibility in this paper, as we assume the underlying lattice to be a
perfect and rigid Penrose tiling, restricting our attention solely to the
ordering of the spins at the lattice sites under thermal fluctuations.

Contrary to the above theory for the Bragg scattering, the situation is far
more complicated for scattering experiments that probe the pair-correlation
function $\langle\sigma_{\bf r}\sigma_{\bf r'}\rangle$ via the
wavevector-dependent susceptibility $\chi({\bf q})$.
This last quantity is defined as
\begin{equation}
k_{\rm B}T\chi({\bf q})\equiv\bar\chi({\bf q})=
\lim_{{\cal L}\to\infty}{\frac 1 {{\cal L}}}
\sum_{\bf r}\sum_{\bf r'} {\rm e}^{{\rm i}{\bf q}\cdot({\bf r'}-{\bf r})}
\big[{\langle\sigma_{\bf r}\sigma_{\bf r'}\rangle}-
\langle\sigma_{\bf r}\rangle\langle\sigma_{\bf r'}\rangle\big],
\label{chi}\end{equation}
where ${\cal L}$ is the number of lattice sites, ${\bf r}$ and ${\bf r}'$
run through all these sites, and ${\bf q}=(q_x,q_y)$. It is the Fourier
transform of the connected pair correlation function, defined by what is
inside the square brackets. Furthermore, $\bar\chi({\bf q})$ is the reduced
${\bf q}$-dependent susceptibility, taking out a trivial factor involving
the absolute temperature $T$.

In the well-known lattice-gas language it becomes proportional to the
structure function, the Fourier transform of the density-density
correlation function, which is also measurable in diffraction experiments,
revealing the symmetry of the lattice. Thus, the $\chi({\bf q})$ in
$Z$-invariant models can indeed be used to show the difference between
quasiperiodic and regular lattices and we expect it to provide a
diffuse scattering pattern both above and below the critical temperature
$T_{\rm c}$, with more structure closer to $T_{\rm c}$.

There is good reason to pick a $Z$-invariant Ising model for our
present study of $\chi({\bf q})$. It is taken from the foremost class of
models with short-range interactions allowing exact
computations. Moreover, comprehensive extensive studies on spin-spin
correlation functions in nontrivial models with short-range interactions
have been done only for Ising
models\cite{KO49,Fisher59a,MPW63,MWbk,WMTB76,Ab78b,Perkd,BaxZI,AP-ZI}.
Cited here is just a fraction of the literature. The accumulative
knowledge of these studies has made the calculations in $Z$-invariant
quasi-periodic Ising models possible. Correlations in other nontrivial
cases are still mostly inaccessible to exact methods of evaluation.

For instance, in 1988, Tracy \cite{Tr1,Tr2} introduced the layered
Fibonacci Ising model, which is not $Z$-invariant. The row correlation
functions in the layered Ising model are known to be block Toeplitz
determinants\cite{AYM}, of which we still have no idea how to evaluate
them exactly in general, except in a few simpler cases\cite{AYM,McWu67a}.
Tracy has shown that the critical exponent of the specific heat remains
unchanged in such quasiperiodic layered models.

In our previous papers \cite{AJPq,APmc1}, we have studied Fibonacci Ising
models whose spins are on regular lattices, but whose nearest-neighbor
interactions are quasi-periodic. They are special cases of
inhomogeneous Ising models whose Hamiltonians are given by
\begin{equation}
-\beta{\cal H}\,=\sum_{m,n}\,({\bar K}_{m,n}\sigma_{m,n}\sigma_{m,n+1}
+ K_{m,n}\sigma_{m,n}\sigma_{m+1,n}),
\label{energy}\end{equation}
with $\beta\equiv1/k_{\rm B}T$.
When the system is periodic, the pair correlations are translationally
invariant. Thus one of the sums in (\ref{chi}) can be carried out. For
$T\ne T_{\rm c}$, the connected correlations decay exponentially as
functions of distance.\footnote{For
$T=T_{\rm c}$, the correlations decay algebraically, so that we need the
Epstein-Ewald summation formula\cite{Kong} to take into account the
long-distance behavior.} Therefore, only a finite number of short-distance
correlation functions are needed for the calculation of the
$\chi({\bf q})$ in Eq.\ (\ref{chi}).

For the Ising models considered previously\cite{AJPq,APmc1}, the couplings
between specific nearest-neighbor spin pairs form Fibonacci sequences
$\{S_{n}\}$ defined recursively \cite{Tr1} by
\begin{equation}
S_{n+1}=S_nS_{n-1},\quad S_0={\rm B},\quad S_1={\rm A},
\end{equation}
so that $S_2={\rm AB}$, $S_3={\rm ABA}$, $S_4={\rm ABAAB}$ and so on.
Since this sequence is quasiperiodic, as arbitrary long subsequences are
repeated infinitely often, the model is also aperiodic. Consequently, the
correlations are no longer translationally invariant. However, the averages
of the correlations for two spins at fixed distance can be evaluated by
using a theorem of Tracy\cite {Tr1}.

In our previous works,\cite{AJPq, APmc1} we have found that the various
${\bf q}$-dependent susceptibilities
$\chi({\bf q})$ of our Fibonacci Ising models are
always periodic. They can have multiple incommensurate everywhere-dense
peaks in each unit cell only if the aperiodic oscillations in the average
correlation functions are not negligibly small. This is true in the
mixed case when the interactions are aperiodic sequences of ferromagnetic
and antiferromagnetic couplings. The number of visible peaks in
$\chi({\bf q})$ increases as the correlation length increases. In
contrast, in the periodic Fibonacci Ising lattices with mixed bonds, the
visible peaks in the ${\bf q}$-dependent susceptibility are at commensurate
positions and their number has a finite maximum.\cite{AJPq}

The ferromagnetic aperiodic Fibonacci Ising lattice, on the other hand,
behaves almost like the regular Ising model---one peak per unit cell,
located at the commensurate position---because the aperiodic oscillations
in its average correlation functions are negligibly small at all
temperatures.\cite{AJPq}

The ${\bf q}$-dependent susceptibility $\chi({\bf q})$ is found to have
almost the same behaviors for both $T<T_{\rm c}$ and
$T>T_{\rm c}$.\cite{AJPq} This is in agreement with the result of Peter
Stephens, who showed that the randomized (disordered) icosahedral system
\cite{Stephens} gives almost the same diffraction pattern as a
quasicrystal---which is in the solid (ordered) phase.

Even though both ferromagnetic and antiferromagnetic edge interactions are
present in the mixed case, the mixed systems in Refs.\ \citen{AJPq} and
\citen{APmc1} are not frustrated. In fact, their partition functions are
equal to the partition functions of the ferromagnetic models, from which
they differ by gauge transformations of signs. Also, the $\chi({\bf q})$
of a fully-frustrated model may not show incommensurate peaks. \cite{Kong}

In this paper, we turn our attention to systems with a quasiperiodic
lattice structure\cite{BGB,GBS,GB}. More specifically we study a
$Z$-invariant Ising model whose spins are on vertices of a Penrose
fat-and-skinny rhombus tiling with the pentagrid as its rapidity
lines\cite{Bruijn1,K1,K2,AK1,AK2}. 

\subsection{Outline}

This paper is organized as follows. In Section \ref{sect2}, we introduce
a $Z$-invariant Ising model on Penrose tiles constructed from a
pentagrid\cite{Bruijn1}, with spins on either the odd or even sublattice.
We show how the pair-correlation functions can be evaluated in Section
\ref{sect3}. Section \ref{sect4} is rather lengthy. In it, a new
method of counting all the spin sites and evaluating the joint
probabilities of the occurrence of two neighborhoods of two spins is
given. In subsection \ref{sect51}, we describe in detail the calculation
for the ${\bf q}$-dependent susceptibility for the odd lattice. The results
are given in subsection \ref{sect52}. In subsection \ref{sect53} we
present a mapping between the odd and the even sublattices. Finally, in
Section \ref{sect6} we present our conclusions.

\setcounter{equation}{0}
\section{Pentagrid and Penrose Tiles}\label{sect2}

\begin{figure}[tbph]
~\vskip0in\hskip0in\epsfclipon
\epsfxsize=0.88\hsize
\centerline{\epsfbox{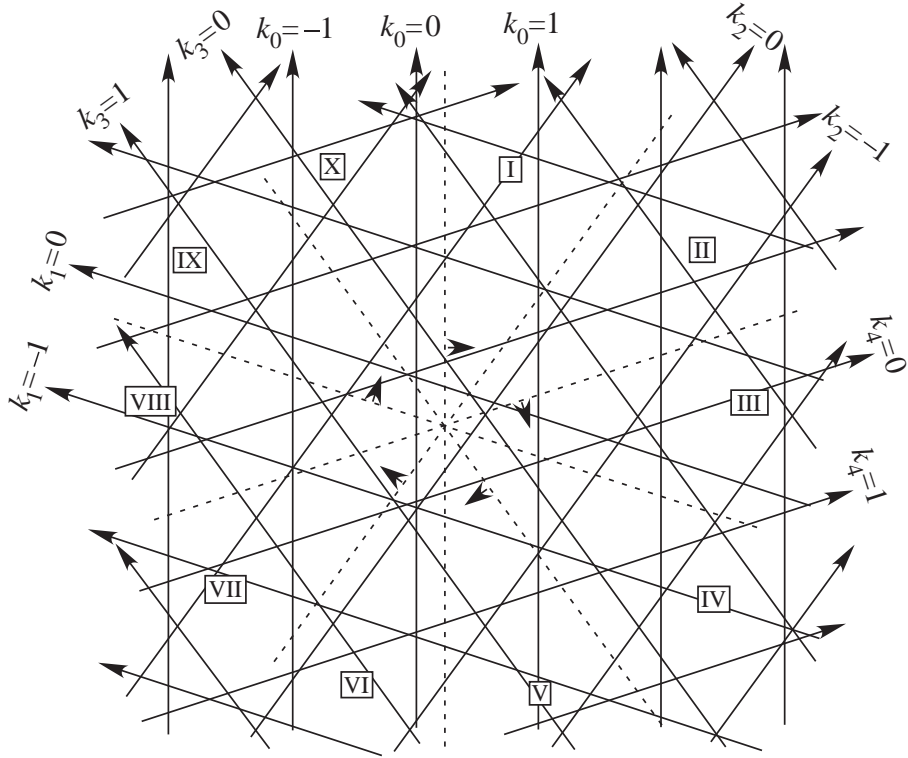}}
\vskip0.2in
\caption{The pentagrid is a superposition of five grids, each of
which consists of parallel equidistanced lines. These grid lines are the
five different kinds of rapidity lines in a {\mysmallit Z}-invariant Ising
model. Arrows are put on the grid lines, and all of them pointing to the
upper half plane. Through a center in one of the meshes (site for a spin
$\sigma$), dashed lines are drawn to split the pentagrid into ten regions
I to X. Correlations between $\sigma$ and $\sigma'$ are different when
$\sigma'$ sits in different regions. Five little arrows perpendicular to
the five dashed lines indicate the directions in which the corresponding
integers $k_j$, $j=0,\cdots,4$, increase.}
\label{fig1}
\end{figure}

In two ingenious papers by the famous Dutch mathematician N.G.\ de Bruijn,
he relates the non-periodic Penrose tilings in a plane to a
pentagrid\cite{Bruijn1}, which is a superposition of five grids. Each grid
consists of parallel lines with equal spacings between the lines; the grids
may be obtained from one another by rotations of angles which are
multiples of $2\pi/5$. This is shown in Fig.~\ref{fig1}. Here some grid
lines of the pentagrid are shown; the arrows on the lines shown in
Fig.~\ref{fig1} should be ignored for the moment, as they will define the
direction of the ``rapidities" which we define later.

To describe the pentagrid in mathematical formulas\cite{Bruijn1}, let
\begin{equation}
\zeta={\rm e}^{2{\rm i}\pi/5},\qquad
\zeta+\zeta^{-1}=2\cos(2\pi/5)=p^{-1}={{\textstyle \frac 1 2}}(\sqrt 5-1),
\label{goldratio}
\end{equation}
in which $p$ is the golden ratio. Then, choose
$\gamma_0,\gamma_1,\gamma_2,\gamma_3,\gamma_4$ to be five real
numbers, satisfying 
\begin{equation}
\gamma_{0}+\gamma_{1}+\gamma_{2}+\gamma_{3}+\gamma_{4}=0.
\label{shift}\end{equation}
Now the $j$th grid in the pentagrid consists of
lines given by 
\begin{equation}
G_j=\{z\in \mbox{\mymsbm C}| {\rm Re} (z\zeta^{-j})
+\gamma_j=k_j, k_j\in
\mbox{\mymsbm Z}\},\qquad j=0,\cdots,4.
\label{grid}\end{equation}
The pentagrid is called regular, if there is no point in the complex plane
$\mbox{\mymsbm C}$ belonging to more than two of the five grids. This also
means, every vertex of the regular pentagrid is an intersection of no more
than two lines. Each vertex is surrounded by four meshes (which are often
called faces in physics).

Now to every point $z$ in the complex plane $\mbox{\mymsbm C}$, de Bruijn
associates an integer vector ${\vec K}(z)=(K_0(z),\cdots,K_4(z))$ whose
five elements are integers given by
\begin{equation}
K_j(z)=\lceil{\rm Re} (z\zeta^{-j})
+\gamma_j\rceil,
\label{mesh}\end{equation}
in which $\lceil x\rceil$ denotes the ``roof of $x$", which is the smallest
integer $\geqslant x$. It is easily seen from (\ref{mesh}) and (\ref{grid})
that  whenever $z$ moves across a line of the $j$th grid, $K_j(z)$ changes
by 1. All points in the same mesh (face) have the same integer vector and
the integer vectors of different meshes are different. From
Fig.~\ref{fig1}, we may already see that some fraction of the meshes (or
faces) becomes infinitesimally small in size as the number of lines in
each grid becomes infinite.

\begin{figure}[tbph]
~\vskip0in\hskip0in\epsfclipon
\epsfxsize=0.8\hsize
\centerline{\epsfbox{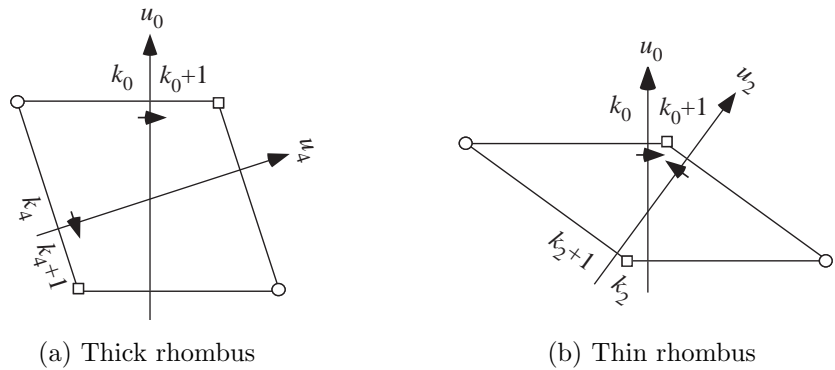}}
\vskip0pt
\hbox to\hsize{\hspace*{50pt}\footnotesize
(a) Thick rhombus\hspace*{50pt}\hfil
(b) Thin rhombus\hfil
\hspace*{8pt}}
\vskip0.3in\caption{Spins interacts along the diagonals. All sides have
length 1. The long diagonal of the thick rhombus has length $p$ and the
short diagonal of the thin rhombus has length $p^{-1}$, with $p$ the
golden ratio, given in (\ref{goldratio}). The little arrows perpendicular
to the grid (rapidity) lines indicate the directions in which the integers
$k_j$, $j=0,\cdots,4$, increase.}
\label{fig2}
\end{figure}

Since every vertex of the regular pentagrid is surrounded by four meshes
(faces), by assigning to each of their four corresponding integer vectors
${\vec K}(z)$ $=$ $(K_0(z),\cdots,K_4(z))$ a complex number
\begin{equation}
f(z)=\sum_{j=0}^4 K_j(z)\zeta^j,
\label{penrose}\end{equation}
these four meshes are now mapped to the vertices of a rhombus. More
specifically, to the intersection of two grid lines $k_r$ and $k_s$,
($r\ne s$), one assigns a rhombus in $\mbox{\mymsbm C}$ whose vertices are
the four complex numbers $f(z)$, $f(z)+\zeta^r$, $f(z)+\zeta^s$ and
$f(z)+\zeta^r+\zeta^s$ assigned to the four surrounding meshes. Clearly,
there are two different kinds of rhombuses: the thick one having angles
$72^{\circ}$ and $108^{\circ}$ for $r=s\pm1$, and the thin one having
angles $36^{\circ}$ and $144^{\circ}$ for $r=s\pm2$. In both rhombuses,
all sides have length 1. They are shown in Fig.~\ref{fig2}.

In these two papers \cite{Bruijn1}, de Bruijn also showed that, even
though there are many different choices of $\gamma_j$ in (\ref{shift}),
many of the resulting pentagrids are shift-equivalent, that is, they can
be obtained from each other by a parallel shift.\footnote{He also proved
that if $\xi\equiv\sum_j\gamma_j\zeta^{2j}$ times a power of $\zeta$ is not
purely imaginary (modulo the principal ideal of $1-\zeta$ given by all
complex numbers of the form $\sum_j n_j\zeta^j$ with integers
$n_0,\cdots,n_4$ satisfying $\sum_j n_j=0$), then the corresponding
pentagrid is regular \cite{Bruijn1}.}

\begin{figure}[tbph]
~\vskip0in\hskip0in\epsfclipon
\epsfxsize=0.5\hsize
\centerline{\epsfbox{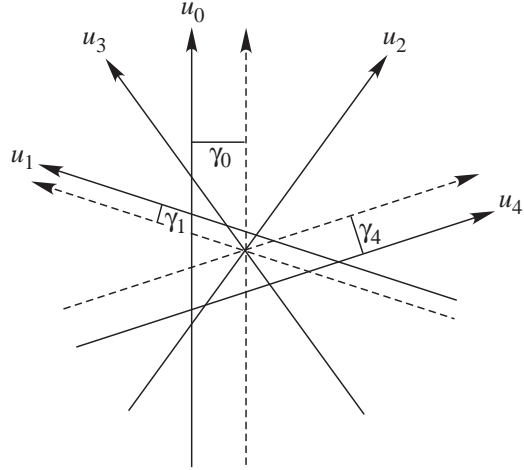}}
\vskip-0.0in
\hskip0pt\caption{The five different kinds of rapidity
lines $u_0,\cdots,u_4$ pointing into the upper half plane. The grid lines
$k_j=0$, for $j=0,\cdots,4$, are shifted from the dashed lines to make the
pentagrid regular.}
\label{fig3}
\end{figure}

We now assign to each grid line in the grid $j$, $j=0,\cdots,4$, of the
pentagrid a rapidity $u_j$ pointing into the upper half plane, as is
shown in Figs.~\ref{fig1} and \ref{fig3}. In Fig.~\ref{fig3}, it is also
shown how the five grid lines $k_j=0$, for $j=0,\cdots,4$, are shifted to
make the grid regular, again following de Bruijn \cite{Bruijn1}.

The usual $Z$-invariant Ising model \cite{BaxZI,AP-ZI} is formed by putting
spins inside the meshes, but here, however, the Ising spins are on the
vertices of the rhombuses. There is an one-to-one mapping given by
(\ref{penrose}) relating a mesh in the pentagrid to a vertex of the Penrose
tiling. After the mapping, the grid lines in the pentagrid---which are also
the rapidity lines---become ``Conway worms" (no longer straight) in the
Penrose tiling \cite{Gardner}. Since the rapidity lines are used to define
commuting transfer matrices, they do not have to be straight lines.

The model so defined is a special case of the inhomogeneous $Z$-invariant
eight-vertex model proposed by Korepin\cite{K1,K2,AK1,AK2}. More
specifically, the four-spin couplings are identically zero, and the
eight-vertex model decomposes into two independent Ising models. The
interactions of the Ising spins are along the diagonals of the rhombuses.
The odd and even sublattices are therefore decoupled.\cite{Wu,KW}

\begin{figure}[tbph]
~\vskip0.05in
\epsfxsize=0.50\hsize
\centerline{\epsfbox{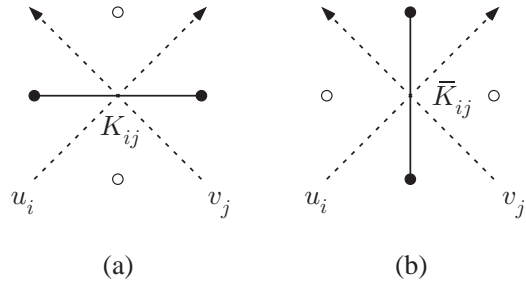}}
\hskip0.1in\caption{(a) The horizontal coupling ${K}(u_i,v_j)$;
and (b) the vertical coupling ${\bar{K}}(u_i,v_j)$.}
\label{fig4}
\end{figure}

As in the earlier works \cite{BaxZI, AP-ZI}, the coupling
${K}(u_i,v_j)$ between two spins is represented by a line
connecting these two spins, with the arrows of the two rapidity lines
$u_{i}$ and $v_{j}$ on the same side of this line, as shown in
Fig.~\ref{fig4}$\,$(a), while the line representing the coupling
${\bar{K}}(u_i,v_j)$ has the arrows of the two rapidity lines $u_{i}$
and $v_{j}$ on opposite sides, as shown in Fig.~\ref{fig4}$\,$(b). The
edge interactions are parametrized by
\begin{eqnarray}
&&\sinh\big(2{K}(u_i,v_j)\big)=
k\,{\rm sc}(u_i-v_j,k')={\rm cs}\big(\lambda+v_j-u_i,k'\big),
\nonumber\\
&&\sinh\big(2{\bar{K}}(u_i,v_j)\big)
={\rm cs}(u_i-v_j,k')=k\,{\rm sc}\big(\lambda+v_j-u_i,k'\big).
\label{couplings}
\end{eqnarray}
Here $\lambda={\rm K}(k')$ is the elliptic integral of the first kind,
and $k$ and $k'=\sqrt{1-k^2}$ are the elliptic moduli; these are convenient
temperature variables, assumed to be the same for all sites.

From Fig.~\ref{fig2}, we can see that the lengths of the four diagonals of
the two rhombuses are different. The interactions between the spins are
chosen to depend on the interparticle spacings only, but not on the
orientations. Consequently, we must have
\begin{equation}
u_{0}-u_{1}=u_{2}-u_{3}=u_{4}-u_{0}
=\lambda+u_{1}-u_{2}=\lambda+u_{3}-u_{4}. 
\end{equation}
From this, we find
\begin{equation}
u_{4}-u_{1}=\frac{4\lambda}{5},\quad u_{2}-u_{1}=\frac{3\lambda}{5},\quad
u_{0}-u_{1}=\frac{2\lambda}{5},\quad u_{3}-u_{1}=\frac{\lambda}{5}.
\label{rapidity}\end{equation}
If we let
\begin{equation}
s_{j}=k\,{\rm sc}(j\lambda/5,k'),
\end{equation}
then for the thick rhombus in Fig.~\ref{fig2}$\,$(a), we assign $s_2$ to
the longer diagonal and $s_3$ to the shorter diagonal, while for the thin
rhombus in Fig.~\ref{fig2}$\,$(b), $s_4$ to the shorter diagonal and $s_1$
to the longer one. Thus, to the four types of diagonals are assigned four
kinds of couplings according to their lengths, with a stronger coupling
for a shorter interparticle distance.

\begin{figure}[tbph]
~\vskip0.1in\hskip0in\epsfclipon
\epsfxsize=0.76\hsize
\centerline{\epsfbox{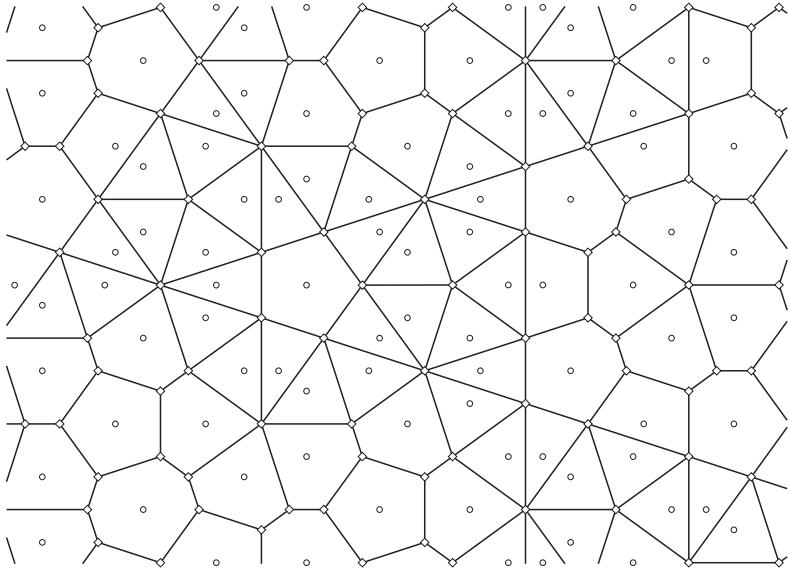}}
\vskip0.15in
\hskip0pt\caption{Two Ising lattices on Penrose tiles. The spins on
the odd sublattice (the first Ising model) are joined by bonds, but the
spins on the even sublattice (the second Ising model on the dual lattice)
are denoted by circles without their bonds. The two lattices are
independent. The coordination number, which is the number of adjacent
spins connected to a given spin in each sublattice, takes values 3, 4, 5,
6, or 7. Incorporating the two models, a third Ising model can be defined
with Ising spins on both the odd and even sublattices, i.e.\ the
decoupling limit of Baxter's eight-vertex model on the Penrose lattice
\cite{K1,BaxZI,K2,AK1,AK2}.}
\label{fig5}
\end{figure}

The two Ising sublattices on the Penrose tiling are indicated in
Fig.~\ref{fig5}. The edges in the even sublattice are omitted. There are
eight types of vertices S, K, Q, D, J, S3, S4, S5 in the Penrose tiling,
which are shown in Fig.~7 of Ref.~\citen{Bruijn1}. The coordination
numbers of spins in the Penrose Ising model are 3 for types Q and D; 4 for
K; 5 for S, J and S5; 6 for S4; 7 for S5.

\setcounter{equation}{0}
\section{Correlations}\label{sect3}

Two spins $\sigma_{\bf r}$ and $\sigma_{\bf r'}$ at two different vertices
of the Ising lattice just defined have different integer vectors
${\vec K}=(K_0,\cdots, K_4)$ and ${\vec K'}=(K'_0,\cdots, K'_4)$. Since
there is a one-to-one mapping between the vertices of the Penrose tiles and
the meshes of the pentagrid, the integer vectors can be used to denote the
positions of the spins: ${\bf r}\leftrightarrow{\vec K}$.
 
Since each grid consists of parallel lines with equal spacings, the
absolute value of the difference $\ell_{j}=K'_j-K_j$ is actually the
number of the $j$th kind of rapidity lines sandwiched between these two
spins. The correlation functions were shown\cite{BaxZI,AP-ZI} to be
\begin{eqnarray}
&&\hspace*{-3.7em}\langle\sigma_{\vec K}\sigma_{\vec K'}\rangle=
\langle\sigma\sigma'\rangle_{[\ell_0,\cdots,\ell_4]}\nonumber\\
&&=g(\overbrace{u'_0,\ldots,u'_0}^{|\ell_0|},
\overbrace{u'_1,\ldots,u'_1}^{|\ell_1|},
\overbrace{u'_2,\ldots,u'_2}^{|\ell_2|},
\overbrace{u'_3,\ldots,u'_3}^{|\ell_3|},
\overbrace{u'_4,\ldots,u'_4}^{|\ell_4|}),
\label{cor}\end{eqnarray}
where $u'_j=u_j$ for rapidity lines of type $j$ with arrows pointing to the
same side of the line joining the two spins, and $u'_j=u_j\pm\lambda$ for
rapidities with arrows pointing to opposite sides of the line. It is as if
the rapidities of lines pointing to the opposite side need to be flipped by
adding $\pm\lambda$, i.e.\ adding $\pm\pi$ to the angle variable
$u_j\pi/\lambda$.\cite{AP-ZI} The functions $g$ have both the ``permutation
symmetry" (which means that they are invariant under all permutations of
the rapidities) and the ``difference property" (which implies a translation
invariance under shifting all the rapidities by the same
amount).\cite{BaxZI}

We next examine in more detail when we have to choose $u'_j=u_j$ or
$u'_j=u_j\pm\lambda$ in (\ref{cor}). Through a point in the mesh where
spin $\sigma$ sits, as shown in Fig.~\ref{fig1}, we draw five dashed lines
parallel to each of the five grids. As the choice of the point in the mesh
is rather arbitrary, the dashed lines should really have been drawn with a
finite thickness, i.e.\ open strips without their boundaries between two
consecutive grid lines of the pentagrid. Five arrows are also drawn
perpendicular to the dashed lines to indicate the directions in which the
integers $k_j$ increase. If the other spin $\sigma'$ is in a mesh crossed
by the $j$th dashed line, i.e.\ $\sigma$ and $\sigma'$ lie within the
same strip, then the two spins have the same $K_j$ ($\ell_j=0$) and their
pair correlation function does not depend on the value of $u'_j$. When
$\sigma'$ moves away from this dashed line in the direction of the arrow,
we have $\ell_j\geqslant 0$, whereas $\ell_j\leqslant 0$ if $\sigma'$
moves away in the direction opposite to the arrow. These dashed lines (or
more precisely strips) divide the entire plane into ten regions, and we
numbered them from I to X. From Fig.~\ref{fig1}, we can see that
$(\ell_0,\cdots,\ell_4)$ have the same signs inside each region.

If $\sigma'$ is in regions III or VIII, we find from Fig.~\ref{fig1} that
the arrows of all the rapidity lines point to the same side of the line
joining the two spins. Thus $u'_j=u_j$ for all five $j$-values. When
$\sigma'$ is in regions II or VII, then the rapidity lines with $u_4$ 
and the other rapidity lines are pointing to opposite sides of the line
joining the spins, so that $u'_4=u_4-\lambda$ and $u'_j=u_j$ for $j\ne 4$.
For regions IV and IX the $u_1$ rapidity lines point in the other
direction with respect to the other rapidity lines, implying
$u'_1=u_1+\lambda$. If $\sigma'$ is in regions I and VI, the
arrows of the $u_2$ and $u_4$ rapidity lines are on the opposite
side and, therefore, $u'_4=u_4-\lambda$ and $u'_2=u_2-\lambda$.
Similarly, for regimes V and X, $u'_1=u_1+\lambda$ and $u'_3=u_3+\lambda$.
For all other $j$-values, $u'_j=u_j$. In summary, our choices for
$u'_j$ are listed in Table \ref{raptab},
\def\bigstrut{\hbox{\vrule height 14pt depth 7pt width 0pt}}
\begin{table}[htb!]
\caption{Rapidities}
\vspace*{7pt}
\begin{center}
\begin{tabular}{|c|c|c|c|c|c|}
\hline
\bigstrut Regions&$u'_4$&$u'_2$&$\;u'_0\;$&$u'_3$&$u'_1$\\
\hline
\bigstrut I and VI&$u_4-\lambda$&$u_2-\lambda$&$u_0$&$u_3$&$u_1$\\\hline
\bigstrut II and VII&$u_4-\lambda$&$u_2$&$u_0$&$u_3$&$u_1$\\\hline
\bigstrut III and VIII&$u_4$&$u_2$&$u_0$&$u_3$&$u_1$\\\hline
\bigstrut IV and IX&$u_4$&$u_2$&$u_0$&$u_3$&$u_1+\lambda$\\\hline
\bigstrut V and X&$u_4$&$u_2$&$u_0$&$u_3+\lambda$&$u_1+\lambda$\\\hline
\end{tabular}
\label{raptab}
\end{center}
\vspace*{7pt}
\end{table}
with the $u_j$'s given in (\ref{rapidity}), where we may set $u_1=0$
without loss of generality in view of the difference property of the pair
correlation function.\cite{BaxZI}

We may even shift the five rapidity values $u'_0,\ldots,u'_4$ in
(\ref{cor}) by the same amount, depending on the choice of region, such
that $\min_j u'_j=0$. We can then also use the permutation
property\cite{BaxZI} of the pair-correlation function $g$, given in
(\ref{cor}), to rearrange the five resulting rapidity values $u'_j$ in
decreasing order as
$\frac{4}{5}\lambda,\frac{3}{5}\lambda,\frac{2}{5}\lambda,
\frac{1}{5}\lambda,0$. Therefore, it suffices to calculate the quantity
\begin{eqnarray}
&&\hspace*{-3.7em}g[m_4,m_3,m_2,m_1,m_0]\equiv\nonumber\\
&&g\bigg(
\overbrace{\frac{4\lambda}{5},\!\ldots\!,\frac{4\lambda}{5}}^{m_4},
\overbrace{\frac{3\lambda}{5},\!\ldots\!,\frac{3\lambda}{5}}^{m_3},
\overbrace{\frac{2\lambda}{5},\!\ldots\!,\frac{2\lambda}{5}}^{m_2},
\overbrace{\frac{\lambda}{5},\!\ldots\!,\frac{\lambda}{5}}^{m_1},
\overbrace{0,\ldots,0\vphantom{\frac{\lambda}{5}}}^{m_0}\bigg),
\label{corems}
\end{eqnarray}
where the $m_j$'s are nonnegative integers depending on the $\ell_j$'s
and the choice of region. To determine this dependence, we can first
use (\ref{rapidity}), from which we find
$u_3+\lambda>u_1+\lambda>u_4>u_2>u_0>u_3>u_1>u_4-\lambda>u_2-\lambda$.
Comparing (\ref{cor}), Table \ref{raptab} and (\ref{rapidity}) we can then
express all pair correlations in the form (\ref{corems}). We list the
results for the ten different regions in Table \ref{corrtab}. This
completes, more or less, the calculation of the pair correlation function,
as we can refer to our previous papers\cite{APmc1,APmc2} for further
details on how to evaluate the $g[m_4,m_3,m_2,m_1,m_0]$ defined in
(\ref{corems}).

\begin{table}[htb!]
\caption{Pair-Correlation Function}
\vspace*{7pt}
\begin{center}
\begin{tabular}{|c|c|c|}
\hline
\bigstrut Regions&Signs of $(\ell_0,\ell_1,\ell_2,\ell_3,\ell_4)$&
$\langle\sigma\sigma'\rangle_{[\ell_0,\cdots,\ell_4]}=$\\
\hline
\bigstrut I \&\ VI&$(+,+,+,-,-)$\ \&\ $(-,-,-,+,+)$&
$g[|\ell_0|,|\ell_3|,|\ell_1|,|\ell_4|,|\ell_2|]$\\\hline
\bigstrut II \&\ VII&$(+,+,-,-,-)$\ \&\ $(-,-,+,+,+)$&
$g[|\ell_2|,|\ell_0|,|\ell_3|,|\ell_1|,|\ell_4|]$\\\hline
\bigstrut III \&\ VIII&$(+,+,-,-,+)$\ \&\ $(-,-,+,+,-)$&
$g[|\ell_4|,|\ell_2|,|\ell_0|,|\ell_3|,|\ell_1|]$\\\hline
\bigstrut IV \&\ IX&$(+,-,-,-,+)$\ \&\ $(-,+,+,+,-)$&
$g[|\ell_1|,|\ell_4|,|\ell_2|,|\ell_0|,|\ell_3|]$\\\hline
\bigstrut V \&\ X&$(+,-,-,+,+)$\ \&\ $(-,+,+,-,-)$&
$g[|\ell_3|,|\ell_1|,|\ell_4|,|\ell_2|,|\ell_0|]$\\
\hline
\end{tabular}
\label{corrtab}
\end{center}
\vspace*{7pt}
\end{table}

Since the pentagrid is invariant under rotations by angles that are integer
multiples of $2\pi/5$, the grids may be relabeled $m \to m+j$, (mod 5).
Then the differences of the five integer vectors of the two spins are also
relabeled $\ell_{m}\to\ell_{m+j}$. This shows that the pair correlation
function must have the cyclic property
\begin{equation}
\langle\sigma\sigma'\rangle_{[\ell_0,\ell_1,\ell_2,\ell_3,\ell_4]}=
\langle\sigma\sigma'\rangle_{[\ell_j,\ell_{j+1},\ell_{j+2},
\ell_{j+3},\ell_{j+4}]},\quad(\mbox{mod 5}).
\label{cyclic}\end{equation}
From Table \ref{corrtab}, we indeed find that this property holds.

\setcounter{equation}{0}
\section{Enumeration of sites}\label{sect4}

As mentioned in the introduction in the context of the Fibonacci Ising
model,\cite{AJPq, APmc1} to calculate the ${\bf q}$-dependent
susceptibility (\ref{chi}) for lattices for which the correlations are not
translationally invariant, one needs to find a way to calculate suitable
averages of the pair-correlation function. Since the meshes in the
pentagrid---and even the distances between them---can be infinitesimally
small in the thermodynamic limit, the problem of counting all the spin
sites must be first solved.

We proceed by considering in detail the parallelograms bounded by two sets
of parallel grid lines in the pentagrid and examining all possible spin
sites in each of these parallelograms. Let $P(k_j,k_{j+1})$ denote the
parallelogram sandwiched between four grid lines $k_j-1$, $k_j$,
$k_{j+1}-1$ and $k_{j+1}$ for any $j$. (Throughout the entire paper,
we let $k_{j+5}\equiv k_j$, i.e.\ the index $j$ is considered mod 5.)
Obviously, for all points $z\in P(k_j,k_{j+1})$, we have $K_j(z)=k_j$ and
$K_{j+1}(z)=k_{j+1}$. The different choices of $j$ give the different
orientations of the parallelograms. Next, we determine how many spin sites
a parallelogram may have, what are the integer vectors for these spins,
etc. Since this section is rather lengthy, we have subdivided it into many
parts, and put the main conclusion at the end.

\subsection{Reference vector for $\myb {P}$}
The vertices of parallelogram $P(k_j,k_{j+1})$ can be
calculated from (\ref{grid}) as the intersections of grid lines in grids
$G_j$ and $G_{j+1}$, i.e.
\begin{equation}
G_j\cap G_{j+1}=\biggl\{z\in \mbox{\mymsbm C}\biggm|
z=\frac{\,{\rm i}\,[\zeta^j(k_{j+1}\!-\!\gamma_{j+1})-
\zeta^{j+1}(k_j\!-\!\gamma_j)]}
{\sin(2\pi/5)}\biggr\}, 
\label{inters}\end{equation}
for $k_{j},k_{j+1}\in\mbox{\mymsbm Z}$. Moreover, any point $z$
in the interior of $P(k_j,k_{j+1})$ may be expressed in terms of
$\myb{\epsilon}=(\epsilon_j,\epsilon_{j+1})$, with
$0\leqslant\epsilon_j,\epsilon_{j+1}\leqslant 1$, as
\begin{equation}
z=\frac{\,{\rm i}\,[\zeta^j(k_{j+1}\!-\!\gamma_{j+1}\!-\!\epsilon_{j+1})-
\zeta^{j+1}(k_j\!-\!\gamma_j\!-\!\epsilon_j)]}{\sin(2\pi/5)}
\equiv z(\myb{\epsilon}),
\label{zinP}\end{equation}
allowing us a change of notation
$K_{j+{\rm m}}(\myb{\epsilon})\equiv K_{j+{\rm m}}(z(\myb{\epsilon}))$
for $z\in P$.

The four corners of parallelogram $P(k_j,k_{j+1})$ are given by
$\myb{\epsilon}\!=\!(0,0),$ $(0,1),$ $(1,0),$ or $(1,1)$ as can be seen
from (\ref{inters}). Now for each $P(k_j,k_{j+1})$, we pick a reference
integer vector $(k_0,\cdots,k_4)$, which is related to the integer vector
of the corner of $P(k_j,k_{j+1})$ with $\myb{\epsilon}\!=\!(0,0)$. Apart
from the obvious identities $k_j=K_{j}(\myb{0})$ and
$k_{j+1}=K_{j+1}(\myb{0})$, we have
\begin{eqnarray}
k_{j+2}&=&K_{j+2}(\myb{0})\nonumber\\
&=&\lceil
p^{-1}(k_{j+1}-\gamma_{j+1})-k_j+\gamma_j+\gamma_{j+2}\rceil
=\lceil\alpha\rceil-k_j,\nonumber\\
k_{j+4}&=&K_{j+4}\left(\myb{0}\right)\nonumber\\ &=&\lceil
p^{-1}(k_j-\gamma_j)-k_{j+1}+\gamma_{j+1}+\gamma_{j+4}\rceil
=\lceil\beta\rceil-k_{j+1},
\label{kj12}\end{eqnarray}
in which
\begin{eqnarray}
&&\alpha\equiv\hat\alpha(k_{j+1})\equiv
p^{-1}(k_{j+1}-\gamma_{j+1})+\gamma_j+\gamma_{j+2},\nonumber\\
&&\beta\equiv\hat\beta(k_j)\equiv
p^{-1}(k_j-\gamma_j)+\gamma_{j+1}+\gamma_{j+4}. 
\label{alpha}\end{eqnarray}
However, for the last component of the reference integer vector we choose
\begin{equation}
k_{j+3}=2-\lceil\alpha\rceil-\lceil\beta\rceil
=-\lfloor\alpha\rfloor-\lfloor\beta\rfloor\not\equiv K_{j+3}(\myb{0}),
\label{kj3}\end{equation}
where $\lfloor x\rfloor$ denotes the ``floor of $x$", which is the largest
integer $\leqslant x$ and $\lfloor x\rfloor=\lceil x\rceil$ if and only if 
$x\in \mbox{\mymsbm Z}$. Since the pentagrid is regular, we find
$\alpha,\beta\notin\mbox{\mymsbm Z}$ and the second equality in the above
equation holds. The index of any mesh, whose integer vector is
${\vec K}(z)$, is defined as $\sum_j K_j(z)$. It is shown by de
Bruijn\cite{Bruijn1} that it has one of the four possible values, 1, 2, 3,
or 4. We associate odd spins to meshes with index 1 or 3, and even spins
to meshes with index 2 or 4. The index of the reference integer vector is
$\sum_j k_j=2$.

From (\ref{mesh}) and (\ref{zinP}) we find
\begin{equation}
K_{j+3}(\myb{0})=
\lceil-p^{-1}(k_{j+1}-\gamma_{j+1}+k_j-\gamma_j)+\gamma_{j+3})\rceil
=\lceil-\alpha-\beta\rceil.
\label{k3}\end{equation}
Using
\begin{equation}
 \{x\}=x-\lfloor x\rfloor,\quad
\lceil-x-y\rceil=
-\lfloor x\rfloor-\lfloor y\rfloor+\lceil-\{x\}-\{y\}\rceil
\label{frac}\end{equation} 
and comparing (\ref{kj3}) with (\ref{k3}), we find
\begin{equation}
K_{j+3}(\myb{0})=\cases{\begin{array}{llr}
k_{j+3}-1 &\hbox{for}&
\{\alpha\}+\{\beta\}\geqslant 1,\\
k_{j+3} &\hbox{for}&
\{\alpha\}+\{\beta\}<1.\upstrut\end{array}}
\label{dkj3}\end{equation}
This shows the mesh below the upper right corner 
with $\myb{\epsilon}=(0,0)$ belongs to the even sublattice
and its integer vector is the reference vector of $P$ only for
$\{\alpha\}+\{\beta\}<1$, but not for
$\{\alpha\}+\{\beta\}\geqslant 1$. 

It would be more natural to choose this corner as our reference and to
compare the integer vectors of other spins inside $P\equiv P(k_j,k_{j+1})$
with it. This was what we did originally. However, we find that the rather
odd choice of the reference vector given by (\ref{kj12}) and (\ref{kj3}),
which may not even be the integer vector of a mesh, has made calculations
much simpler. We next examine the differences between the integer vectors
of the other spins in $P(k_j,k_{j+1})$ with respect to this reference
vector.

\subsection{Integer vectors for $\myb {z\in P}$}

Substituting (\ref{zinP}) into (\ref{mesh}) and using (\ref{kj12}) and
(\ref{kj3}), we evaluate the integer vectors $\vec{K}(\myb{\epsilon})$
for every point in $P$ that is not on a grid line of the pentagrid. We
find that its components for
${\rm m}=2,3,4$ are given by
\begin{eqnarray}
&&K_{j+{\rm m}}(\myb{\epsilon})=
k_{j+{\rm m}}+\partial K_{j+{\rm m}}(\myb{\epsilon}),\nonumber\\
&&\partial K_{j+{\rm m}}(\myb{\epsilon})=
\lceil\lambda_{j+{\rm m}}(\myb{\epsilon})-1\rceil=
\lfloor\lambda_{j+{\rm m}}(\myb{\epsilon})\rfloor,
\label{kz}\end{eqnarray}
with
\begin{equation}
\lambda_{j+2}(\myb{\epsilon})
=\{\alpha\}+\epsilon_j-p^{-1}\epsilon_{j+1},\quad
\lambda_{j+4}(\myb{\epsilon})=\{\beta\}+\epsilon_{j+1}-p^{-1}\epsilon_j.
\label{dk2}\end{equation}
and
\begin{equation}
\lambda_{j+3}(\myb{\epsilon})=p^{-1}(\epsilon_j+\epsilon_{j+1})-\{\alpha\}
-\{\beta\}+1.
\label{dk3}\end{equation}
The last equality in (\ref{kz}) does not hold if
$\lambda_{j+{\rm m}}(\myb{\epsilon})$ is an integer, i.e.\ if the point
is on a grid line of grid $G_{j+{\rm m}}$. If we define the  difference
vector $\partial\vec{K}(\myb{\epsilon})$ for each mesh in $P$
as\footnote{In our notation we suppress the two trivial components
$\partial K_{j}(\myb{\epsilon})\equiv
\partial K_{j+1}(\myb{\epsilon})\equiv0$, since in parallelogram $P$ we
have by definition $K_{j}(\myb{\epsilon})\equiv k_j$,
$K_{j+1}(\myb{\epsilon})\equiv k_{j+1}$.}
\begin{equation}
\partial\vec{K}(\myb{\epsilon})=\bigl[\partial
K_{j+2}(\myb{\epsilon}),\partial K_{j+3}(\myb{\epsilon}),\partial
K_{j+4}(\myb{\epsilon})\bigr],
\label{dv}\end{equation}
then for $\{\alpha\}+\{\beta\}>1$, we have
$\partial\vec{K}(\myb{0})=[0,-1,0]$, and for $\{\alpha\}+\{\beta\}<1$,
$\partial\vec{K}(\myb{0})=[0,0,0]$.

For fixed $\gamma_j$'s with $j=0,\cdots,4$, which are the shifts of the
pentagrid, the $\hat\alpha(k_{j+1})$ and $\hat\beta(k_{j})$ in
(\ref{alpha}) are uniquely determined for each $P(k_j,k_{j+1})$.
Consequently, the number of meshes in $P$ and the difference vectors
$\partial\vec{K}$ for each mesh are uniquely determined by (\ref{kz}) to
(\ref{dk3}). The configurations of two parallelograms $P$ and $P'$ are the
same, if they have the same number of meshes (spin sites), and the same
sets of difference vectors. The difference in the configurations does not
depend on the exact locations of the relevant grid lines or their
intersections. However, whenever a grid line or an intersection moves in
or out of the parallelogram $P(k_j,k_{j+1})$, the configuration changes.

\subsection{Relevant grid lines}

It is easy to see from (\ref{mesh}) that
$\partial K_{j+{\rm m}}(\myb{\epsilon})$ changes its value whenever lines
in the $(j\!+\!{\rm m})$th grid are crossed. Because
$0\leqslant\{x\},\epsilon_j,\epsilon_{j+1}<1$, we find from (\ref{dk2}) and
(\ref{dk3}) that
\begin{equation}
-1<\lambda_{j+2}(\myb{\epsilon}),\lambda_{j+4}(\myb{\epsilon})<2,
\quad -1<\lambda_{j+3}(\myb{\epsilon})<3.
\label{ineq}\end{equation} 
Consequently, the only relevant grid lines for the parallelogram
$P(k_j,k_{j+1})$ are those having integer labels $k_{j+2}-1+n'$,
$k_{j+3}-1+m$, or $k_{j+4}-1+n$, with $n,n'=0,1$ and $m=0,1,2$.
Indeed, these are the only integer values that the
$K_{j+{\rm m}}(\myb{\epsilon})$ in (\ref{kz}), or
$k_{j+{\rm m}}-1+\lceil\lambda_{j+{\rm m}}(\myb{\epsilon})\rceil$,
can assume. The loci of these lines are given by linear equations in
$\epsilon_j$ and $\epsilon_{j+1}$ as
\begin{equation}
\lambda_{j+2}(\myb{\epsilon})=n',\quad
\lambda_{j+3}(\myb{\epsilon})=m,\quad \lambda_{j+4}(\myb{\epsilon})=n.
\label{linear}\end{equation}
From (\ref{dk2}), we find 
\begin{eqnarray}
&0<\lambda_{j+2}(\myb{\epsilon})<2 &\quad\hbox{if}\quad
\{\alpha\}>p^{-1},\nonumber\\
&0<\lambda_{j+4}(\myb{\epsilon})<2 &\quad\hbox{if}\quad
\{\beta\}>p^{-1}.
\label{line12}\end{eqnarray}
Therefore the equation $\lambda_{j+2}=0$ ($\lambda_{j+4}=0$) cannot be
satisfied if $\{\alpha\}>p^{-1}$ ($\{\beta\}>p^{-1}$), while equations
$\lambda_{j+2}=1$ and $\lambda_{j+4}=1$ always have solutions in $P$. This
means that the only grid lines in grids $G_{j+2}$ and $G_{j+4}$ crossing
$P$ are given by
\begin{eqnarray}
&k_{j+2},k_{j+4}\in P &\quad\hbox{always},\nonumber\\
&k_{j+2}-1\in P &\quad\hbox{if}\quad\{\alpha\}<p^{-1},\nonumber\\
&k_{j+4}-1\in P &\quad\hbox{if}\quad\{\beta\}<p^{-1}.
\label{lcon12}\end{eqnarray}
From (\ref{dk3}) we find that the grid lines $k_{j+3}+m$ are
parallel to the diagonal $\epsilon_j+\epsilon_{j+1}=1$, and
\begin{eqnarray}
&1\!-\!p^{-3}<\lambda_{j+3}(\myb{\epsilon})<2p^{-1}\!+\!1
&\quad\hbox{if}\quad 0<\{\alpha\}+\{\beta\}<p^{-3},\nonumber\\
&0<\lambda_{j+3}(\myb{\epsilon})<2 &\quad\hbox{if}\quad
p^{-3}<\{\alpha\}+\{\beta\}<1,\nonumber\\
&-p^{-3}<\lambda_{j+3}(\myb{\epsilon})<2p^{-1} &\quad\hbox{if}\quad
1<\{\alpha\}+\{\beta\}<2p^{-1},\nonumber\\
&-1<\lambda_{j+3}(\myb{\epsilon})<1 &\quad\hbox{if}\quad
2p^{-1}<\{\alpha\}+\{\beta\}<2,
\label{line3}\end{eqnarray}
where we used the identity $p^{-3}=2p^{-1}-1$.
This means, there can be at most two lines of grid $G_{j+3}$ going
through the inside of the parallelogram. We find that $\lambda_{j+3}=2$
has a solution in $P$ if
$0<\{\alpha\}+\{\beta\}<p^{-3}$;
$\lambda_{j+3}=1$ has a solution in $P$ for
$0<\{\alpha\}+\{\beta\}<2p^{-1}$;
$\lambda_{j+3}=0$ has a solution in $P$ for $1<\{\alpha\}+\{\beta\}$.
These facts can be summarized as
\begin{eqnarray}
&k_{j+3}+1\in P\quad&\hbox{if}\quad\{\alpha\}+\{\beta\}<p^{-3}
=\sqrt{5}-2,\nonumber\\
&k_{j+3}\in
P\phantom{\;\,+1}\quad&\hbox{if}\quad\{\alpha\}+\{\beta\}<2p^{-1}
=\sqrt{5}-1,\nonumber\\
&k_{j+3}-1\in P\quad&\hbox{if}\quad\{\alpha\}+\{\beta\}>1.
\label{lcon3}\end{eqnarray}
At this point one may note the symmetry under
$k_{j+2}\leftrightarrow k_{j+4}$ and $\alpha\leftrightarrow\beta$ in
conditions (\ref{lcon12}) and (\ref{lcon3}).

\subsection{Intersections of the grid lines}

Next, we need to calculate the positions of the intersections. We let
${\myb a}^{n,m}$ denote the intersection of a pair of grid lines in
$G_{j+3}$ and $G_{j+4}$ numbered $k_{j+3}\!-\!1\!+\!m$ and
$k_{j+4}\!-\!1\!+\!n$, ${\myb b}^{n'\!,m}$ the intersection of lines
$k_{j+3}\!-\!1\!+\!m$ and $k_{j+2}\!-\!1\!+\!n'$ in $G_{j+3}$ and
$G_{j+2}$, while ${\myb c}^{n,n'}$ the intersection of lines
$k_{j+4}\!-\!1\!+\!n$ and $k_{j+2}\!-\!1\!+\!n'$ in $G_{j+4}$ and
$G_{j+2}$. The locations of these intersections are found by solving the
corresponding linear equations given in (\ref{linear}). We find
\begin{eqnarray}
&&{\myb a}^{n,m}=
\big(p(\{\beta\}\!-\!n)\!+\!m\!+\!n\!-\!1\!+\!\{\alpha\},\,
p^{-1}(\{\alpha\}\!+\!n\!+\!m\!-\!1)\big),\hspace{0.5in}\label{pa}\\
&&{\myb b}^{n'\!,m}=
\big(p^{-1}(\{\beta\}\!+\!m\!+\!n'\!\!-\!1),\,
p(\{\alpha\}\!-\!n')\!+\!m\!+\!n'\!\!-\!1\!+\!\{\beta\}\big),\label{pb}\\
&&{\myb c}^{n,n'}=\big(pn'\!+\!n\!-\!p\{\alpha\}\!-\!\{\beta\},\,
pn\!+\!n'\!-\!\{\alpha\}\!-\!p\{\beta\}\big).
\label{pc}\end{eqnarray}
Clearly, whenever both components ($\epsilon_j,\epsilon_{j+1}$) of an
intersection are positive and less than 1, it is inside the parallelogram
$P(k_j,k_{j+1})$. This way, we find from (\ref{pa}) the conditions for
the three possible cases, namely 
\begin{eqnarray}
&{\myb a}^{0,1}\in P(k_j,k_{j+1})\quad&\hbox{if}\quad
\{\beta\}<p^{-1}-p^{-1}\{\alpha\},\qquad\qquad\nonumber\\
&{\myb a}^{1,0}\in P(k_j,k_{j+1})\quad&\hbox{if}\quad
\{\beta\}>1-p^{-1}\{\alpha\},\qquad\qquad\nonumber\\
&{\myb a}^{1,1}\in P(k_j,k_{j+1})\quad&\hbox{if}\quad
\cases{\begin{array}{l}
0<\{\alpha\}<p^{-1}\quad\hbox{and}\\
p^{-2}\!-\!p^{-1}\{\alpha\}\!<\!\{\beta\}\!<\!1\!-\!p^{-1}
\{\alpha\}\upstrut.
\end{array}}
\label{cona}\end{eqnarray}
Similarly, from (\ref{pb}) we obtain
\begin{eqnarray}
&{\myb b}^{0,1}\in P(k_j,k_{j+1})\quad&\hbox{if}\quad
\{\beta\}<1-p\{\alpha\},\nonumber\\
&{\myb b}^{1,0}\in P(k_j,k_{j+1})\quad&\hbox{if}\quad
\{\beta\}>p(1-\{\alpha\}),\nonumber\\
&{\myb b}^{1,1}\in P(k_j,k_{j+1})\quad&\hbox{if}\quad
\cases{\begin{array}{l}
0<\{\beta\}<p^{-1}\quad\hbox{and}\\
p^{-2}\!-\!p^{-1}\{\beta\}\!<\!\{\alpha\}\!<\!1\!-\!p^{-1}
\{\beta\}\upstrut,
\end{array}}
\label{conb}\end{eqnarray}
and from (\ref{pc}) we get
\begin{eqnarray}
&{\myb c}^{0,1}\in P(k_j,k_{j+1})\quad&\hbox{if}\quad
p^{-1}\!-\!p\{\alpha\}\!<\!\{\beta\}\!<\!p^{-1}(1\!-\!\{\alpha\}),
\nonumber\\
&{\myb c}^{1,0}\in P(k_j,k_{j+1})\quad&\hbox{if}\quad
p^{-2}\!-\!p^{-1}\{\alpha\}\!<\!\{\beta\}\!<\!1\!-\!p\{\alpha\},
\nonumber\\
&{\myb c}^{1,1}\in P(k_j,k_{j+1})\quad&\hbox{if}\quad
  \{\beta\}>\max(1\!-\!p^{-1}\{\alpha\},p\!-\!p\{\alpha\}).
\label{conc}\end{eqnarray}
Note the symmetry between (\ref{cona}) and (\ref{conb}) and between the
first two lines of (\ref{conc}) under $\alpha\leftrightarrow\beta$ and
implicitly $k_{j+2}\leftrightarrow k_{j+4}$.

Three lines cannot exactly meet in a common intersection, as was shown
by de Bruijn \cite{Bruijn1} for a regular pentagrid. However, they can
meet arbitrarily close and the theoretical limiting conditions of triple
intersection
\begin{eqnarray}
&``\,{\myb a}^{1,0}={\myb b}^{1,0}={\myb c}^{1,1}\,"\quad
&\Longleftrightarrow\quad
\{\alpha\}+\{\beta\}=p,\nonumber\\
&``\,{\myb a}^{0,1}={\myb b}^{1,1}={\myb c}^{0,1}\,"\quad
&\Longleftrightarrow\quad
\{\alpha\}+\{\beta\}=p^{-1},\nonumber\\
&``\,{\myb a}^{1,1}={\myb b}^{0,1}={\myb c}^{1,0}\,"\quad
&\Longleftrightarrow\quad
\{\alpha\}+\{\beta\}=p^{-1},
\label{contri}\end{eqnarray}
play a role in the following subsection.

\subsection{The twenty-four allowed configurations}

Next, we use (\ref{lcon12}), (\ref{lcon3}), (\ref{cona})--(\ref{contri})
to study how the configuration $C(m)$ of parallelogram
$P(k_j,k_{j+1})$ depends on the values of
$\{\alpha\}=\{\hat\alpha(k_{j+1})\}$ and $\{\beta\}=\{\hat\beta(k_j)\}$. We
show the various cases in Fig.~\ref{fig6} for $j\!=\!0$. For $j\!\ne\!0$
we need to rotate each picture $j$ times $72^\circ$.
\begin{figure}[tbph]
\vskip0in\hskip0pt\epsfclipon
\epsfxsize=0.22\hsize\hfil
\epsfbox{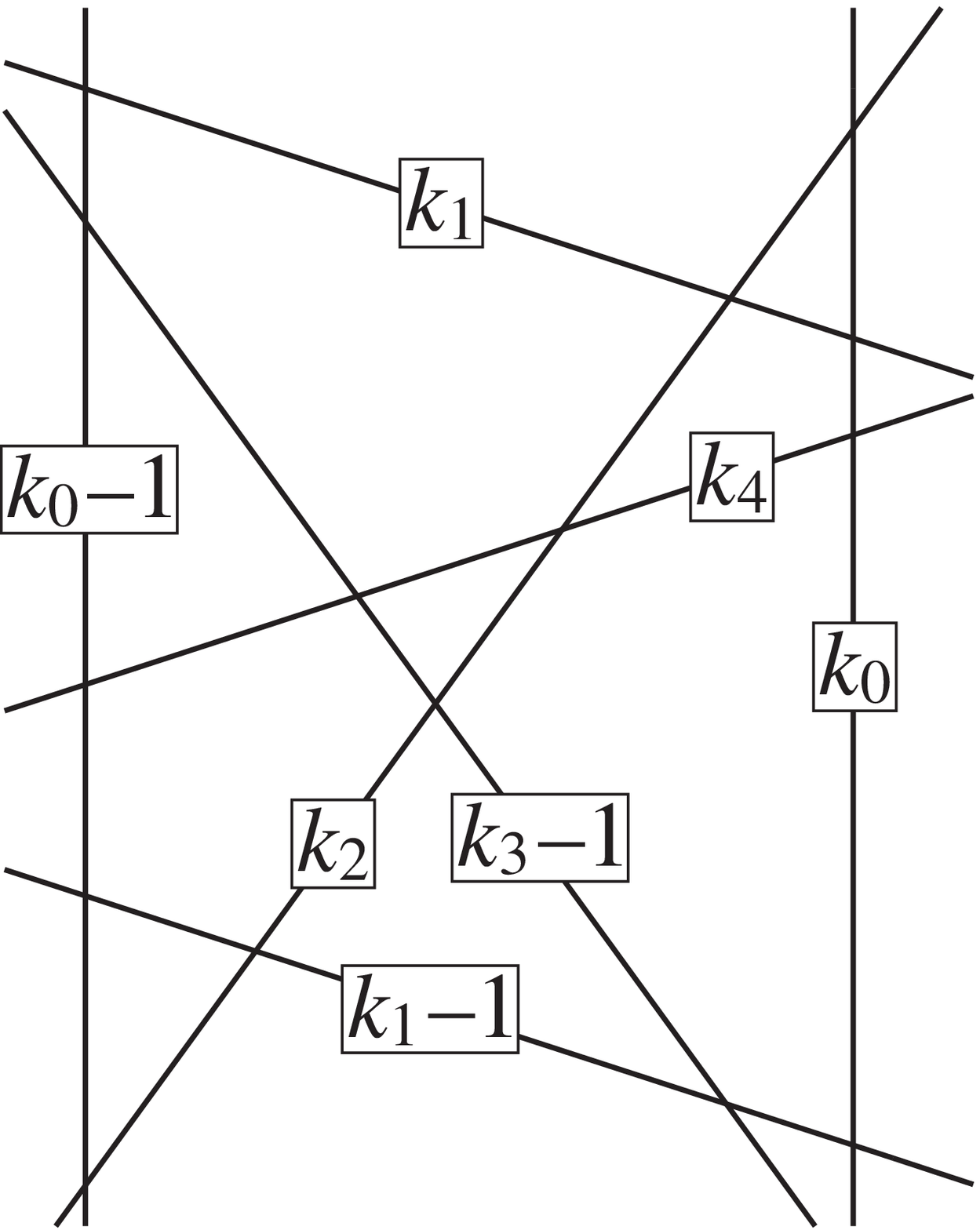}
\epsfxsize=0.22\hsize\hfil
\epsfbox{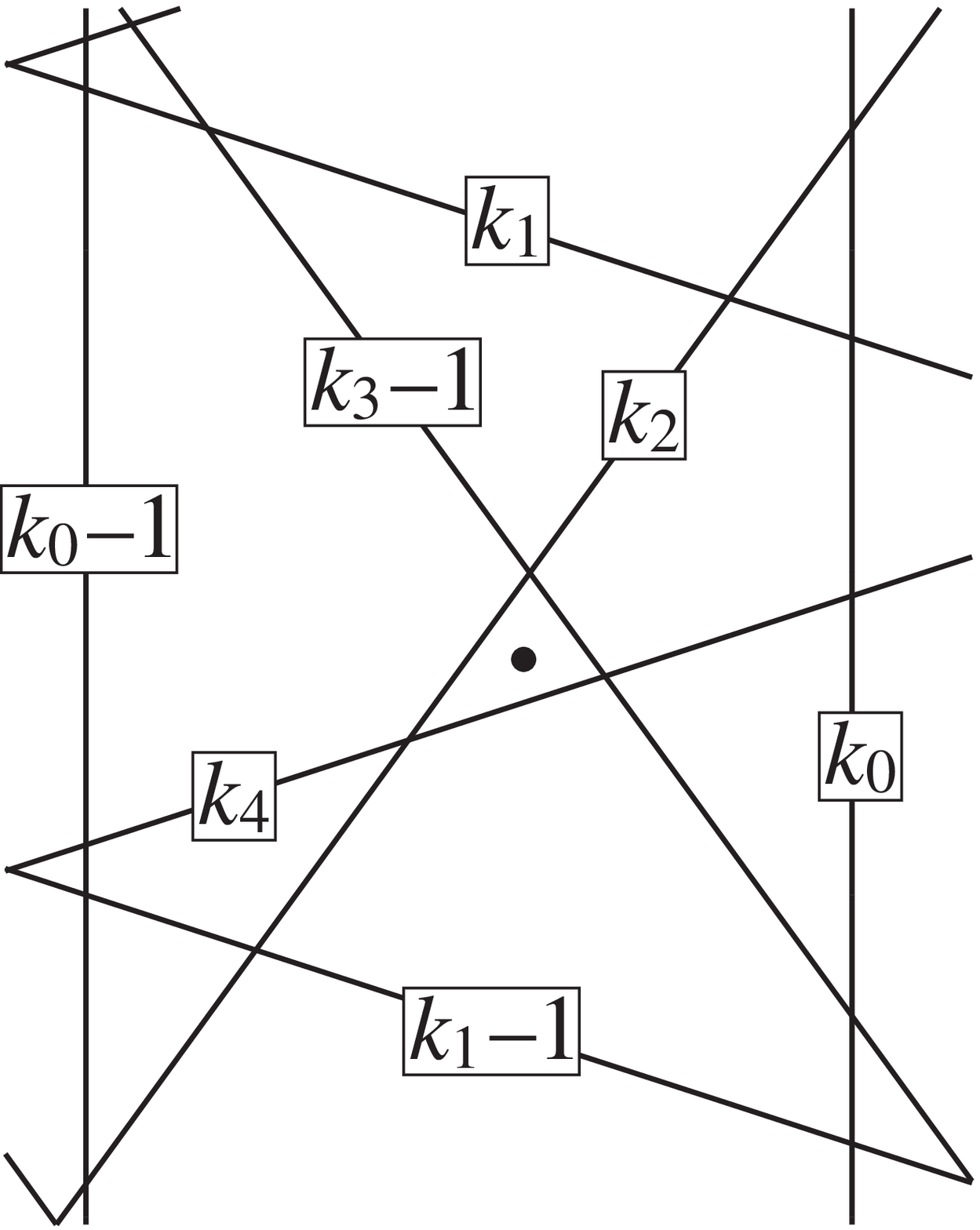}
\epsfxsize=0.22\hsize\hfil
\epsfbox{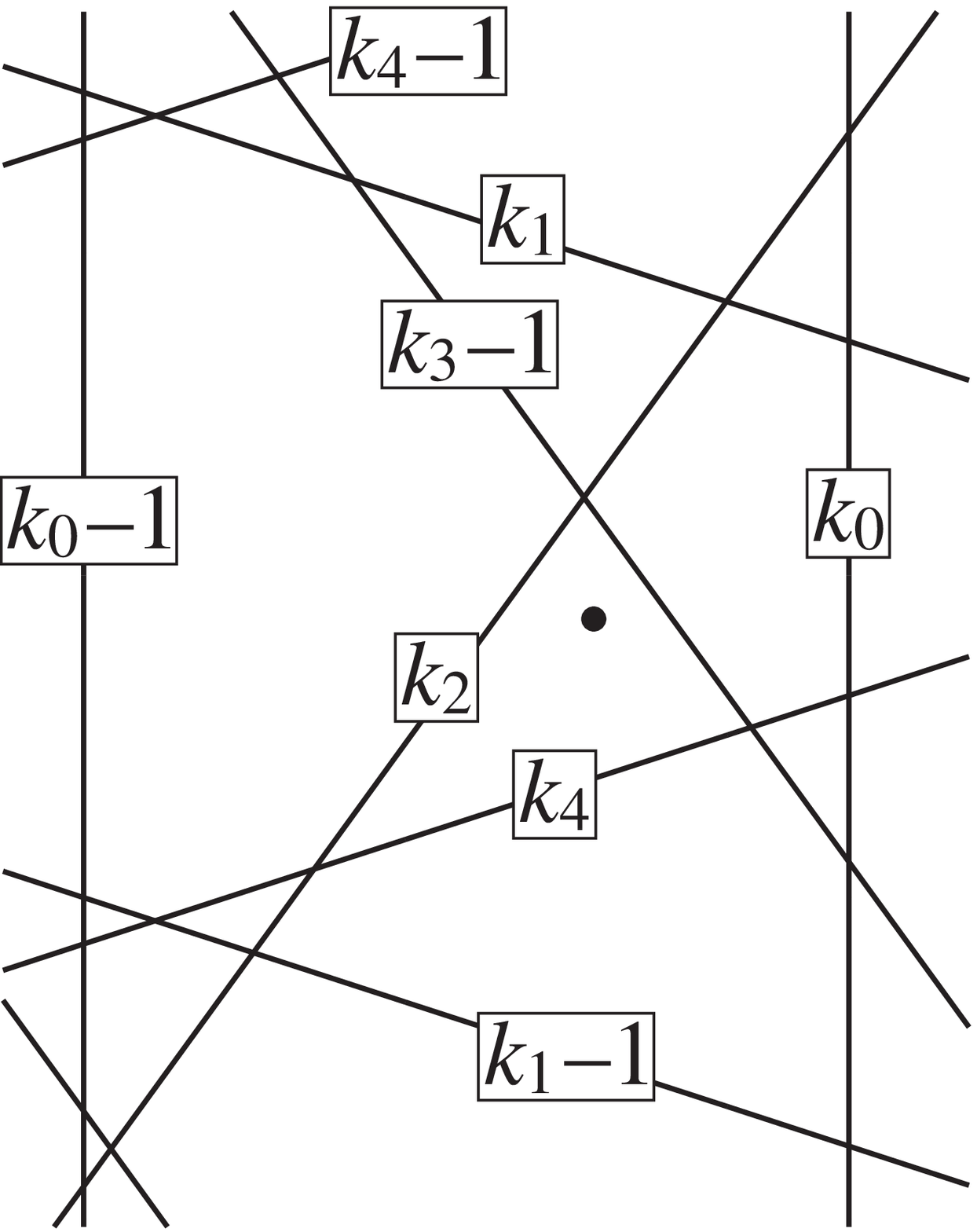}
\epsfxsize=0.22\hsize\hfil
\epsfbox{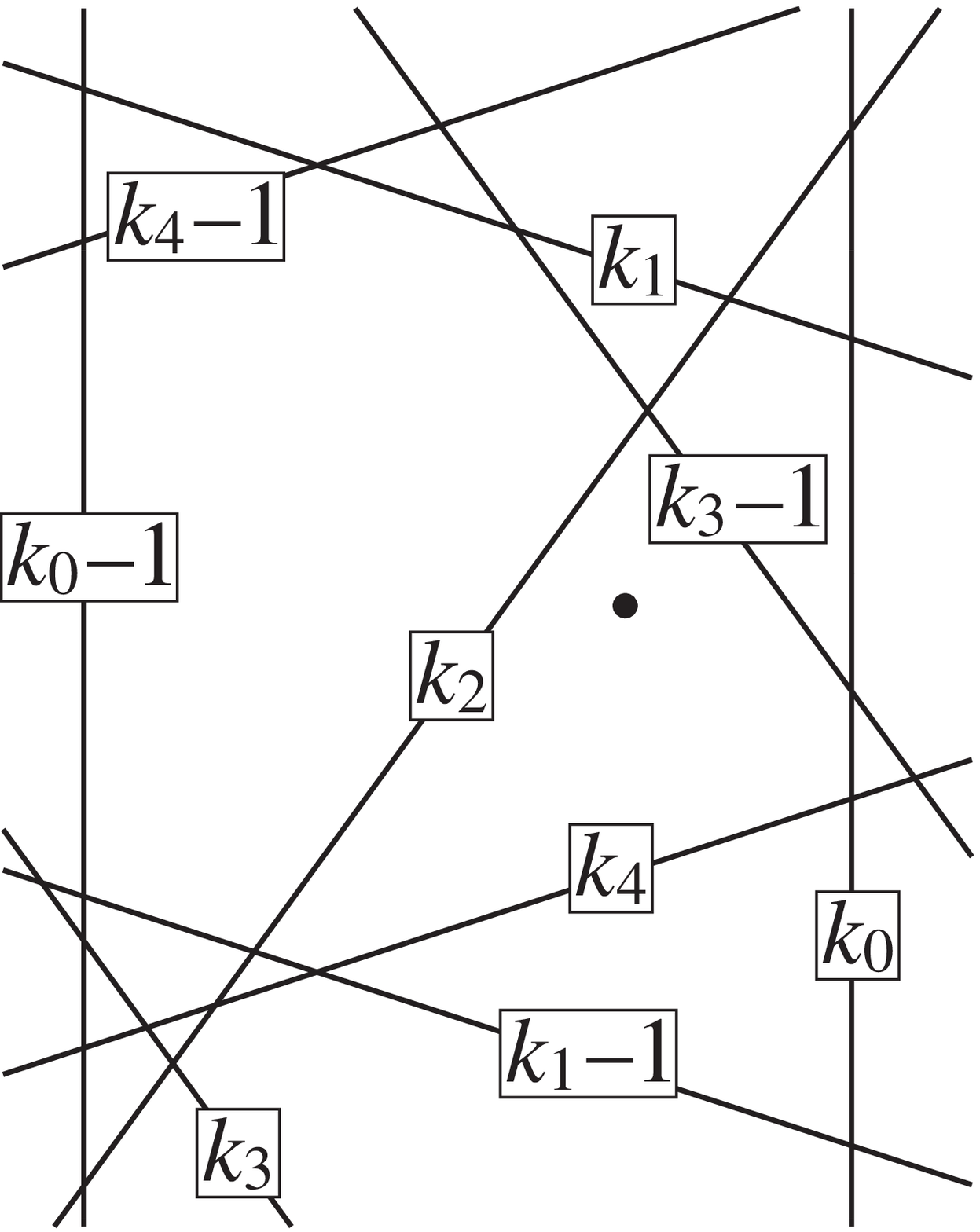}
\vskip0pt
\hbox to\hsize{\hspace*{20pt}\tiny
(a) \vtop{\hbox{C(1): $N\!=\!7$}
\hbox{$N_{\rm o}\!=\!4$, $N_{\rm e}\!=\!3$}}\hfil
(b) \vtop{\hbox{C(2): $N\!=\!7$}
\hbox{$N_{\rm o}\!=\!3$, $N_{\rm e}\!=\!4$}}\hfil
(c) \vtop{\hbox{C(3): $N\!=\!8$}
\hbox{$N_{\rm o}\!=\!3$, $N_{\rm e}\!=\!5$}}\hfil
(d) \vtop{\hbox{C(4): $N\!=\!6$}
\hbox{$N_{\rm o}\!=\!3$, $N_{\rm e}\!=\!3$}}\hfil
\hspace*{-18pt}}
\vskip0.1in\hskip0pt
\epsfxsize=0.22\hsize\hfil
\epsfbox{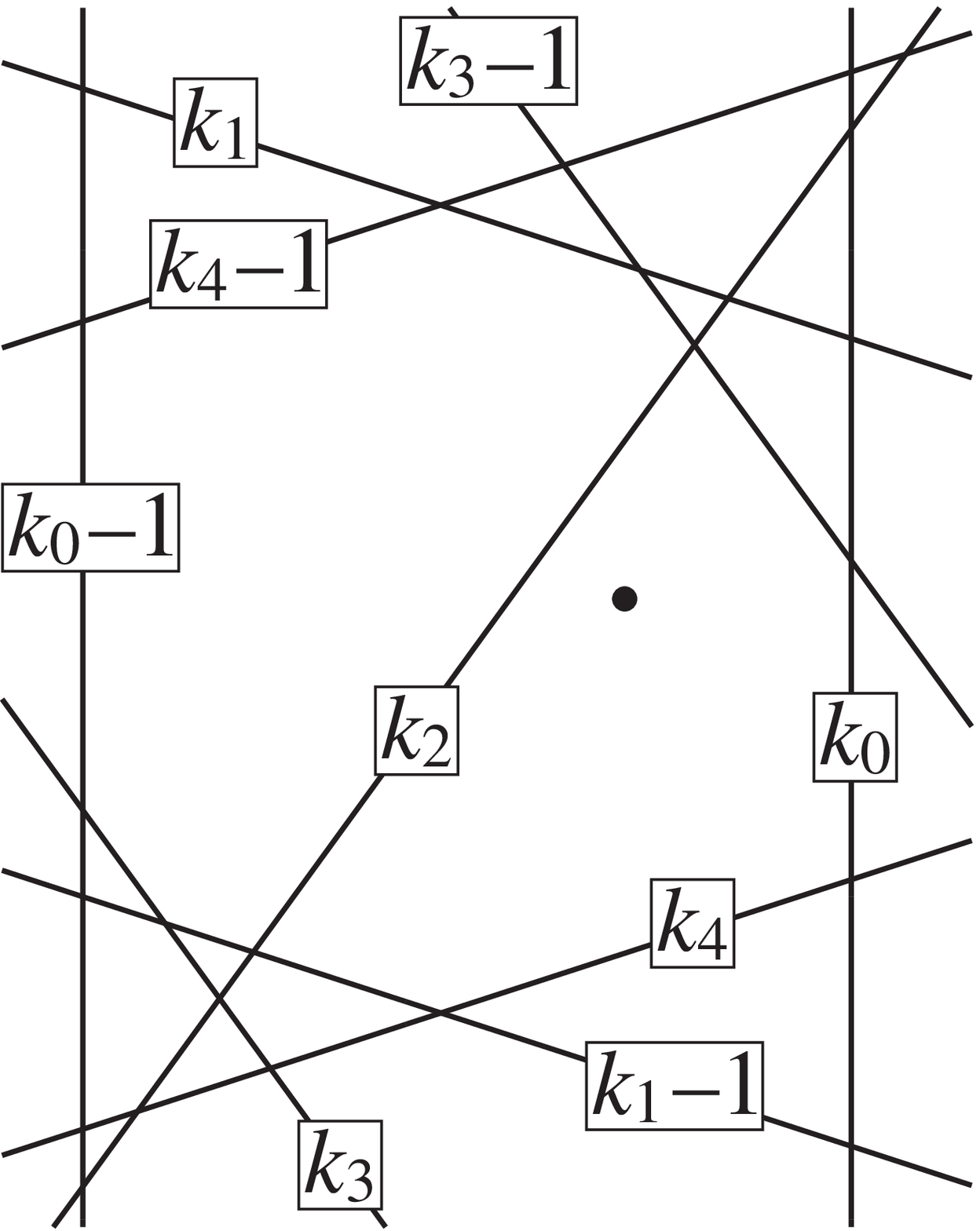}
\epsfxsize=0.22\hsize\hfil
\epsfbox{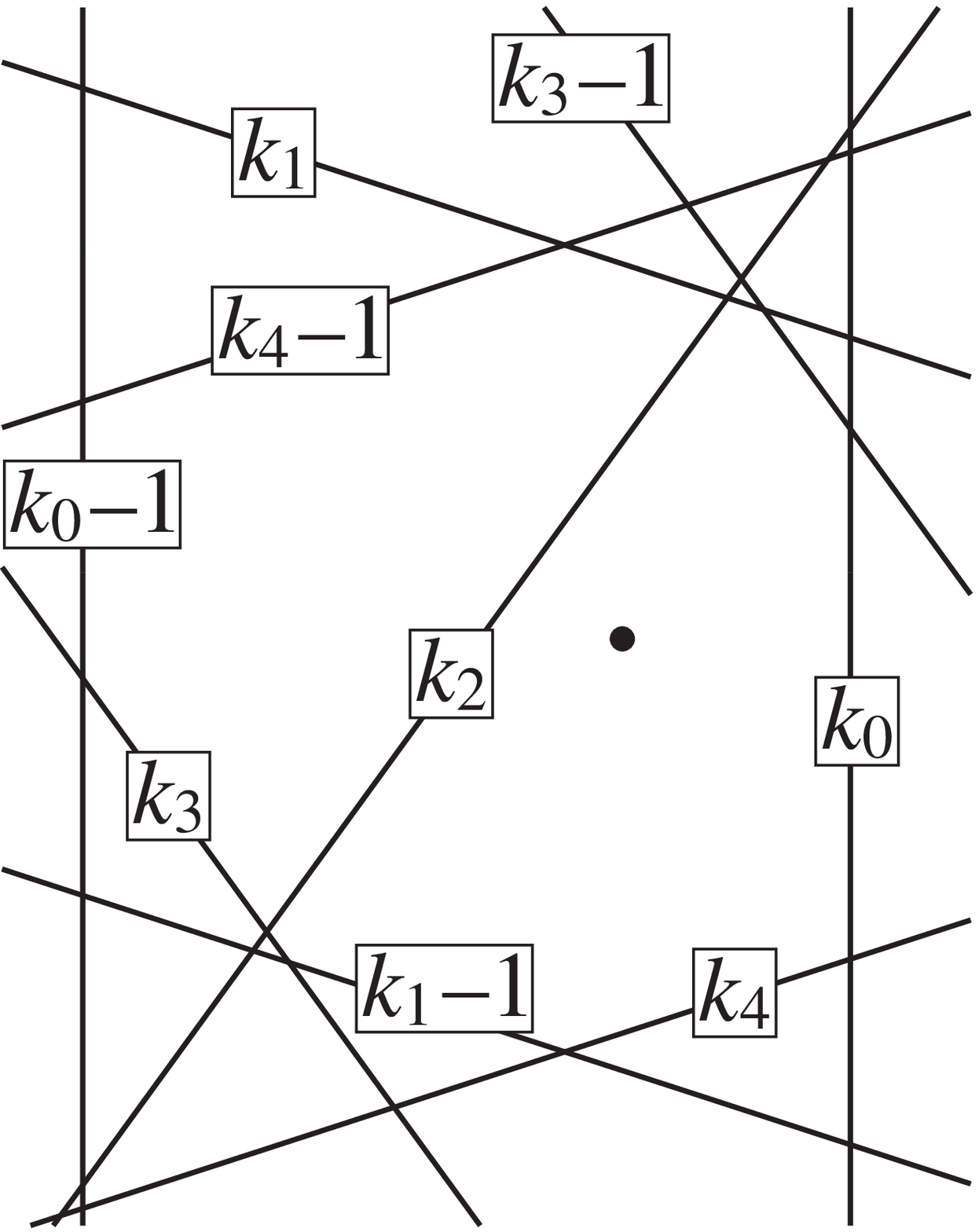}
\epsfxsize=0.22\hsize\hfil
\epsfbox{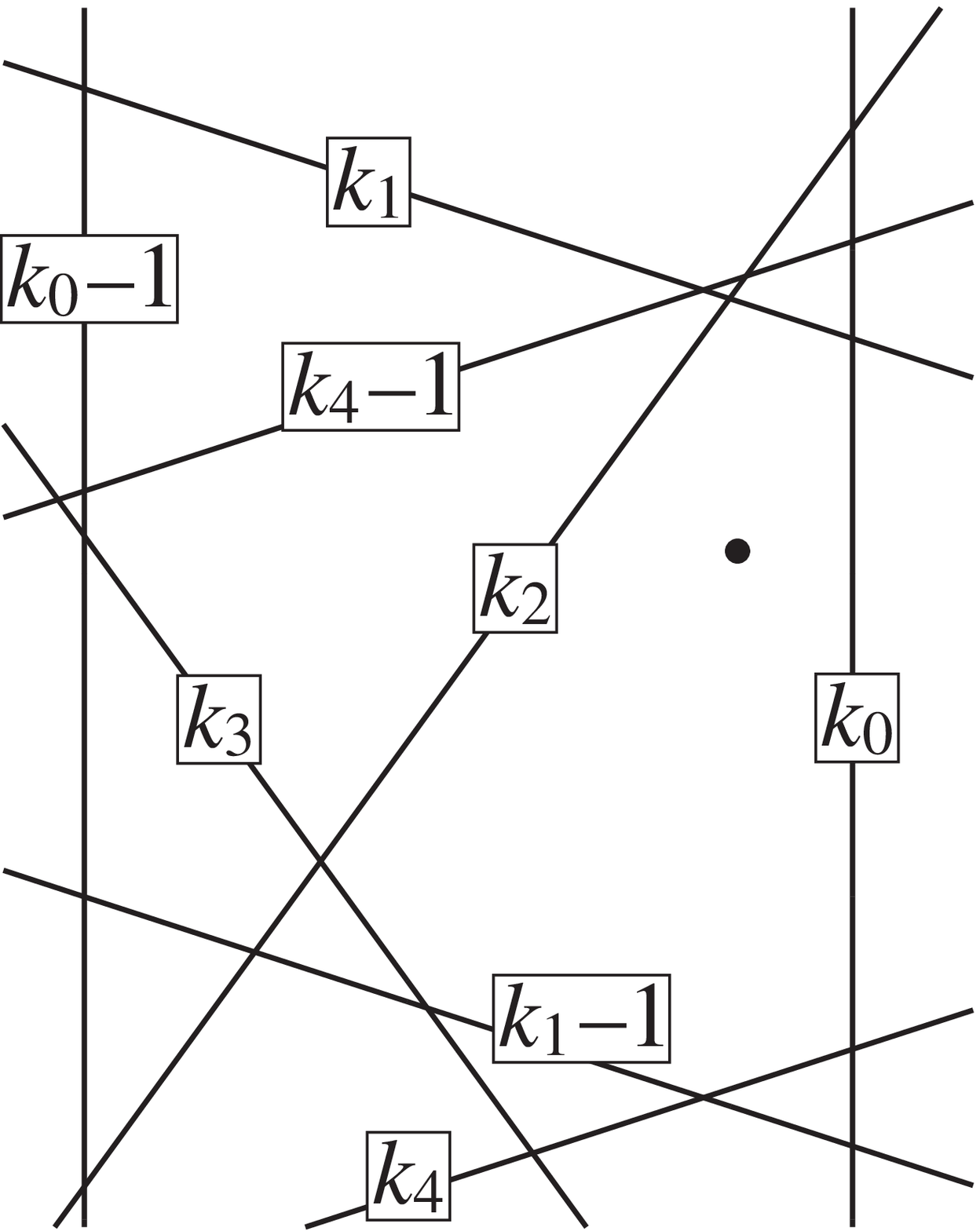}
\epsfxsize=0.22\hsize\hfil
\epsfbox{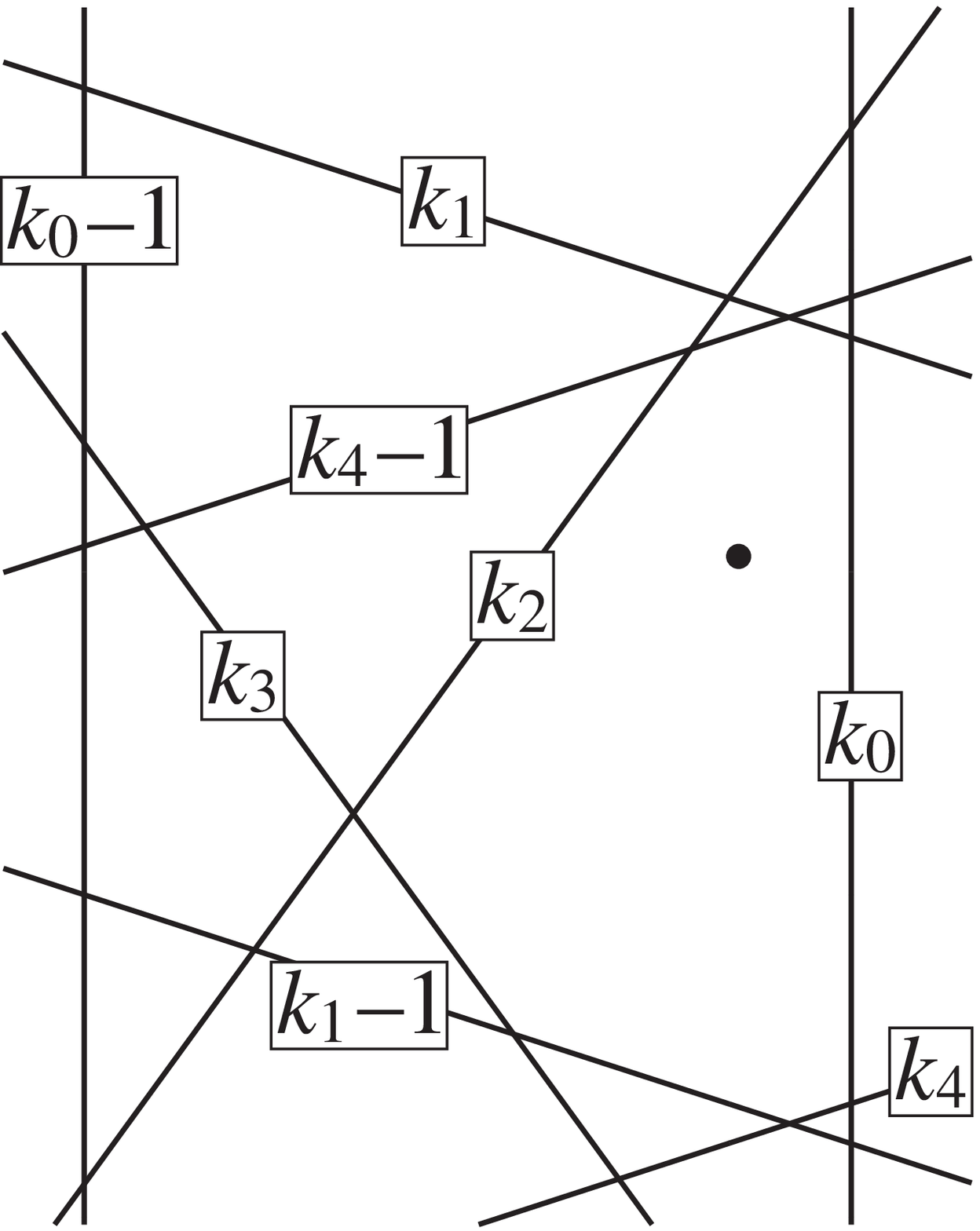}
\vskip0pt
\hbox to\hsize{\hspace*{20pt}\tiny
(e) \vtop{\hbox{C(5): $N\!=\!7$}
\hbox{$N_{\rm o}\!=\!3$, $N_{\rm e}\!=\!4$}}\hfil
(f) \vtop{\hbox{C(6): $N\!=\!7$}
\hbox{$N_{\rm o}\!=\!4$, $N_{\rm e}\!=\!3$}}\hfil
(g) \vtop{\hbox{C(7): $N\!=\!6$}
\hbox{$N_{\rm o}\!=\!3$, $N_{\rm e}\!=\!3$}}\hfil
(h) \vtop{\hbox{C(8): $N\!=\!8$}
\hbox{$N_{\rm o}\!=\!5$, $N_{\rm e}\!=\!3$}}\hfil
\hspace*{-18pt}}
\vskip0.1in\hskip0pt
\epsfxsize=0.22\hsize\hfil
\epsfbox{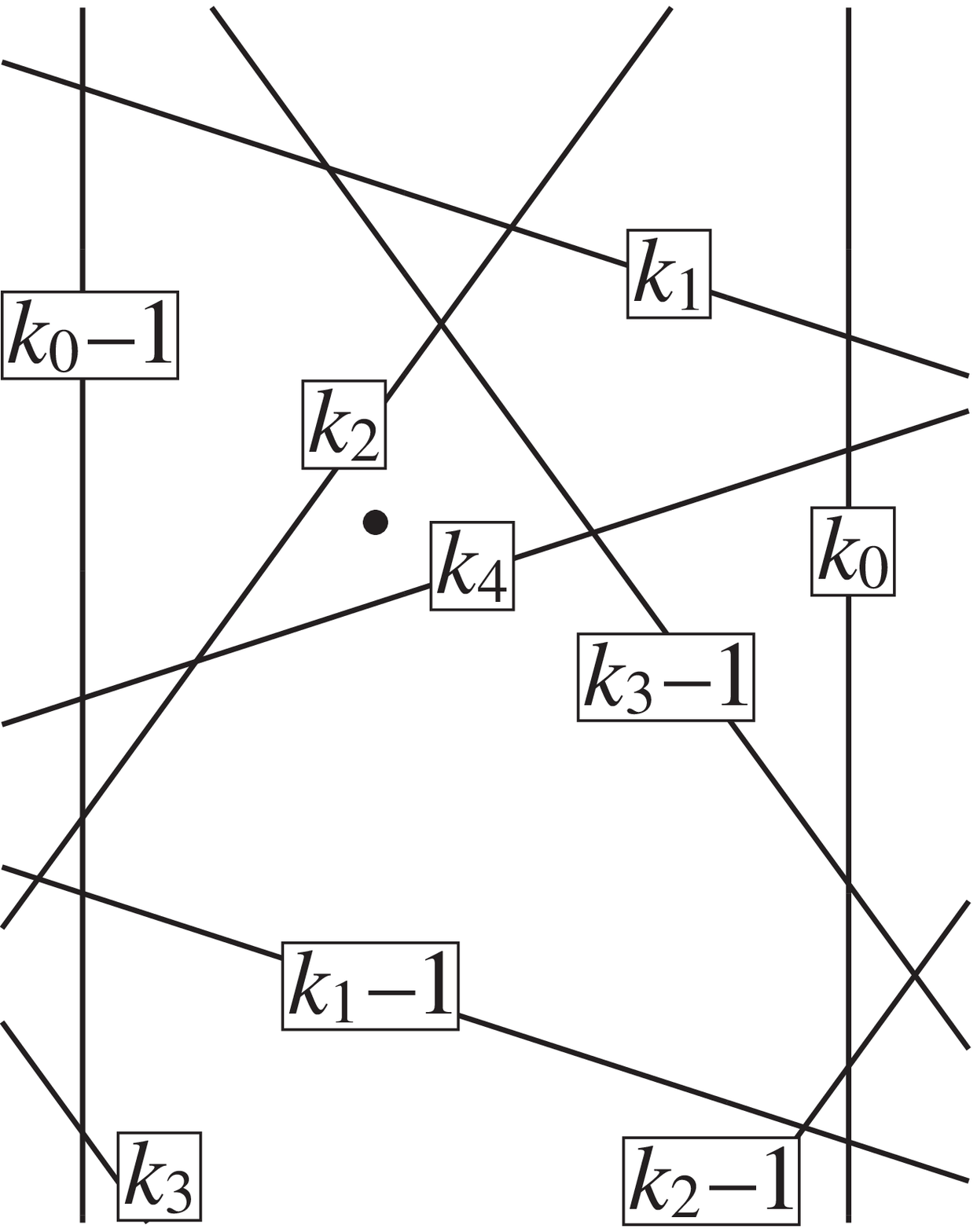}
\epsfxsize=0.22\hsize\hfil
\epsfbox{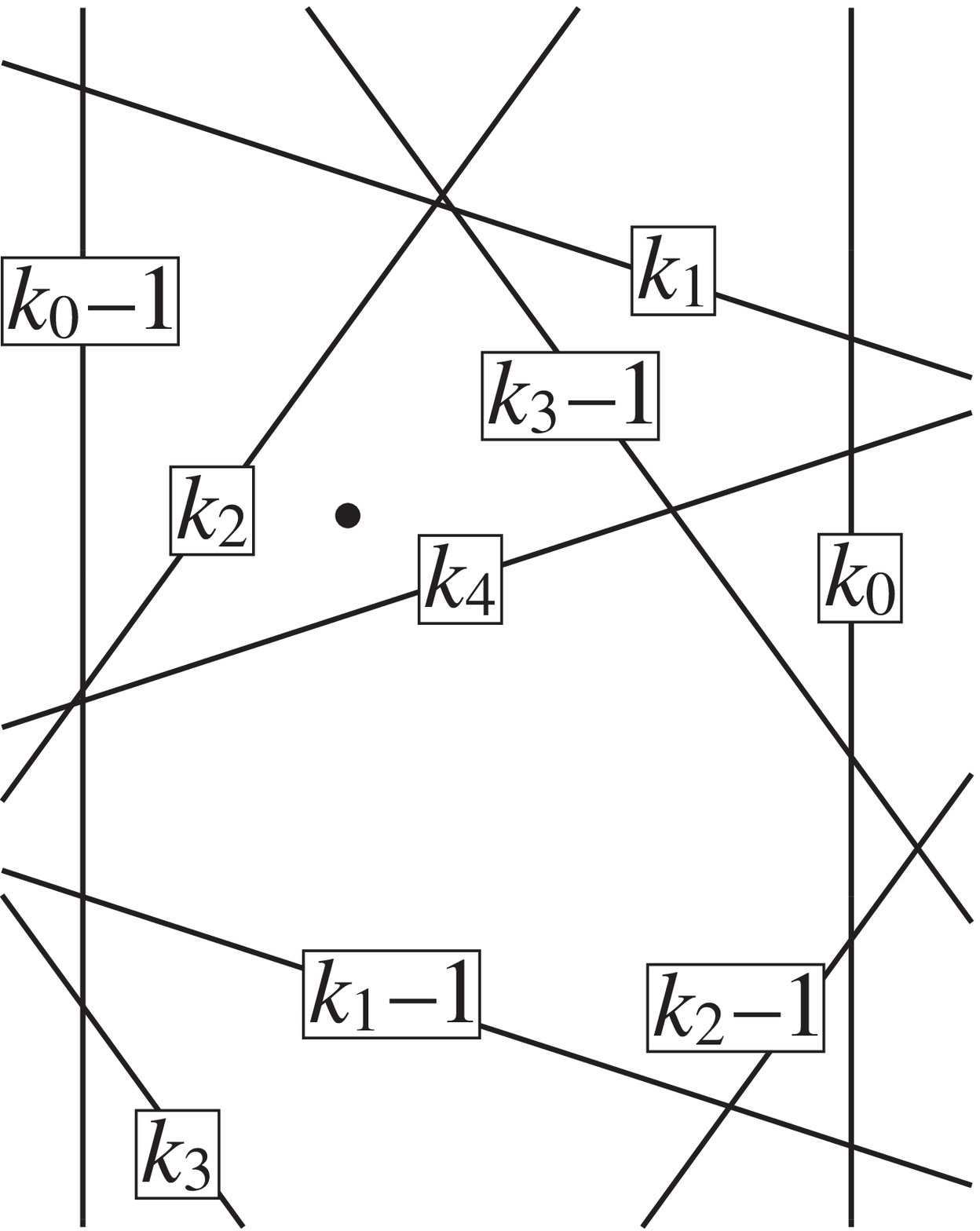}
\epsfxsize=0.22\hsize\hfil
\epsfbox{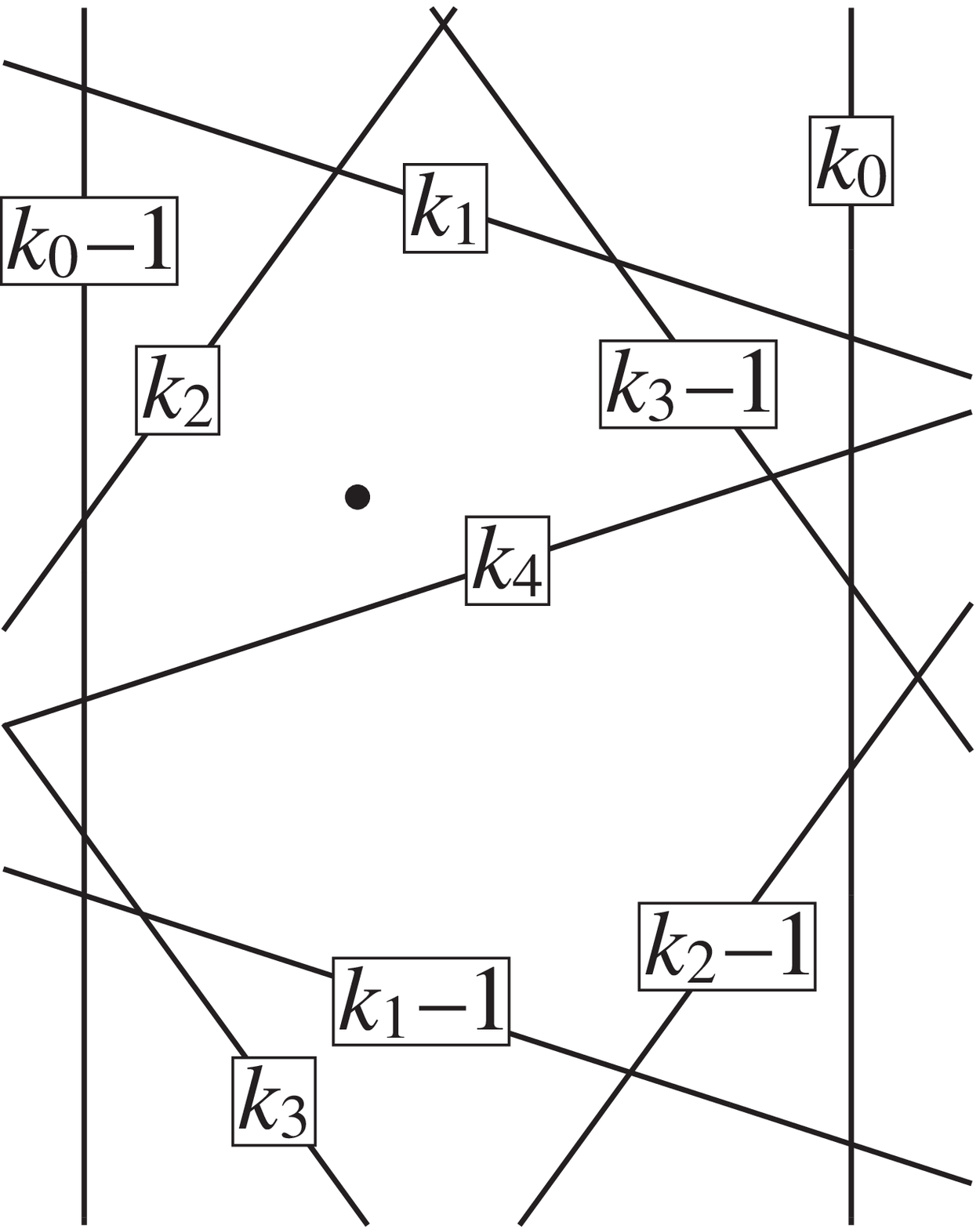}
\epsfxsize=0.22\hsize\hfil
\epsfbox{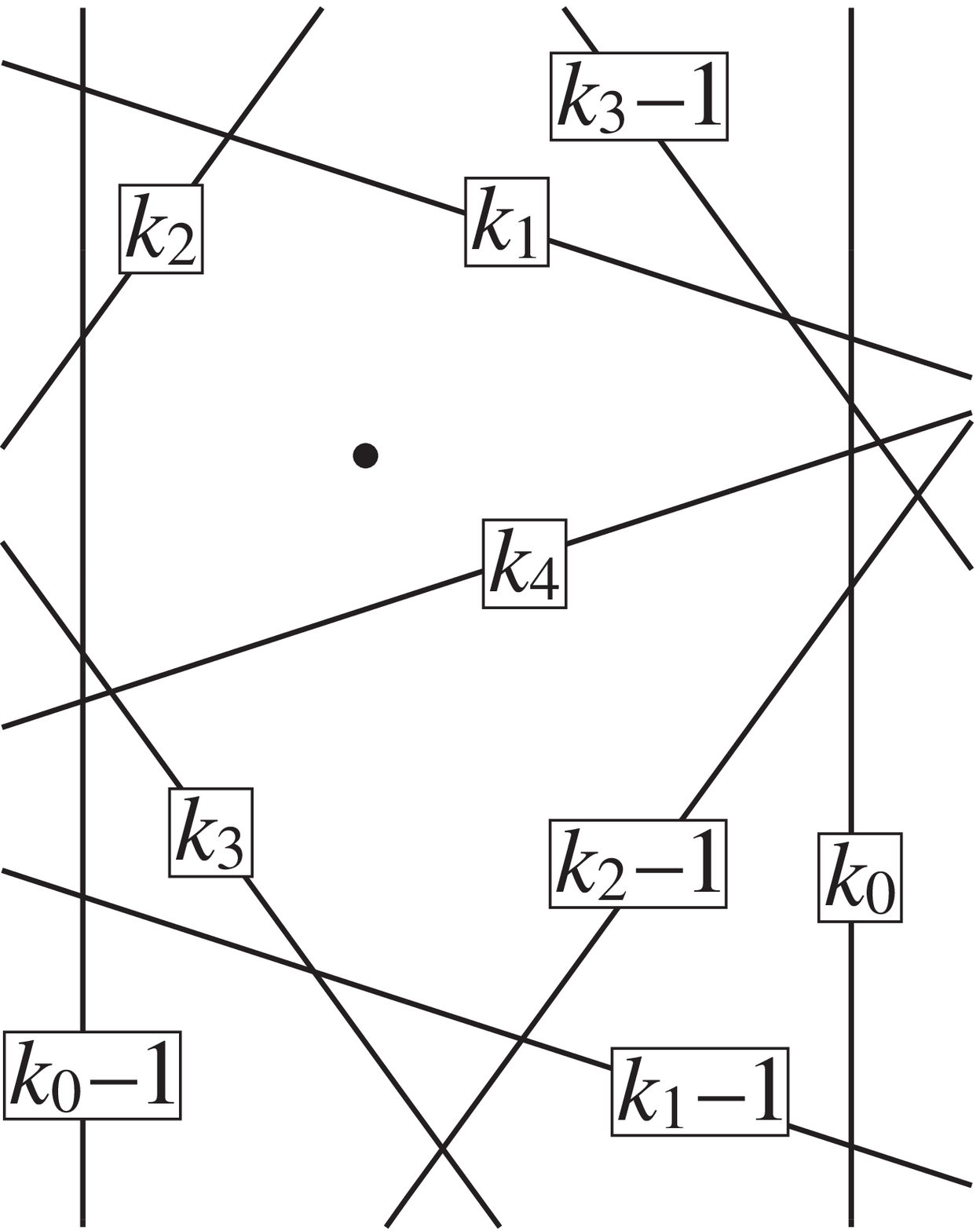}
\vskip0pt
\hbox to\hsize{\hspace*{20pt}\tiny
(i) \vtop{\hbox{C(9): $N\!=\!8$}
\hbox{$N_{\rm o}\!=\!3$, $N_{\rm e}\!=\!5$}}\hfil
(j) \vtop{\hbox{C(10): $N\!=\!6$}
\hbox{$N_{\rm o}\!=\!3$, $N_{\rm e}\!=\!3$}}\hfil
(k) \vtop{\hbox{C(11): $N\!=\!7$}
\hbox{$N_{\rm o}\!=\!3$, $N_{\rm e}\!=\!4$}}\hfil
(l) \vtop{\hbox{C(12): $N\!=\!7$}
\hbox{$N_{\rm o}\!=\!4$, $N_{\rm e}\!=\!3$}}\hfil
\hspace*{-18pt}}
\vskip0.1in
\hskip0pt\caption{The spin configurations C($m$), $m=1,\ldots,24$,
for parallelogram $P(k_0,k_1)$ are shown in (a) through (x), where $N$
denotes the total number of spin sites inside $P(k_0,k_1)$, $N_{\rm o}$
the number of odd spins sites, and $N_{\rm e}$ the number of even spin
sites. The mesh that contains the dot is the special even site whose
integer vector is the reference integer vector for $P(k_0,k_1)$, such that
$\partial{\vec K}=(0,0,0)$. For $\{\alpha\}+\{\beta\}>p$, such a site does
not exist, as seen in 6(a). For $\{\alpha\}+\{\beta\}<1$, this site is the
mesh right below the upper right corner of $P$.}
\label{fig6}
\end{figure}
\addtocounter{figure}{-1}
\begin{figure}[tbph]
\vskip0.1in\hskip0pt\epsfclipon
\epsfxsize=0.22\hsize\hfil
\epsfbox{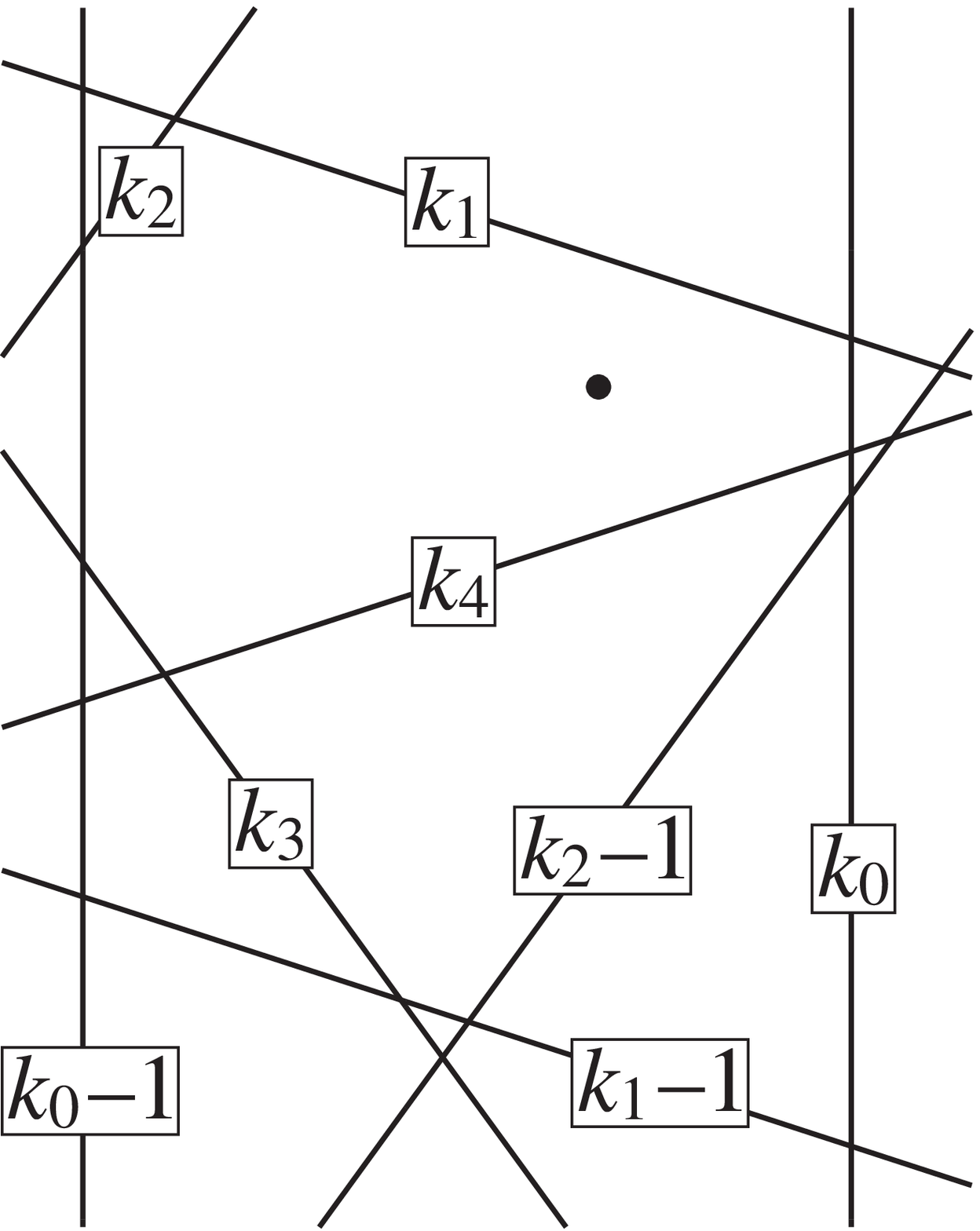}
\epsfxsize=0.22\hsize\hfil
\epsfbox{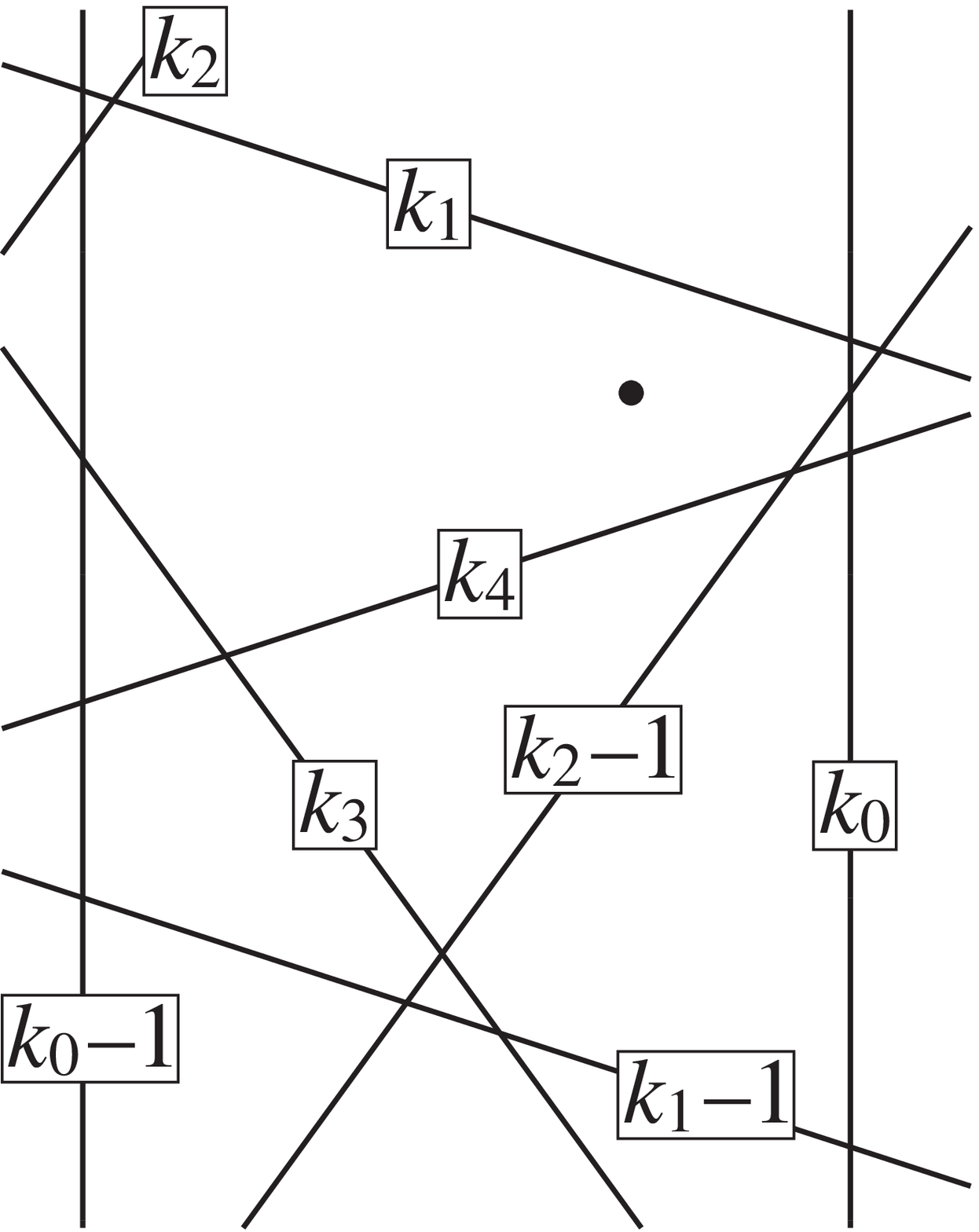}
\epsfxsize=0.22\hsize\hfil
\epsfbox{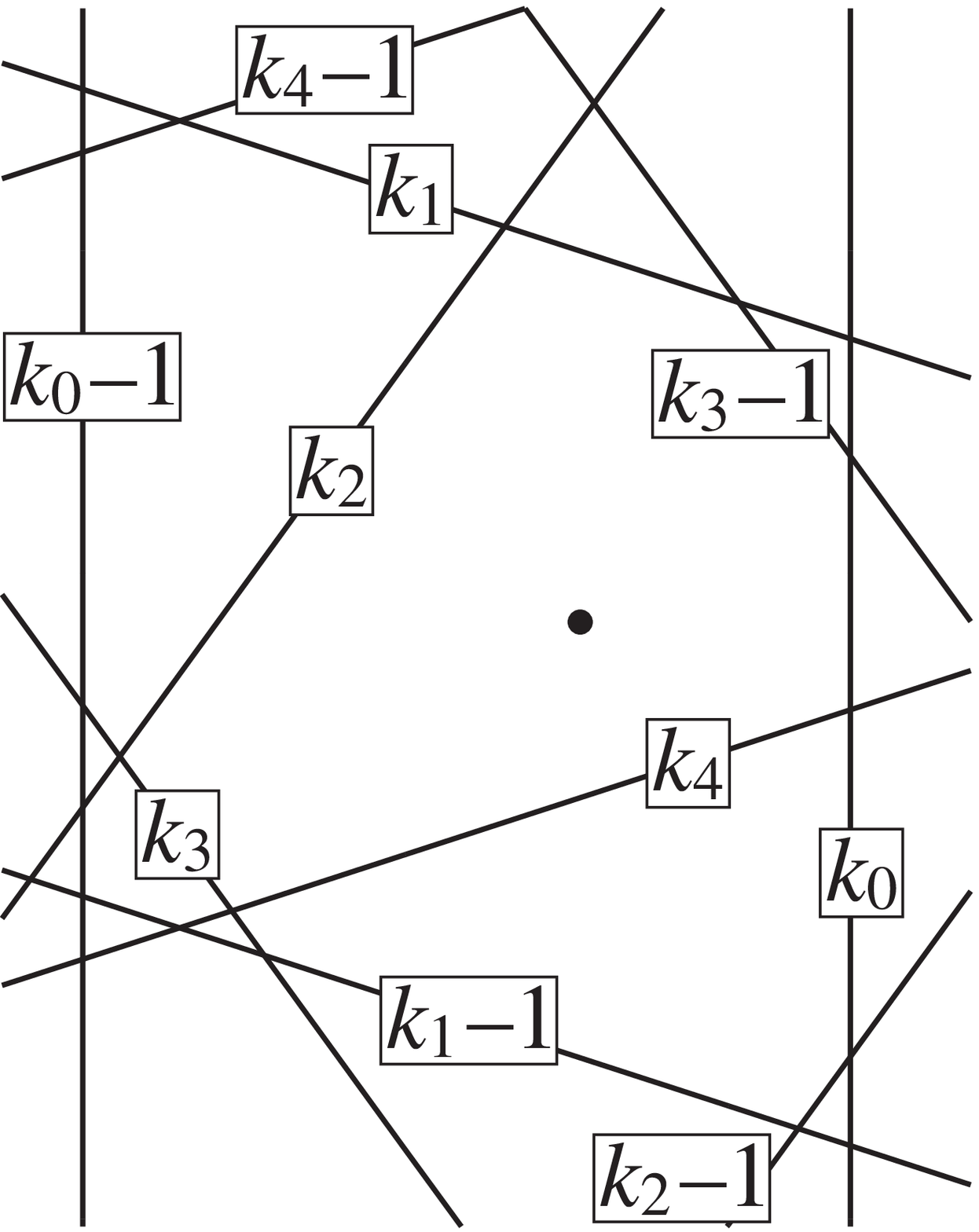}
\epsfxsize=0.22\hsize\hfil
\epsfbox{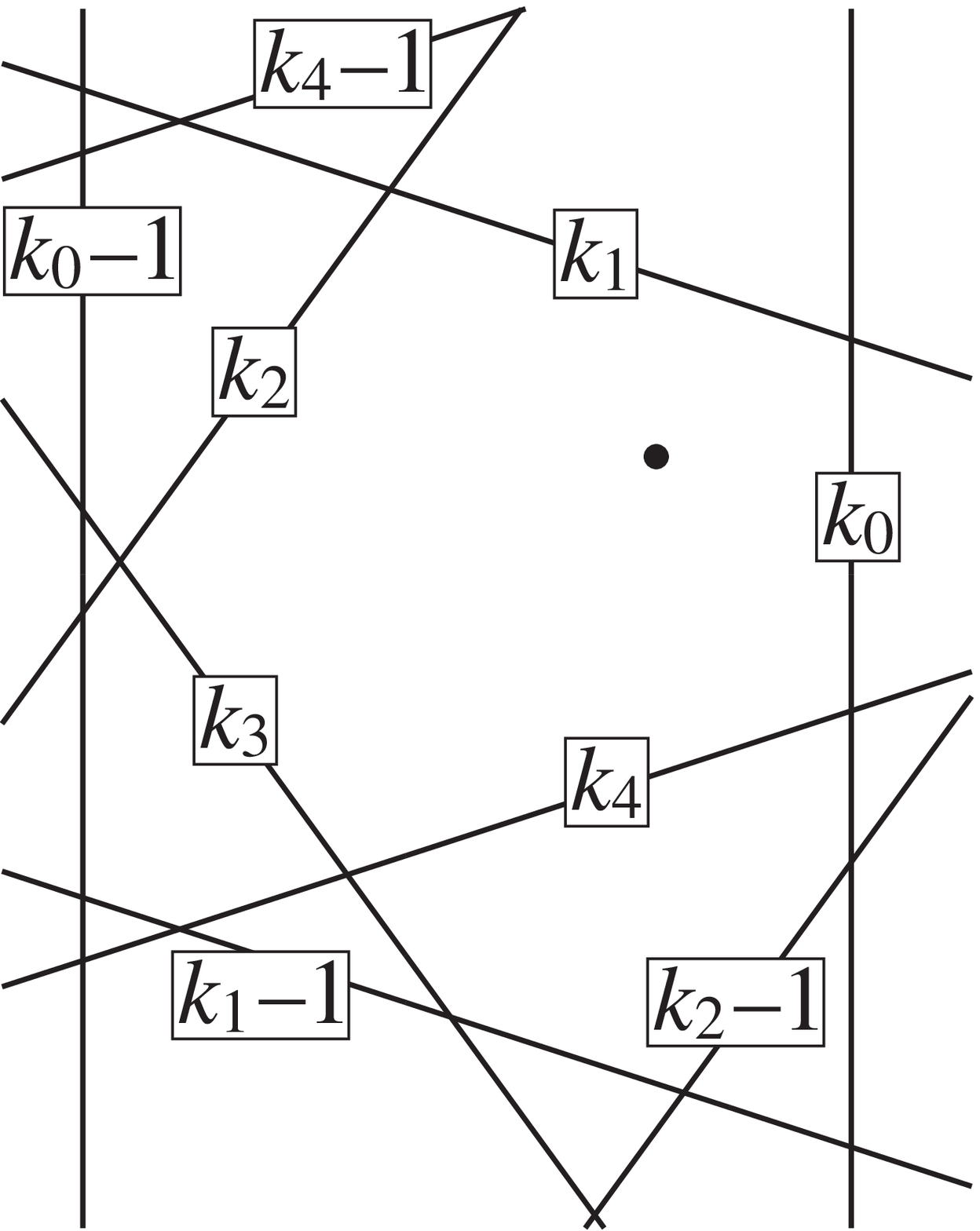}
\vskip0pt
\hbox to\hsize{\hspace*{20pt}\tiny
(m) \vtop{\hbox{C(13): $N\!=\!6$}
\hbox{$N_{\rm o}\!=\!3$, $N_{\rm e}\!=\!3$}}\hfil
(n) \vtop{\hbox{C(14): $N\!=\!8$}
\hbox{$N_{\rm o}\!=\!5$, $N_{\rm e}\!=\!3$}}\hfil
(o) \vtop{\hbox{C(15): $N\!=\!9$}
\hbox{$N_{\rm o}\!=\!4$, $N_{\rm e}\!=\!5$}}\hfil
(p) \vtop{\hbox{C(16): $N\!=\!8$}
\hbox{$N_{\rm o}\!=\!3$, $N_{\rm e}\!=\!5$}}\hfil
\hspace*{-18pt}}
\vskip0.1in\hskip0pt
\epsfxsize=0.22\hsize\hfil
\epsfbox{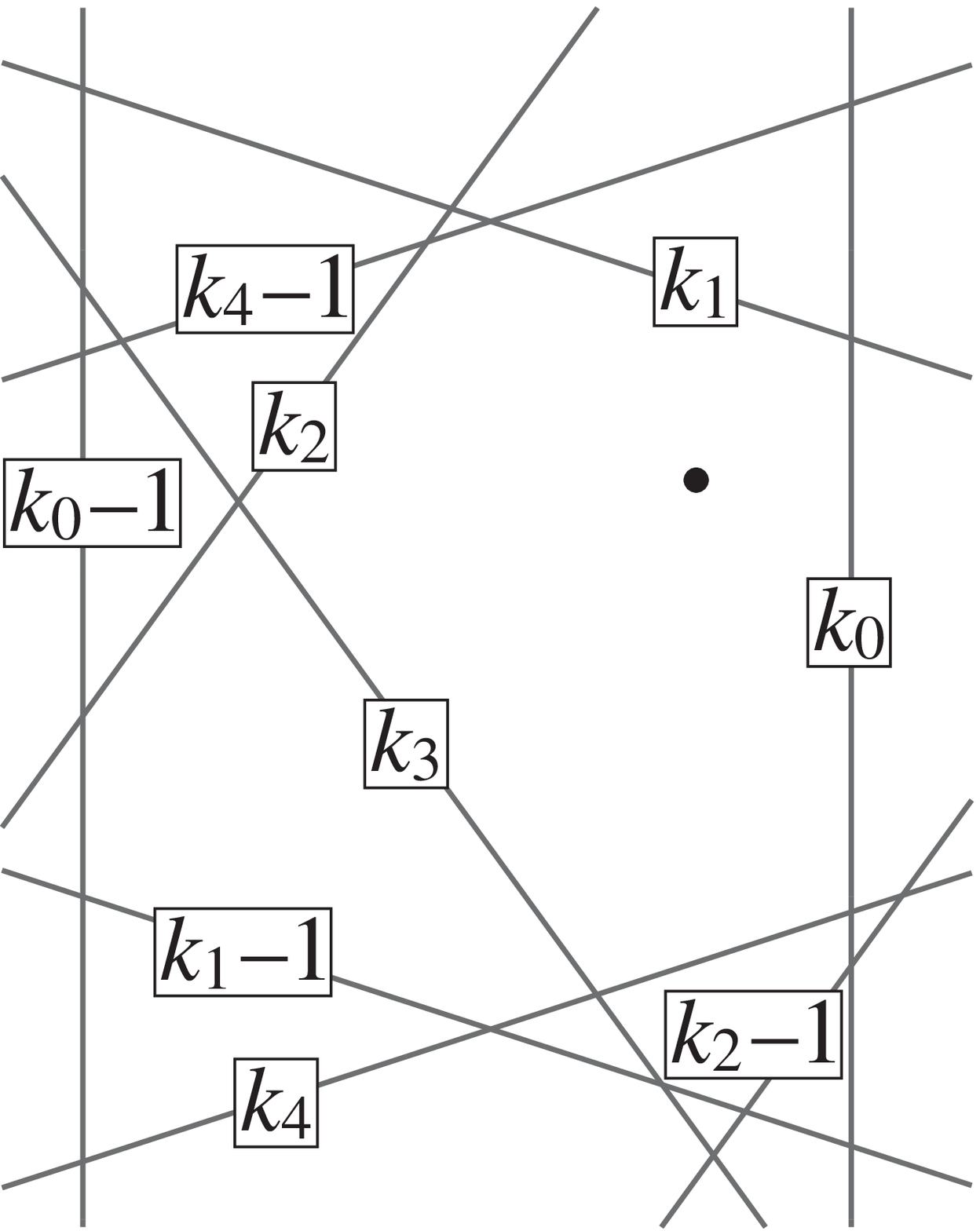}
\epsfxsize=0.22\hsize\hfil
\epsfbox{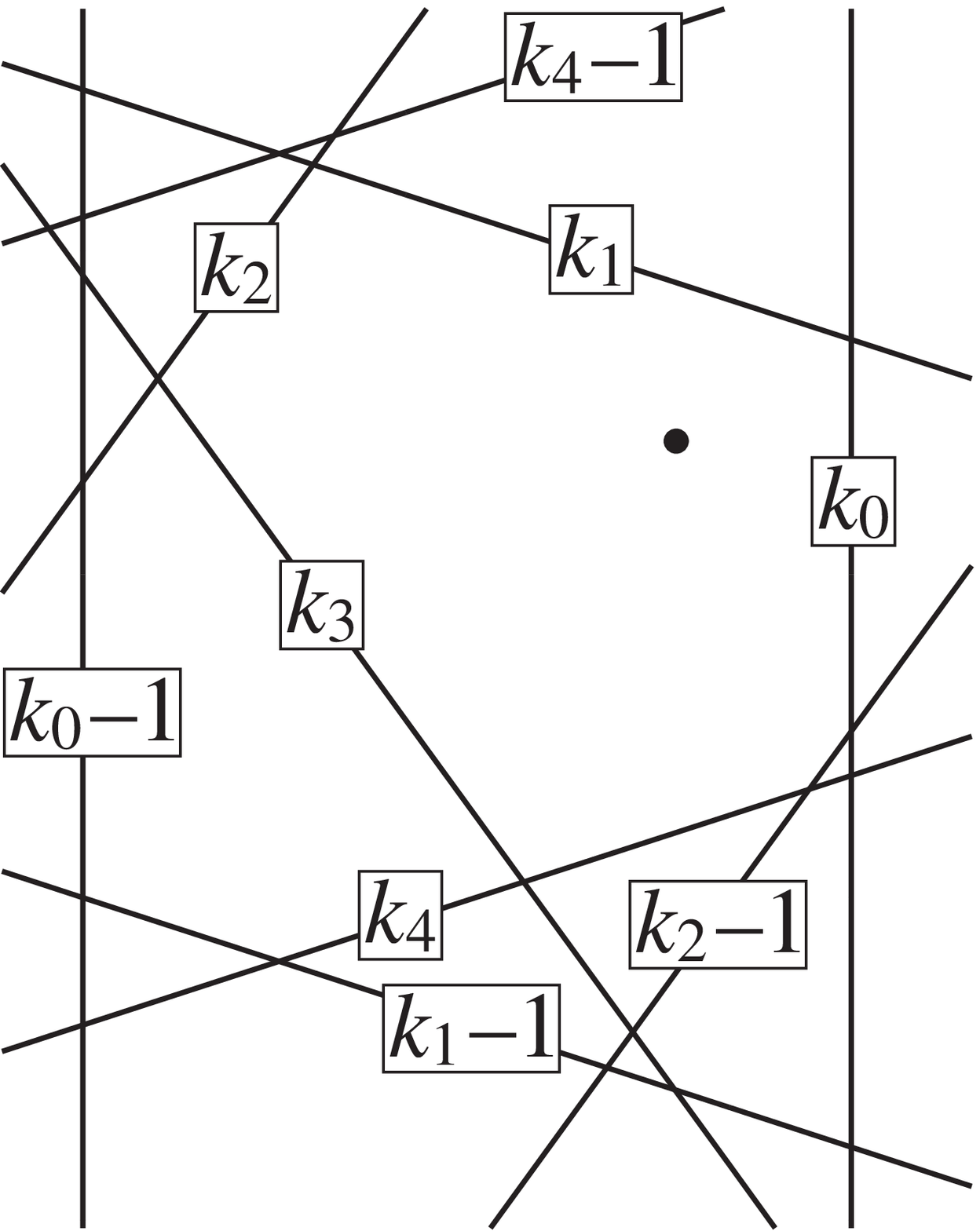}
\epsfxsize=0.22\hsize\hfil
\epsfbox{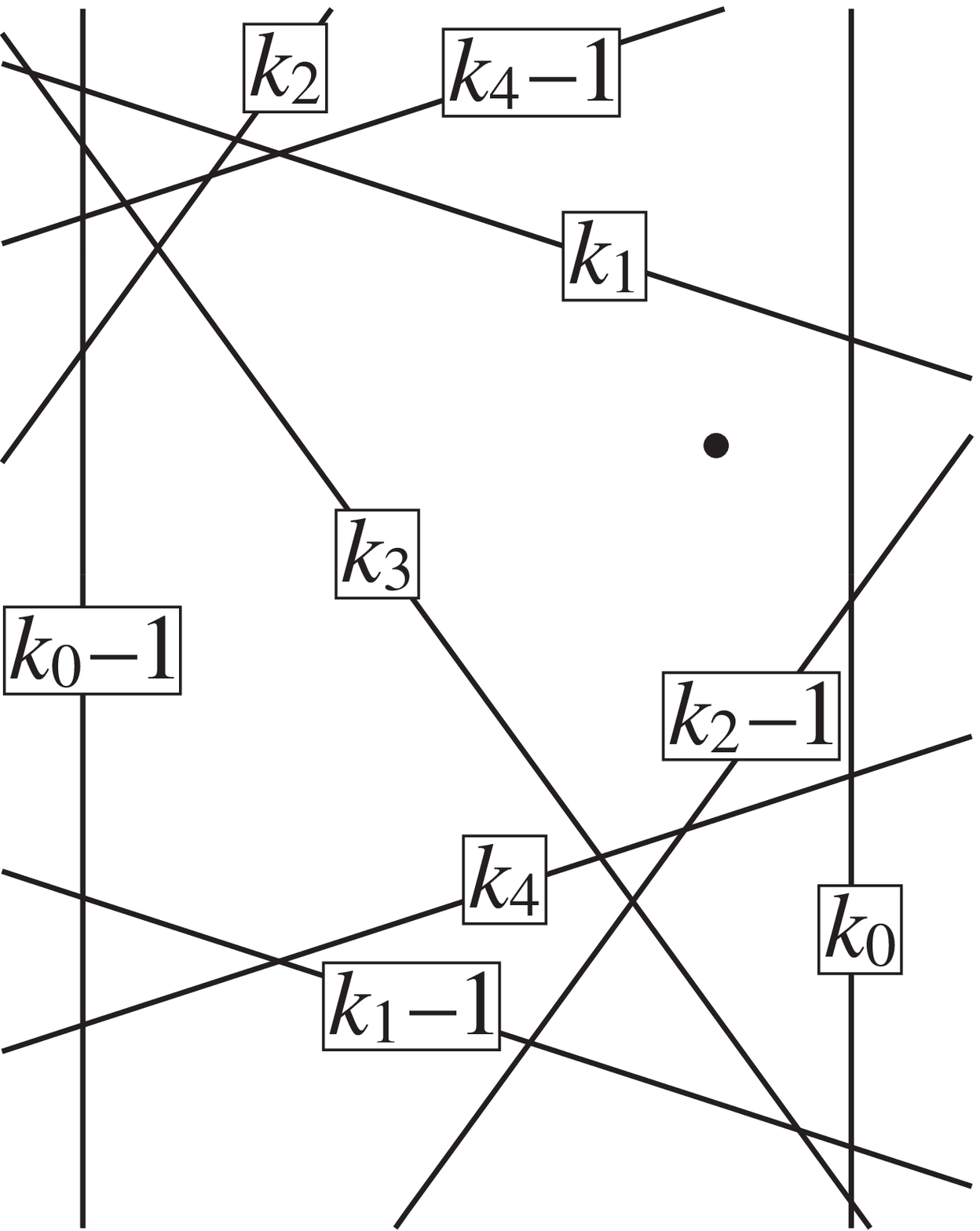}
\epsfxsize=0.22\hsize\hfil
\epsfbox{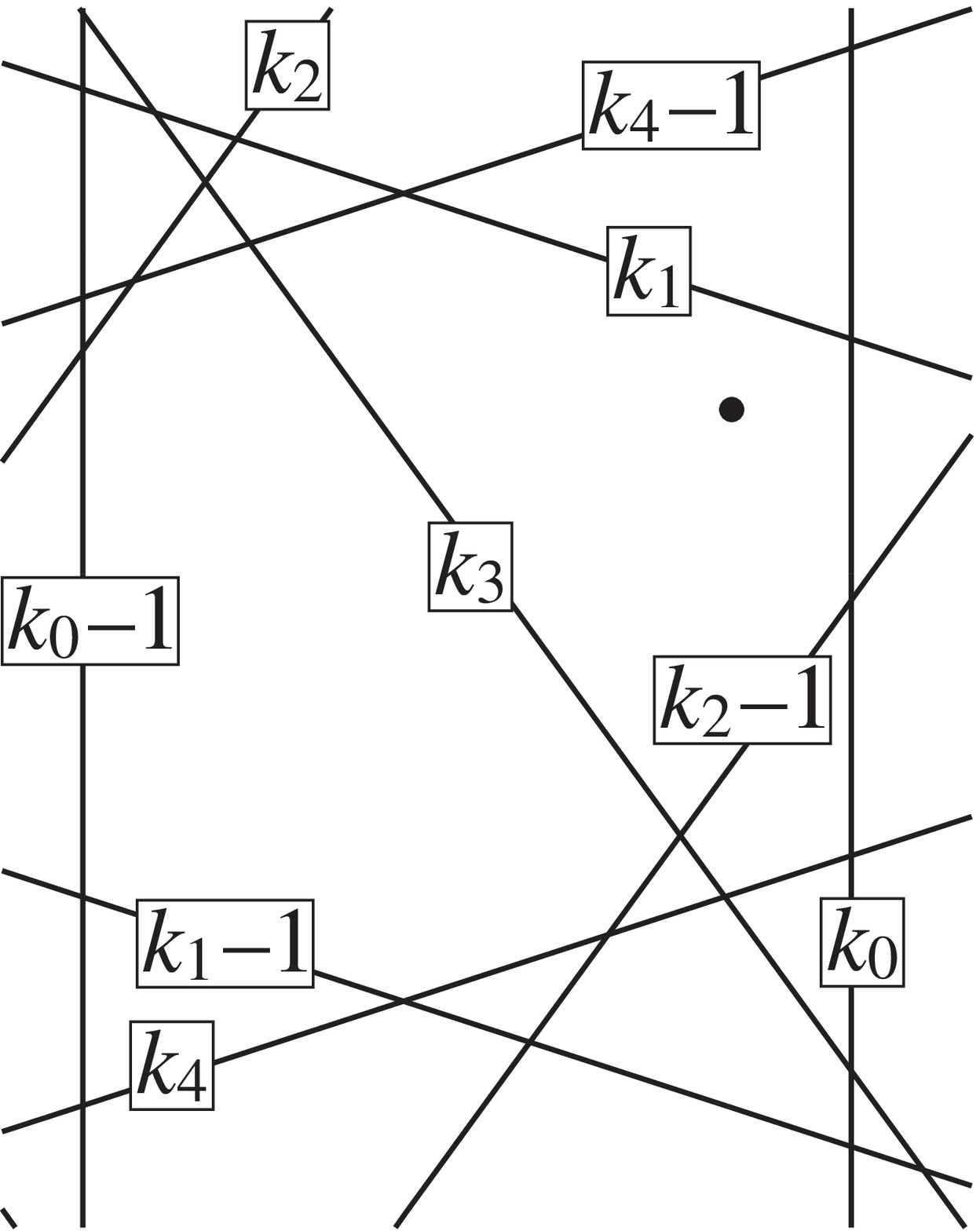}
\vskip0pt
\hbox to\hsize{\hspace*{20pt}\tiny
(q) \vtop{\hbox{C(17): $N\!=\!10$}
\hbox{$N_{\rm o}\!=\!5$, $N_{\rm e}\!=\!5$}}\hfil
(r) \vtop{\hbox{C(18): $N\!=\!10$}
\hbox{$N_{\rm o}\!=\!5$, $N_{\rm e}\!=\!5$}}\hfil
(s) \vtop{\hbox{C(19): $N\!=\!12$}
\hbox{$N_{\rm o}\!=\!7$, $N_{\rm e}\!=\!5$}}\hfil
(t) \vtop{\hbox{C(20): $N\!=\!12$}
\hbox{$N_{\rm o}\!=\!5$, $N_{\rm e}\!=\!7$}}\hfil
\hspace*{-18pt}}
\vskip0.1in\hskip0pt
\epsfxsize=0.22\hsize\hfil
\epsfbox{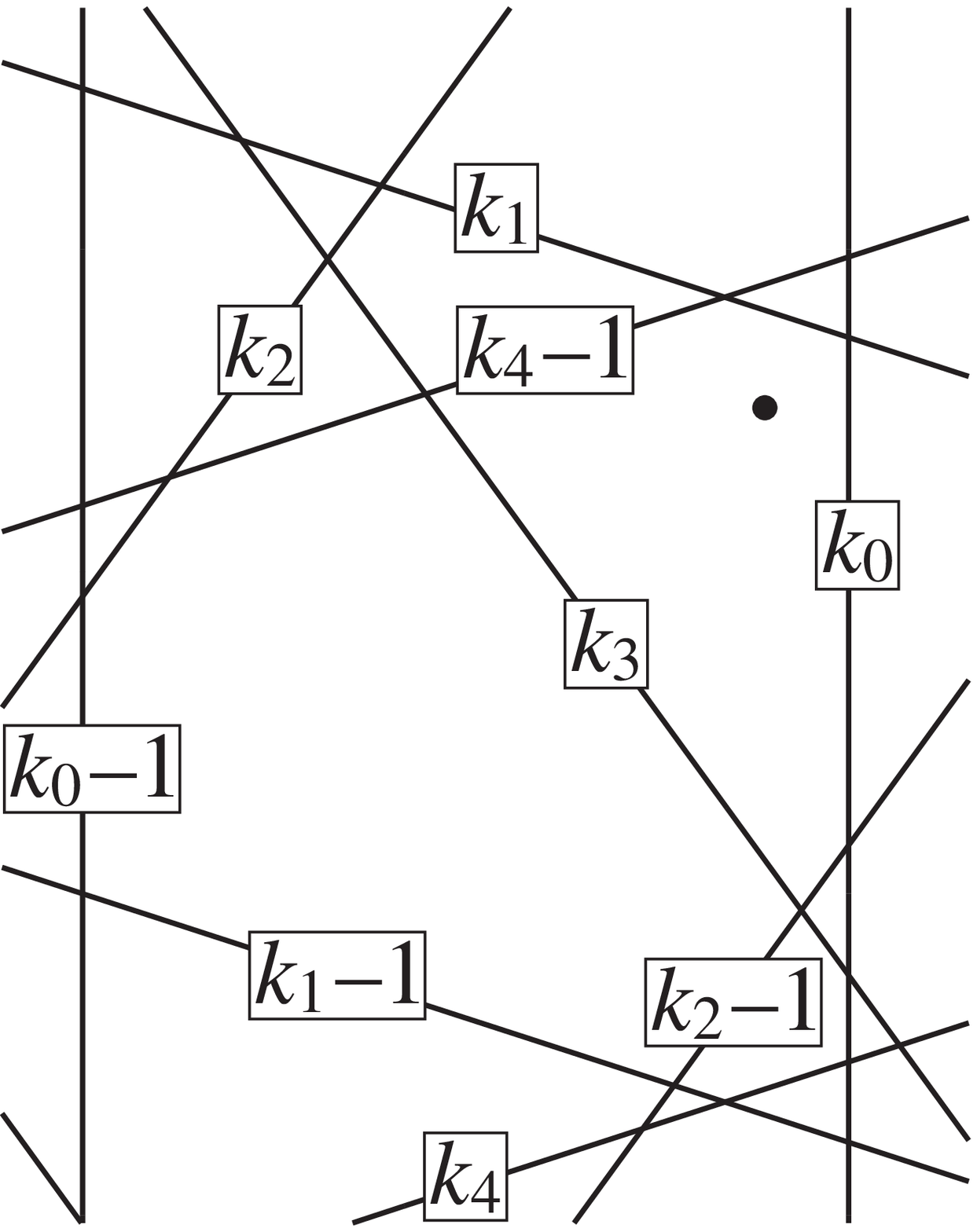}
\epsfxsize=0.22\hsize\hfil
\epsfbox{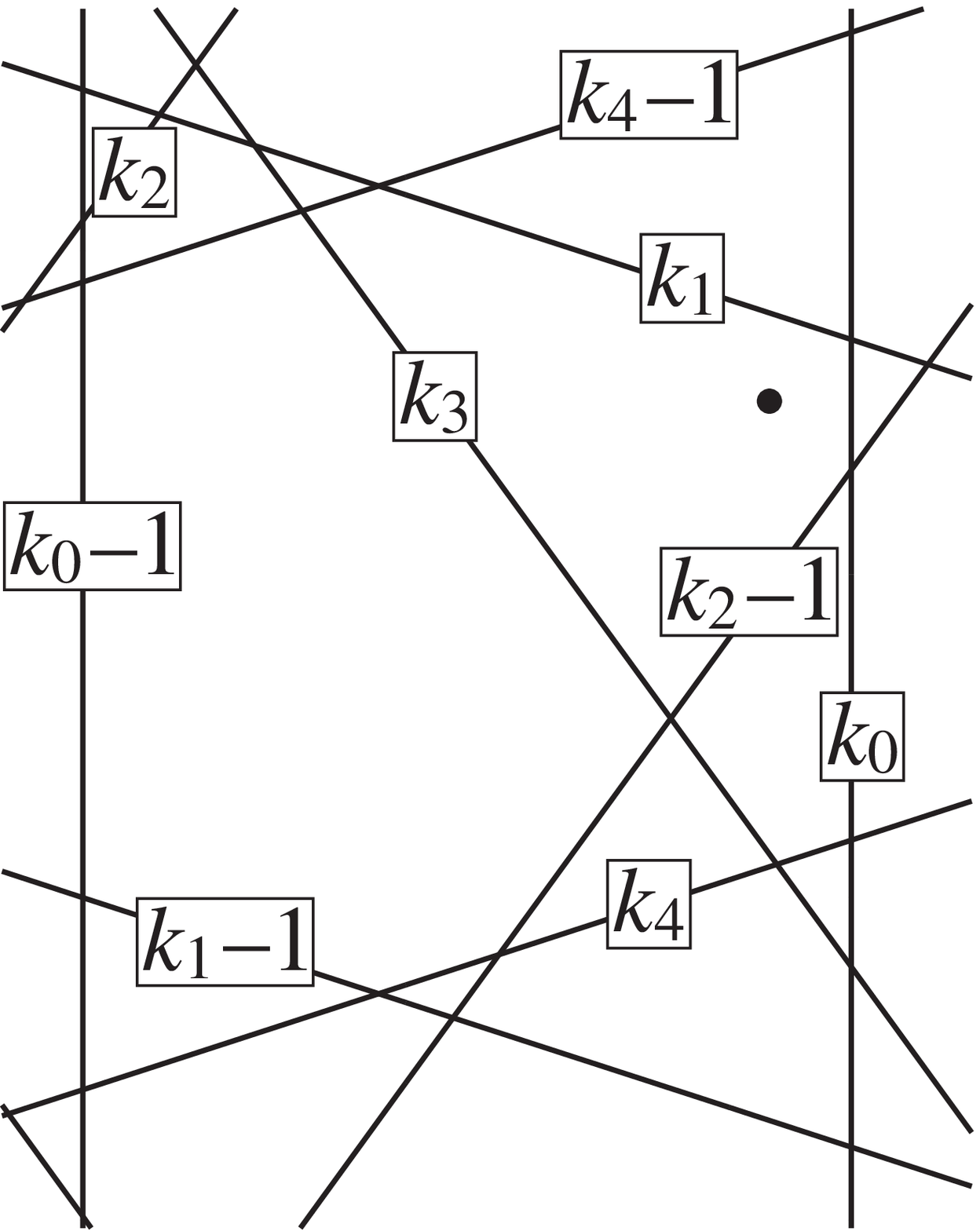}
\epsfxsize=0.22\hsize\hfil
\epsfbox{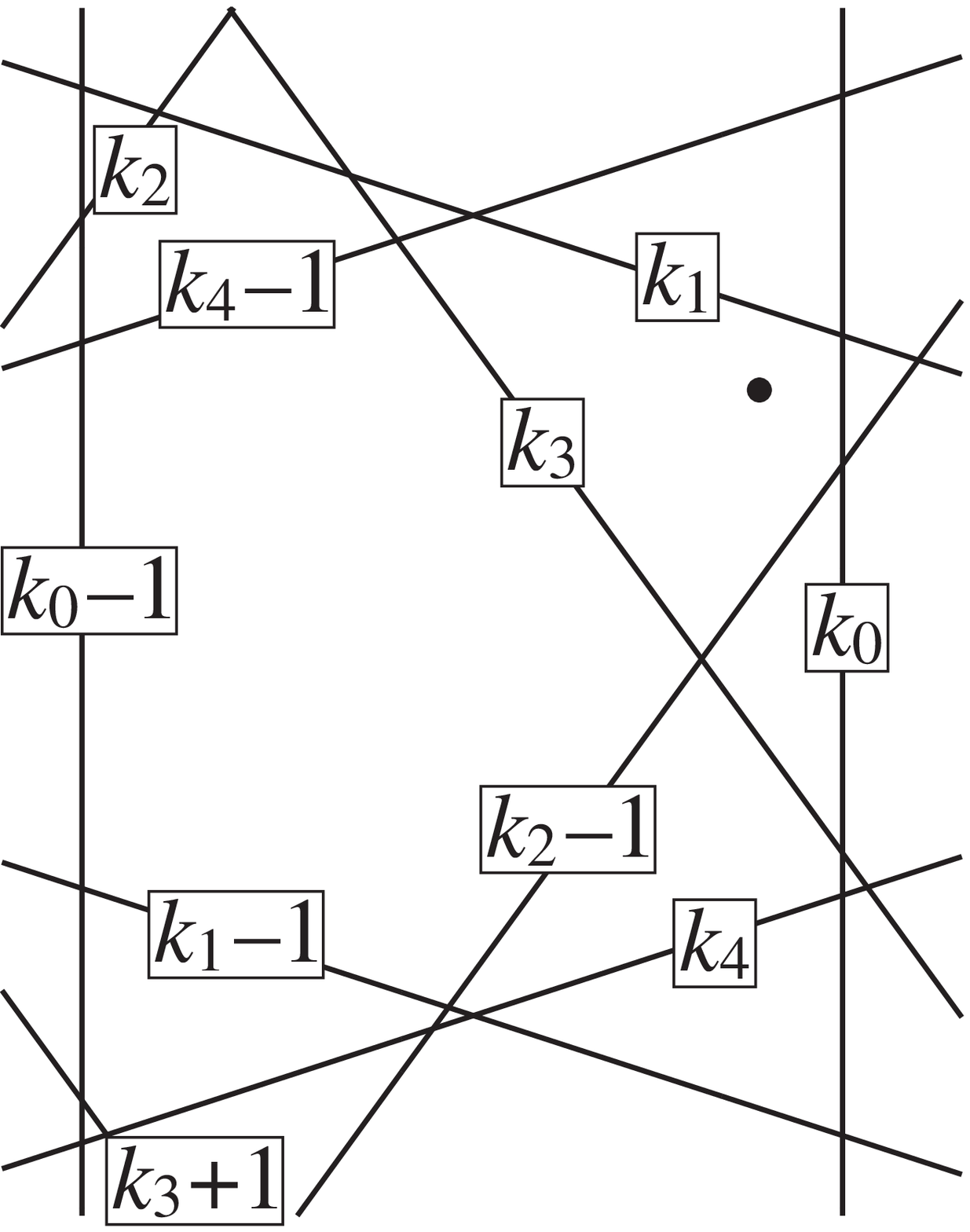}
\epsfxsize=0.22\hsize\hfil
\epsfbox{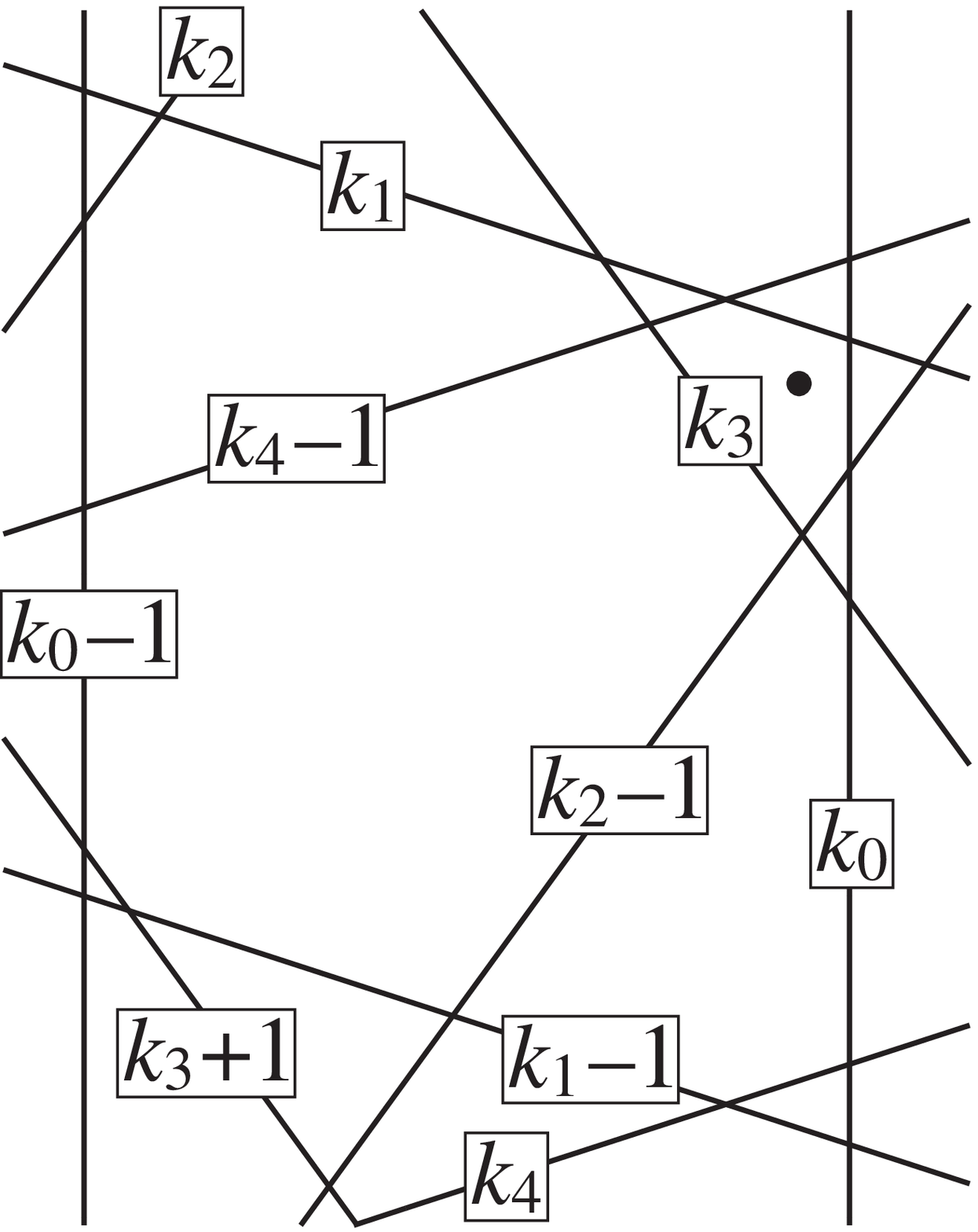}
\vskip0pt
\hbox to\hsize{\hspace*{20pt}\tiny
(u) \vtop{\hbox{C(21): $N\!=\!10$}
\hbox{$N_{\rm o}\!=\!5$, $N_{\rm e}\!=\!5$}}\hfil
(v) \vtop{\hbox{C(22): $N\!=\!10$}
\hbox{$N_{\rm o}\!=\!5$, $N_{\rm e}\!=\!5$}}\hfil
(w) \vtop{\hbox{C(23): $N\!=\!8$}
\hbox{$N_{\rm o}\!=\!5$, $N_{\rm e}\!=\!3$}}\hfil
(x) \vtop{\hbox{C(24): $N\!=\!9$}
\hbox{$N_{\rm o}\!=\!5$, $N_{\rm e}\!=\!4$}}\hfil
\hspace*{-18pt}}
\vskip0.1in
\hskip0pt\caption{{\em Continued}: The remaining twelve configurations. The
number of spin sites $N$ varies between 6 and 12, whereas $N_{\rm o}$
and $N_{\rm e}$ vary between 3 and 7.}
\end{figure}

For $\{\alpha\},\{\beta\}>p^{-1}$, the three grid lines $k_{j+2}$,
$k_{j+3}\!-\!1$, $k_{j+4}$ and their three intersections ${\myb a}^{1,0}$,
${\myb b}^{1,0}$, ${\myb c}^{1,1}$ are inside the parallelogram, producing
seven meshes in $P(k_j,k_{j+1})$, as shown in Figs.~\ref{fig6}$\,$(a) and
\ref{fig6}$\,$(b). The difference between the two cases is that the
intersection ${\myb c}^{1,1}$ is on opposite sides of the grid line
$k_{j+3}\!-\!1$, as the sign of $\{\alpha\}+\{\beta\}-p$ changes,
cf.\ (\ref{contri}). The grid line $k_{j+3}\!-\!1$ moves upward toward the
upper right corner as $\{\beta\}$ decreases. It is below the diagonal
$\epsilon_j+\epsilon_{j+1}=1$ for $\{\alpha\}+\{\beta\}>p$, corresponding
to Fig.~\ref{fig6}$\,$(a), and above the diagonal for
$\{\alpha\}+\{\beta\}<p$, as shown in Fig.~\ref{fig6}$\,$(b). The index of
the inner triangle changes from odd to even in view of (\ref{dk2}),
(\ref{dk3}) and (\ref{linear}). Hence, the spin configurations for the two
cases are different: C(1) in Fig.~\ref{fig6}$\,$(a) has 4 odd sites and 3
even sites, and C(2) shown in Fig.~\ref{fig6}$\,$(b) has 3 odd sites and 4
even sites. Also, C(1) is the only configuration for which the reference
integer vector does not correspond to an actual mesh.

For $\{\alpha\}>p^{-1}$ and $0<\{\beta\}<p^{-1}$, we have six cases
denoted by C(3) to C(8) arranged in the decreasing order of $\{\beta\}$.
For $1-p^{-1}\{\alpha\}<\{\beta\}$, the configuration C(3) is almost the
same as C(2) except having one more even site because line $k_{j+4}\!-\!1$
is now inside the parallelogram, as shown in Fig.~\ref{fig6}$\,$(c). As
$\{\beta\}$ decreases, line $k_{j+4}$ moves downward and for
$2p^{-1}\!-\!\{\alpha\}<\{\beta\}<1-p^{-1}\{\alpha\}$, the intersections
of line $k_{j+4}$ with lines $k_{j+3}\!-\!1$ (${\myb a}^{1,0}$) and
$k_{j+2}$ (${\myb c}^{1,1}$) are seen from (\ref{cona}) and (\ref{conc})
to be outside of $P(k_j,k_{j+1})$. As a result, two of the even sites are
now outside, and C(4), shown in Fig.~\ref{fig6}$\,$(d), has only six
sites, three of which are odd, and three even.

As $\{\beta\}$ decreases further to
$p(1\!-\!\{\alpha\})<\{\beta\}<2p^{-1}\!-\!\{\alpha\}$, line
$k_{j+3}$ is now inside $P(k_j,k_{j+1})$, as seen from (\ref{lcon3}). Thus,
configuration C(5), shown in Fig.~\ref{fig6}$\,$(e), has one more even
site than C(4). Both lines $k_{j+3}$ and $k_{j+3}\!-\!1$ move upward as
$\{\beta\}$ decreases. For $1\!-\!\{\alpha\}<\{\beta\}<p(1-\{\alpha\})$,
the intersection ${\myb b}^{1,1}$ of line $k_{j+3}$ with $k_{j+2}$ moves
inside of $P(k_j,k_{j+1})$, while ${\myb b}^{1,0}$, which is the
intersection of $k_{j+3}\!-\!1$ and $k_{j+2}$, moves out, giving rise to
configuration C(6) shown in Fig.~\ref{fig6}$\,$(f), with three even sites
and four odd sites.

For $p^{-1}(1\!-\!\{\alpha\})<\{\beta\}<1\!-\!\{\alpha\}$, line
$k_{j+3}\!-\!1$ moves out of $P$, as shown in Fig.~\ref{fig6}$\,$(g). Its
configuration C(7) has 3 odd sites and 3 even sites differing from C(6) in
that the odd site with $\partial{\vec K}({\myb 0})=(0,-1,0)$ is now outside
$P(k_j,k_{j+1})$. For $0<\{\beta\}<p^{-1}(1\!-\!\{\alpha\})$, the
intersections ${\myb a}^{0,1}$ and ${\myb c}^{0,1}$ of line $k_{j+4}-1$
with $k_{j+3}$ and $k_{j+2}$ are now seen from (\ref{cona}) and
(\ref{conc}) to be inside of $P(k_j,k_{j+1})$ adding 2 more odd sites to
C(8), which is shown in Fig.~\ref{fig6}$\,$(h). 

In Fig.~\ref{fig6}$\,$(i) through Fig.~\ref{fig6}$\,$(n), the six cases
C(9) through C(14) are shown for $\{\alpha\}<p^{-1}$ and
$\{\beta\}>p^{-1}$. Because of (\ref{kj12}) and (\ref{alpha}) we can use
the reflection symmetry $\{\alpha\}\leftrightarrow\{\beta\}$,
$k_{j}\leftrightarrow k_{j+1}$, $k_{j+2}\leftrightarrow k_{j+4}$, which was
noted also in the previous subsections. Thus these cases are similar to
the configurations C(3) to C(8), and obtainable simply by replacing
$k_{j+4}+n$ by $k_{j+2}+n$ and vice versa. To summarize, we find C(9) has
3 odd sites and 5 even sites; C(10) 3 odd sites and 3 even sites; C(11) 3
odd sites and 4 even sites; C(12) 4 odd sites and 3 even sites; C(13) has
3 odd sites and 3 even sites, and C(14) 5 odd sites and 3 even sites.

For $\{\alpha\},\{\beta\}<p^{-1}$, at least five grid lines $k_{j+2}$,
$k_{j+2}\!-\!1$, $k_{j+3}$, $k_{j+4}$ and $k_{j+4}\!-\!1$ are inside
$P(k_j,k_{j+1})$. In  Fig.~\ref{fig6}$\,$(o), we show configuration C(15)
valid for $1<\{\alpha\}+\{\beta\}$, when both lines $k_{j+3}-1$ and
$k_{j+3}$ and the intersections ${\myb b}^{1,1}$ and ${\myb a}^{1,1}$ are
inside $P(k_j,k_{j+1})$, as seen from (\ref{cona}) and (\ref{conb}).
Configuration C(15) has 4 odd sites and 5 even sites.
 
For the region satisfying the three inequalities
$\{\beta\}<1\!-\!\{\alpha\}$, $\{\beta\}>1\!-\!p\{\alpha\}$ and
$\{\beta\}>p^{-1}(1\!-\!\{\alpha\})$, grid line $k_{j+3}$ and the
corresponding odd site with $\partial{\vec K}=(0,-1,0)$ are now outside
$P(k_j,k_{j+1})$, such that configuration C(16) shown in
Fig.~\ref{fig6}$\,$(p) has one site less than C(15). It has 3 odd and 5
even sites.

For $1\!-\!p\{\alpha\}<\{\beta\}<p^{-1}(1\!-\!\{\alpha\})$, the
intersections ${\myb a}^{0,1}$ and ${\myb c}^{0,1}$ of line $k_{j+4}\!-\!1$
are both also inside $P(k_j,k_{j+1})$, as seen from (\ref{cona}) and
(\ref{conc}), adding two odd sites to C(16). Thus, C(17) in
Fig.~\ref{fig6}$\,$(q) has 5 odd sites and 5 even. However, for
$p^{-1}(1\!-\!\{\alpha\})<\{\beta\}<1\!-\!p\{\alpha\}$ the intersections
${\myb b}^{0,1}$ and ${\myb c}^{1,0}$ of lines $k_{j+2}\!-\!1$ are now
inside instead, adding two different odd sites to configuration C(16). The
resulting configuration C(18) shown in Fig.~\ref{fig6}$\,$(r) also has ten
sites and relates to C(17) by the above reflection symmetry.

When both conditions $p^{-1}\!-\!p\{\alpha\}<\{\beta\}<1\!-\!p\{\alpha\}$
and $p^{-2}\!-\!p^{-1}\{\alpha\}<\{\beta\}<p^{-1}(1\!-\!\{\alpha\})$ are
satisfied, we find six intersections ${\myb a}^{1,1}$, ${\myb b}^{1,1}$,
${\myb a}^{0,1}$, ${\myb b}^{0,1}$, ${\myb c}^{0,1}$ and ${\myb c}^{1,0}$
inside $P(k_j,k_{j+1})$. As a result, there are 12 sites for the two cases
C(19) and C(20) shown in Fig.~\ref{fig6}$\,$(s) and Fig.~\ref{fig6}$\,$(t)
respectively. The difference between the two cases is that the
intersections ${\myb c}^{0,1}$ and ${\myb c}^{1,0}$ are on opposite sides
of the grid line $k_{j+3}$, as the sign of $\{\alpha\}+\{\beta\}-p^{-1}$
changes, cf.\ (\ref{contri}). For $\{\alpha\}+\{\beta\}>p^{-1}$, line
$k_{j+3}$ lies below the diagonal, as can be seen from (\ref{dk3}) and
(\ref{linear}), so that its configuration C(19) has 7 odd sites and 5 even
sites; for $\{\alpha\}+\{\beta\}<p^{-1}$, $k_{j+3}$ is above the diagonal,
and C(20) has 5 odd sites and 7 even sites.

For $p^{-1}\!-\!p\{\alpha\}<\{\beta\}<p^{-2}\!-\!p^{-1}\{\alpha\}$, 
intersections ${\myb a}^{1,1}$ and ${\myb c}^{1,0}$ are no longer inside
$P(k_j,k_{j+1})$. As a consequence two of the even sites are now outside,
leaving C(21) shown in Fig.~\ref{fig6}$\,$(u) with 5 odd sites and 5 even.
For $p^{-2}\!-\!p^{-1}\{\alpha\}<\{\beta\}<p^{-1}\!-\!p\{\alpha\}$,
however, intersections ${\myb b}^{1,1}$ and ${\myb c}^{0,1}$ are outside
$P(k_j,k_{j+1})$ instead, such that two different even sites are now
outside. The resulting configuration C(22) shown in Fig.~\ref{fig6}$\,$(v)
has the same 5 odd sites as C(21), but a different set of 5 even sites.
Again, C(21) and C(22) are related by the aforementioned reflection
symmetry.

For $\{\beta\}<p^{-2}\!-\!p^{-1}\{\alpha\}$ and
$\{\beta\}<p^{-1}\!-\!p\{\alpha\}$, only two intersections
${\myb a}^{0,1}$ and ${\myb b}^{0,1}$ are still inside $P(k_j,k_{j+1})$.
These are the cases C(23) and C(24) shown in Fig.~\ref{fig6}$\,$(w) and
Fig.~\ref{fig6}$\,$(x). For $\{\beta\}+\{\alpha\}>p^{-3}$, configuration
C(23), shown in Fig.~\ref{fig6}$\,$(w), has eight sites: 5 odd and 3 even.
Finally, for $0<\{\beta\}<p^{-3}\!-\!\{\alpha\}$, the grid line $k_{j+3}+1$
is inside $P$, so that C(24) shown in Fig.~\ref{fig6}$\,$(x) has nine
sites, of which the 5 odd sites are identical to those of C(20) through
C(23).

In summary, the boundaries for the above 24 regions are the 13 lines given
by 
\begin{eqnarray}
&&\{\alpha\}=p^{-1},\quad \{\beta\}=p^{-1},\nonumber\\
&&\{\beta\}+\{\alpha\}=p^{-3},\;p^{-1},\;1,\;2p^{-1},
\hbox{ or }p,\nonumber\\
&&\{\beta\}+p^{-1}\{\alpha\}=p^{-2},\;p^{-1},\hbox{ or }1,\nonumber\\
&&\{\beta\}+p\{\alpha\}=p^{-1},\;1,\hbox{ or }p,
\label{bound}\end{eqnarray}
which are also the boundaries of the inequalities in (\ref{lcon12}),
(\ref{lcon3}), (\ref{cona}), (\ref{conb}) and (\ref{conc}), together with
the two conditions in (\ref{contri}). These are exactly all  conditions
for three grid lines to meet at a corner of $P$,  on an edge of $P$, or
inside of $P$, respectively, i.e.\ the only conditions under which some
mesh can appear or disappear in $P$ under shifts of grid lines.

\begin{figure}[tbph]
~\vskip0.03in\hskip0.1in
\epsfxsize=0.45\hsize\epsfclipon
\epsfbox{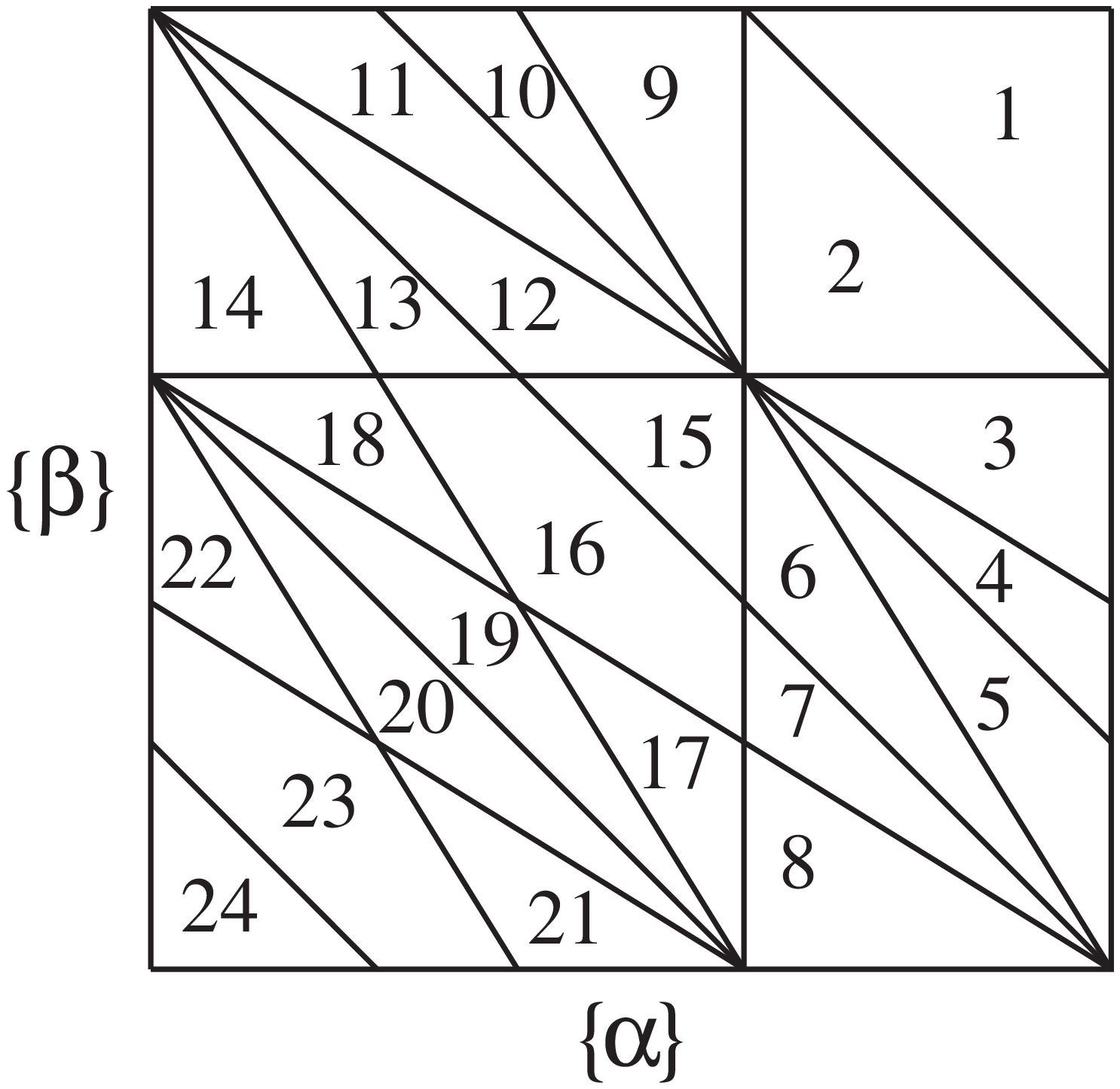}\hfil
\epsfxsize=0.45\hsize
\epsfbox{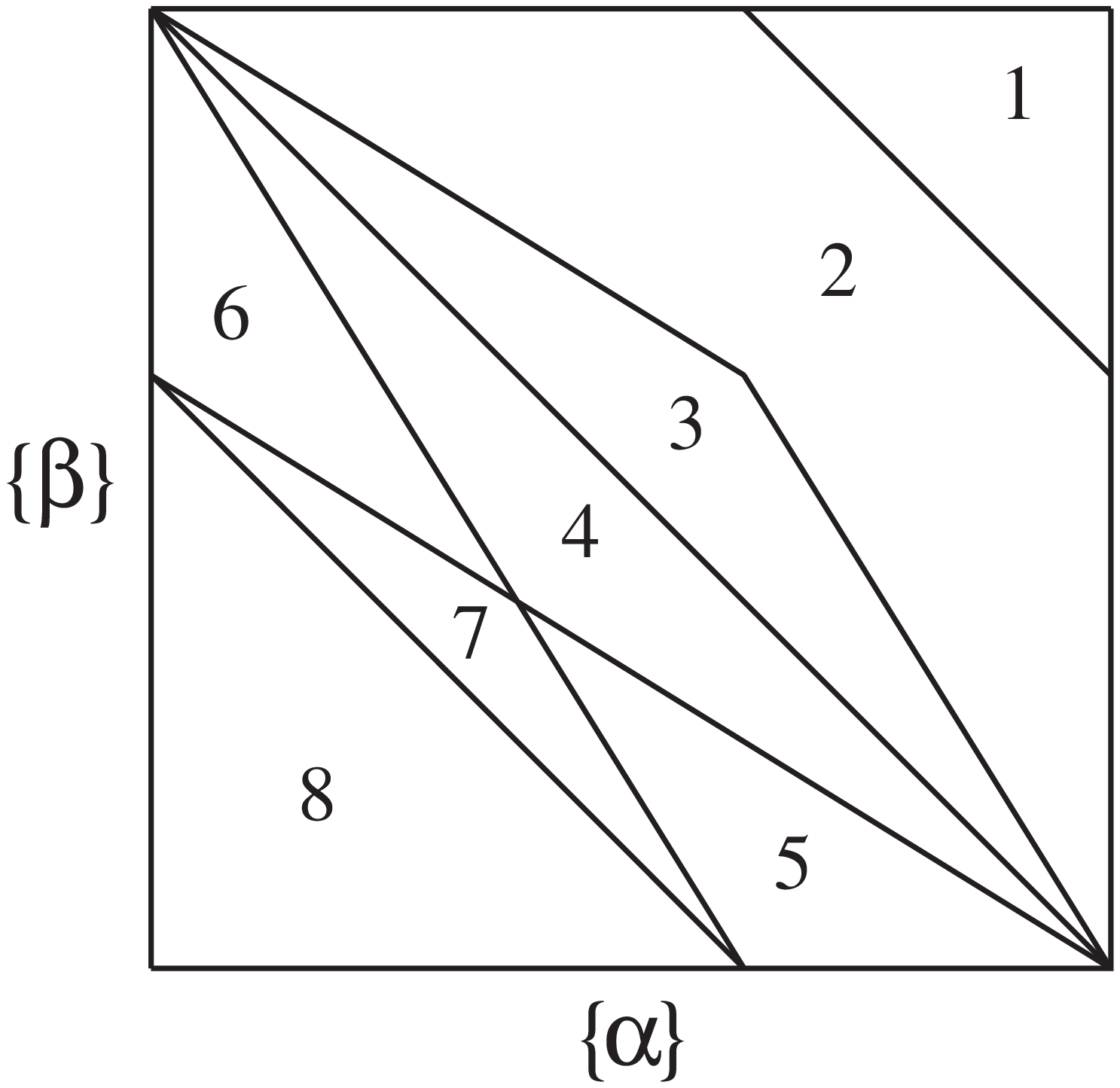}
\vskip6pt
\hbox to\hsize{\hspace*{12pt}\footnotesize
\hfil (a) \hfil\hfil (b)\hfil\hspace*{-2pt}}
\vskip-0.2in
\hskip0pt\caption{(a) The 24 regions for the 24 different configurations
are shown within the unit square with $\{\alpha\}$ and $\{\beta\}$ along
the horizontal and vertical axes. The integer $m$ denotes the $m$th region
for the $m$th configuration. (b) The 8 regions for the 8 different odd
configurations are shown.}
\label{fig7}
\end{figure}

In Fig.~\ref{fig7}$\,$(a), we plot the boundaries lines given by
(\ref{bound}) in the unit square with $\{\alpha\}$ and $\{\beta\}$ along
the horizontal and vertical axes. These lines indeed divide the unit
square into 24 regions. Each of these 24 regions corresponds to a
different configuration of the parallelogram $P(k_j,k_{j+1})$. The above
analysis shows that the parallelograms can only have 6, 7, 8, 9, 10 or 12
sites inside. The position of $\{\hat\alpha(k_{j+1})\}$ and
$\{\hat\beta(k_j)\}$ in the unit square shown in Fig.~\ref{fig7}$\,$(a)
determines which configuration the parallelogram $P(k_j,k_{j+1})$ is in.

The areas of the 24 regions in Fig.~\ref{fig7}$\,$(a) can be easily
calculated. We find, after using the formula of the area of a triangle in
terms of the coordinates of its three vertices,
\begin{eqnarray}
&&A(1)=A(2)={\textstyle\frac{5}{2}}-{\textstyle\frac{3}{2}}p=
{\textstyle\frac{1}{2}}p^{-4},\nonumber\\
&&A(3)=A(5)=A(6)=A(8)=A(9)=A(11)=A(12)\nonumber\\
&&\qquad=A(14)=A(19)=A(20)={\textstyle\frac{5}{2}}p-4=
{\textstyle\frac{1}{2}}p^{-5},\nonumber\\
&&A(4)=A(7)=A(10)=A(13)=A(15)=A(17)=A(18)\nonumber\\
&&\qquad=A(21)=A(22)=A(24)={\textstyle\frac{13}{2}}-4p=
{\textstyle\frac{1}{2}}p^{-6},\nonumber\\
&&A(16)=A(23)=9p-{\textstyle\frac{29}{2}}=
{\textstyle\frac{1}{2}}p^{-3}-p^{-6}.
\label{areacalc}
\end{eqnarray}
The areas of all 22 triangular areas differ by powers of the golden ratio
$p$.

For later reference, we display in Fig.~\ref{fig7}$\,$(b) the eight
different regions with equivalent odd configurations. Two odd
configurations are equivalent, if they have an equal number of odd sites
with the same sets of integer vectors for these odd sites. This
identification can most easily be made using the information in Tables
\ref{conftab} and \ref{conftabb} below.

\subsection{Probability} 

From the definition (\ref{alpha}), we find that the fractional parts
$\{\hat\alpha(k_{j+1})\}$ and $\{\hat\beta(k_j)\}$ are related to the
golden ratio $p$ which is irrational. From the well-known theorem of
Kronecker \cite{HW}, we conclude that $\{\hat\alpha(k_{j+1})\}$ and
$\{\hat\beta(k_j)\}$ are everywhere dense and uniformly distributed in the
interval $(0,1)$, as the integers $k_j$ and $k_{j+1}$ vary from $-\infty$
to $\infty$. As a consequence, every point in the unit square in
Fig.~\ref{fig7} is equally probable. Therefore, the frequency or
probability for a parallelogram to be in one of the twenty-four
configurations, say $m$, is given by the area $A(m)$ of the $m$th region.

Although the pentagrids are different for different choices of the shifts
$\gamma_j$, and the values of $\{\hat\alpha(k_{j+1})\}$ and
$\{\hat\beta(k_j)\}$ are also different for different $\gamma_j$'s,
this does not change the probability distributions of
$\{\hat\alpha(k_{j+1})\}$ and $\{\hat\beta(k_j)\}$ in the thermodynamic
limit of $k_j$ and $k_{j+1}$ varying from $-\infty$ to $\infty$. In other
words, the area $A(m)$ for the $m$th configuration is independent of the
$\gamma_j$'s, and is the same for all regular pentagrids.

\subsection{Difference vectors}

For $\{\alpha\}+\{\beta\}>1$ the grid line $k_{j+3}\!-\!1$ is seen from
(\ref{lcon3}) to be inside $P(k_j,k_{j+1})$. Consequently, the mesh below
corner ${\myb \epsilon}=(0,0)$ is odd and $\partial{\vec K}=(0,-1,0)$. For
$\{\alpha\}+\{\beta\}<1$, line $k_{j+3}-1$ is outside $P(k_j,k_{j+1})$, and
the mesh at corner ${\myb \epsilon}=(0,0)$ is even. In this case, its
integer vector is identical to the reference vector for
$P(k_j,k_{j+1})$ with $\partial{\vec K}({\myb 0})=(0,0,0)$. The difference
vectors $\partial{\vec K}$ for all other meshes in the parallelograms in
each of the twenty-four configurations can be easily obtained from
Fig.~\ref{fig6}, as $\partial K_{j+{\rm m}}$ changes its values only when
a lines in the $(j+{\rm m})$th grid is crossed.

In Tables \ref{conftab} and \ref{conftabb}, we list for each of the 24
configurations all the difference vectors $\partial{\vec K}=(\partial
K_{j+2},\partial K_{j+3},\partial K_{j+4})$, with Table \ref{conftab}
for $\{\alpha\}+\{\beta\}>1$ and Table \ref{conftabb} for
$\{\alpha\}+\{\beta\}<1$. From these tables one can also immediately read
off which regions have equivalent odd (or even) configurations.
\renewcommand{\thetable}{\Roman{table}(a)}
\begin{sidewaystable}
\caption{\boldmath $\{\alpha\}+\{\beta\}>1$}
\vspace{0.5em}
\mysmall
\begin{tabular}{|p{0.18in}|c|c|c|c|c|c|c|c|c|c|c|}
\hline
&C(1)&C(2)&C(3)&C(4)&C(5)&C(6)&C(9)&C(10)&C(11)&C(12)&C(15)\cr
\hline
&[0,-1,\,0]&[0,-1,\,0]&[0,-1,\,0]&[0,-1,\,0]&[0,-1,\,0]
&[0,-1,\,0]&[0,-1,\,0]&[0,-1,\,0]&[0,-1,\,0]&[0,-1,\,0]&[0,-1,\,0]
\cr
${\rm o}$&[1,\,0,\,0]&[1,\,0,\,0]&[1,\,0,\,0]&[1,\,0,\,0]&[1,\,0,\,0]
&[1,\,0,\,0]&[1,\,0,\,0]&[1,\,0,\,0]&[1,\,0,\,0]&[1,\,0,\,0]&[1,\,0,\,0]
\cr
${\rm d}$&[0,\,0,\,1]&[0,\,0,\,1]&[0,\,0,\,1]&[0,\,0,\,1]&[0,\,0,\,1]
&[0,\,0,\,1]
&[0,\,0,\,1]&[0,\,0,\,1]&[0,\,0,\,1]&[0,\,0,\,1]&[0,\,0,\,1]
\cr
${\rm d}$&[1,-1,\,1]&&&&&&&&&&
\cr
&&&&&&[0,\,1,\,0]&&&&[0,\,1,\,0]&[0,\,1,\,0]
\cr
\hline
&[1,-1,\,0]&[1,-1,\,0]&[1,-1,\,0]&[1,-1,\,0]&[1,-1,\,0]&&[1,-1,\,0]
&&&&
\cr
${\rm e}$&&[0,\,0,\,0]&[0,\,0,\,0]&[0,\,0,\,0]&[0,\,0,\,0]&[0,\,0,\,0]
&[0,\,0,\,0]&[0,\,0,\,0]&[0,\,0,\,0]&[0,\,0,\,0]&[0,\,0,\,0]
\cr
${\rm v}$&[0,-1,\,1]&[0,-1,\,1]&[0,-1,\,1]&&&&[0,-1,\,1]&[0,-1,\,1]
&[0,-1,\,1]&&
\cr
${\rm e}$&[1,\,0,\,1]&[1,\,0,\,1]&[1,\,0,\,1]&&&&[1,\,0,\,1]&&&&
\cr
${\rm n}$&&&[1,\,0,-1]&[1,\,0,-1]&[1,\,0,-1]&[1,\,0,-1]&&&&&[1,\,0,-1]
\cr
&&&&&&&[-1,\,0,\,1]&[-1,\,0,\,1]&[-1,\,0,\,1]&[-1,\,0,\,1]&[-1,\,0,\,1]
\cr
&&&&&[1,\,1,\,0]&[1,\,1,\,0]&&&&&[1,\,1,\,0]
\cr
&&&&&&&&&[0,\,1,\,1]&[0,\,1,\,1]&[0,\,1,\,1]\cr
\hline
\end{tabular}
\label{conftab}
\end{sidewaystable}

\renewcommand{\thetable}{\Roman{table}(b)}
\addtocounter{table}{-1}
\begin{sidewaystable}
\caption{\boldmath $\{\alpha\}+\{\beta\}<1$}
\vspace{0.5em}
\mysmall
\begin{tabular}{|p{0.085in}|c|c|c|c|c|c|c|c|c|c|c|c|c|}
\hline
&C(7)&C(8)&C(13)&C(14)&C(16)&C(17)&C(18)&C(19)&C(20)
&C(21)&C(22)&C(23)&C(24)\cr
\hline
&[1,\,0,\,0]&[1,\,0,\,0]&[1,\,0,\,0]&[1,\,0,\,0]&[1,\,0,\,0]
&[1,\,0,\,0]&[1,\,0,\,0]&[1,\,0,\,0]&&&&&
\cr
&[0,\,1,\,0]&[0,\,1,\,0]&[0,\,1,\,0]&[0,\,1,\,0]&[0,\,1,\,0]
&[0,\,1,\,0]&[0,\,1,\,0]&[0,\,1,\,0]&[0,\,1,\,0]&[0,\,1,\,0]
&[0,\,1,\,0]&[0,\,1,\,0]&[0,\,1,\,0]
\cr
${\rm o}$&[0,\,0,\,1]&[0,\,0,\,1]&[0,\,0,\,1]&[0,\,0,\,1]&[0,\,0,\,1]
&[0,\,0,\,1]&[0,\,0,\,1]&[0,\,0,\,1]&&&&&
\cr
${\rm d}$&&[0,\,0,\,-1]&&&&[0,\,0,\,-1]&&[0,\,0,\,-1]&[0,\,0,\,-1]&
[0,\,0,\,-1]&[0,\,0,\,-1]&[0,\,0,\,-1]&[0,\,0,\,-1]
\cr
${\rm d}$&&[1,\,1,\,-1]&&&&[1,\,1,\,-1]&&[1,\,1,\,-1]&[1,\,1,\,-1]
&[1,\,1,\,-1]&[1,\,1,\,-1]&[1,\,1,\,-1]&[1,\,1,\,-1]
\cr
&&&&[-1,\,0,\,0]&&&[-1,\,0,\,0]&[-1,\,0,\,0]&[-1,\,0,\,0]
&[-1,\,0,\,0]&[-1,\,0,\,0]&[-1,\,0,\,0]&[-1,\,0,\,0]
\cr
&&&&[-1,\,1,\,1]&&&[-1,\,1,\,1]&[-1,\,1,\,1]&[-1,\,1,\,1]
&[-1,\,1,\,1]&[-1,\,1,\,1]&[-1,\,1,\,1]&[-1,\,1,\,1]
\cr
\hline
&[0,\,0,\,0]&[0,\,0,\,0]&[0,\,0,\,0]&[0,\,0,\,0]&[0,\,0,\,0]
&[0,\,0,\,0]&[0,\,0,\,0]&[0,\,0,\,0]&[0,\,0,\,0]&[0,\,0,\,0]
&[0,\,0,\,0]&[0,\,0,\,0]&[0,\,0,\,0]
\cr
&&&[0,\,1,\,1]&[0,\,1,\,1]&[0,\,1,\,1]&[0,\,1,\,1]
&[0,\,1,\,1]&[0,\,1,\,1]&[0,\,1,\,1]&&[0,\,1,\,1]&&
\cr
${\rm e}$&[1,\,1,\,0]&[1,\,1,\,0]&&&[1,\,1,\,0]&[1,\,1,\,0]
&[1,\,1,\,0]&[1,\,1,\,0]&[1,\,1,\,0]&[1,\,1,\,0]&&&
\cr
${\rm v}$&&&&&&&&&[-1,\,1,\,0]&[-1,\,1,\,0]&[-1,\,1,\,0]&[-1,\,1,\,0]
&[-1,\,1,\,0]
\cr
${\rm e}$&&&&&&&&&[0,\,1,\,-1]&[0,\,1,\,-1]&[0,\,1,\,-1]&[0,\,1,\,-1]
&[0,\,1,\,-1]
\cr
${\rm n}$&[1,\,0,\,-1]&[1,\,0,\,-1]&&&[1,\,0,\,-1]&[1,\,0,\,-1]
&[1,\,0,\,-1]&[1,\,0,\,-1]&[1,\,0,\,-1]&[1,\,0,\,-1]&&&
\cr
&&&[-1,\,0,\,1]&[-1,\,0,\,1]&[-1,\,0,\,1]&[-1,\,0,\,1]
&[-1,\,0,\,1]&[-1,\,0,\,1]&[-1,\,0,\,1]&&[-1,\,0,\,1]
&&\cr
&&&&&&&&&&&&&[0,\,2,\,0]\cr
\hline
\end{tabular}
\label{conftabb}
\end{sidewaystable}
\renewcommand{\thetable}{\Roman{table}}

\subsection{Average Number $\myb{\cal N}$}

We shall now calculate ${\cal N}$, which is the average number of spin
sites in a parallelogram. The total number of sites can be evaluated by
counting all the sites in each parallelograms $P(k_j,k_{j+1})$, and then
adding all of them together for all the $P$'s. This is equivalent to
splitting the summation over all sites into two parts---first summing over
all sites in $P$ represented by their integer vectors
${\vec K}(\myb{\epsilon})$ and then adding all of them for all the
parallelograms. Let there be ${\cal M}$ lines in each of the five grids,
so that there are ${\cal M}^2$ parallelograms, ignoring boundary effects
that cancel in the thermodynamic limit. The average ${\cal N}$ then equals
the total number of lattice sites divided by ${\cal M}^2$.

We have already shown that each parallelogram $P(k_j,k_{j+1})$ is in one
of 24 configurations $C(m)$ uniquely determined by the values of
$\{\hat\alpha(k_{j+1})\}$ and $\{\hat\beta(k_j)\}$. The allowed
configurations $C(m)$ have $N(m)=6$, 7, 8, 9, 10 or 12 sites inside $P$.
The frequency or probability $A(m)$ is defined as the number of
parallelograms in the $m$th configuration divided by the total number of
parallelograms. If we let $k_j$ and $k_{j+1}$ in $P(k_j,k_{j+1})$ run over
the ${\cal M}$ values, each of the parallelograms is counted once.

As ${\cal M}$ approaches $\infty$, so that $-\infty<k_j,k_{j+1}<\infty$,
the values of $\{\hat\alpha(k_{j+1})\}$ and
$\{\hat\beta(k_j)\}$ are everywhere dense and uniformly distributed
\cite{HW} between $0$ and $1$. The frequency $A(m)$ is the area of the
$m$th region in the unit square in Fig.~\ref{fig7}$\,$(a). The values of
these $A(m)$'s are listed in (\ref{areacalc}). Denoting the number of sites
in the $m$th configuration by $N(m)$, with values also given in the
captions of Fig.~\ref{fig6}, then the average number of sites per
parallelogram is
\begin{eqnarray} 
{\cal N}&=&\frac 1 {{\cal M}^2}\!
\sum_{{\vec K}(z\in\mbox{\mymsbms C})}1=\frac 1 {{\cal M}^2}
\sum_{{\rm all}\,{\vphantom{{\vec K}}} P}\,\sum_{{\vec K}(z\in P)}1
\label{split}\\
&&\stackrel{{\cal M}\to\infty}{\longrightarrow}
\sum_{m=1}^{24} A(m)\sum_{n=1}^{N(m)}1=\sum_{m=1}^{24}
A(m)N(m)=5p.
\label{spsum}\end{eqnarray}
where we have used the notation ${\vec K}(z\!\in\!\mbox{\mymsbm C})$ to
denote the integer vectors of all the meshes in the pentagrid, while
${\vec K}(z\!\in\!P)$ denotes only those meshes in parallelogram $P$.

\subsection{Penrose Tiles}

The above method provides an alternative way to draw the Penrose tiles. To
illustrate this, we let $j=0$. For some fixed shifts $\gamma_j$, we let
$-J\leqslant k_0,k_1\leqslant J$ for some positive integer $J$. For any
values $k_0$ and $k_1$ in this set, $\{\hat\alpha(k_1)\}$,
$\{\hat\beta(k_0)\}$ are uniquely determined from (\ref{alpha}). The
elements $k_{\rm m}$, for ${\rm m}=2$, 3 ,4, of the reference vector in
$P(k_0,k_1)$ are given by (\ref{kj12}) and (\ref{kj3}). From the values of
$\{\hat\alpha(k_1)\}$ and $\{\hat\beta(k_0)\}$, we can determine from
Fig.~\ref{fig7}$\,$(a), what configuration $C(m)$ the parallelogram
$P(k_0,k_1)$ is in. Then we can use Tables \ref{conftab} and \ref{conftabb}
to obtain the difference vectors $\partial{\vec K}$ for the $N(m)$ sites
inside $P(k_0,k_1)$. The integer vectors for these different sites in
$P(k_0,k_1)$ are then given by
\begin{equation} 
{\vec K}(\myb\epsilon)=
( k_0, k_1, k_2+\partial K_2, k_3+\partial K_3, k_4+\partial K_4 ).
\label{intvec}\end{equation}
We use (\ref{penrose}) and (\ref{intvec}) to obtain the positions of the
spins in the complex plane for the meshes in $P(k_0,k_1)$. Hence, as $k_0$
and $k_1$ run over the values from $-J$ to $J$ we obtain the positions of
the spins in both odd and even sublattices shown in Fig.~\ref{fig5}. This
figure has been plotted using Maple.

\subsection{Summary}

Consider the parallelograms $P(k_j,k_{j+1})$ which contains all points
$z\in\mbox{\mymsbm C}$ such that $K_j(z)=k_j$ and $K_{j+1}(z)=k_{j+1}$,
cf.\ (\ref{mesh}). The configurations of two such parallelograms are
considered to be the same, if they contain the same number of spin sites
and the corresponding sites have the same difference vectors. The
different configurations do not depend on the exact locations of the
relevant grid lines or their intersections. However, whenever a grid line
or an intersection moves in or out of the parallelogram $P$, the
configuration changes. The above analysis shows that there are only 24
allowed configurations, with 6, 7, 8, 9, 10 and 12 sites inside $P$.

The configuration of $P(k_j,k_{j+1})$ is uniquely determined by the values
of $\{\hat\alpha(k_{j+1})\}$ and $\{\hat\beta(k_j)\}$ defined in
(\ref{alpha}). By examining the locations of the relevant grid lines and
their intersections, we find that the unit square, with $\{\alpha\}$ and
$\{\beta\}$ along the horizontal and vertical axes, is divided into 24
regions, corresponding to the 24 possible configurations of the
parallelogram $P$. This is shown Fig.~\ref{fig7}$\,$(a), and the 24
configurations of $P$ are shown in Fig.~\ref{fig6}. Using the theorem of
Kronecker \cite{HW}, we find that the area $A(m)$ of the $m$th region is
actually proportional to the probability for the $m$th configuration to
occur. Even though the pentagrids and configurations of the
$P(k_j,k_{j+1})$'s are different for different choices of the shifts
$\gamma_j$, the area $A(m)$ is independent of these shifts and is the same
for all regular pentagrids.

In each parallelogram $P(k_j,k_{j+1})$, a reference integer vector is
chosen whose components are given by (\ref{kj12}) and (\ref{kj3}). The
difference vectors with respect to this reference vector are defined in
(\ref{kz}) and calculated for all of the sites inside $P$. The number of
sites $N(m)$, the area $A(m)$, and the difference vectors
$\partial{\vec K}$ for the $N(m)$ spin sites for $m=1...24$ are listed
in Fig.~\ref{fig6}, Eq.~(\ref{areacalc}) and Tables \ref{conftab}
and \ref{conftabb}, respectively.

\setcounter{equation}{0}
\section{Susceptibility}\label{sect5}

There are three $Z$-invariant Ising models that can be defined on the
vertices of the Penrose rhombus tiles using the prescriptions of Section
\ref{sect2}. Model 1 has spins on all odd sites only, interacting along
the diagonals of the tiles, as is illustrated in Fig.~\ref{fig5}. Model 2
is defined similarly with spins only on the even sites. Model 3 has all
sites of the Penrose tiling, but the even and odd sites are decoupled,
with the odd spins interacting as in model 1 and the even spins as in
model 2. We will see that the three models have the identical
wavevector-dependent susceptibility $\chi({\bf q})$ per spin site in the
thermodynamic limit.

The physical positions of the spins have been expressed in
(\ref{penrose}) as complex numbers depending on the integer vectors
${\vec K}(z)$ of the meshes. Let $q$ be a complex number and
$q=q_x+{\rm i}q_y$ so that $q^*$ denotes its complex conjugate (while
${\bf q}=(q_x,q_y)$), then the ${\bf q}$-dependent susceptibility is
\begin{eqnarray}
k_{\rm B}T\chi({\bf q})\!=\!\!
\lim_{{\cal M}\to\infty}{\frac{1}{{\cal N}{\cal M}^{2}}}\!
\sum_{{\vec K}(z\in\mbox{\mymsbms C})}
\sum_{{\vec K}(z'\in\mbox{\mymsbms C})}
{\rm cos}\,{\rm Re}\Bigl\{q^*\sum_{j=0}^4[K_j(z')-K_j(z)]\zeta^j\Bigr\}
\nonumber\\
\times\bigr[{\langle\sigma_{{\vec K}(z)}\sigma_{{\vec K}(z')}\rangle}-
\langle\sigma_{{\vec K}(z)}\rangle\langle\sigma_{{\vec K}(z')}
\rangle\bigl]=
\left\{\begin{array}{ll}
2{\hat\chi}^{\rm o}({\bf q}),&\text{(model 1)},\quad\cr
2{\hat\chi}^{\rm e}({\bf q}),&\text{(model 2)},\upstrut\quad\cr
{\hat\chi}^{\rm o}({\bf q})+{\hat\chi}^{\rm e}({\bf q}),\upstrut
&\text{(model 3)}.\quad
\end{array}\right.
\label{chipt}\end{eqnarray}
Here the double sums denoted by ${\vec K}(z\in\mbox{\mymsbm C})$ are over
all the odd spin sites in model 1, over all the even sites in model 2, and
over all sites in model 3. In the last case, since the spins on the odd
sublattice do not interact with those on the even sublattice, the
${\bf q}$-dependent susceptibility $\chi({\bf q})$ becomes the sum of two
parts: ${\hat\chi}^{\rm o}({\bf q})$ denoting the contribution from the odd
sublattice and ${\hat\chi}^{\rm e}({\bf q})$ from the even sublattice. For
model 3 the average number of spin sites per parallelogram was given in
(\ref{spsum}) as ${\cal N}=5p$. For models 1 and 2 this number becomes
${\cal N}=5p/2$, explaining the extra factors 2 in the last member of
(\ref{chipt}).

We shall first consider ${\hat\chi}^{\rm o}({\bf q})$ and show later that
${\hat\chi}^{\rm e}({\bf q})$ is equal to it, implying that the
susceptibilities of the three models are indeed equal.

\subsection{Calculation of $\myb{\chi^{\rm o}({\bf q})}$}\label{sect51}

We again split the sum over all odd spin sites into two parts as in
(\ref{split}), and let ${\vec K}^{\rm o}(z\in P)$ run over all odd spins
in parallelogram $P$. Consequently,
\begin{equation}
{\hat\chi}^{\rm o}({\bf q})=
\lim_{{\cal M}\to\infty}{\frac{1}{{\cal N}{\cal M}^{2}}}
\sum_{{\rm all}\,{\vphantom{{\vec K}}} P}\,
{\sum_{{\vec K}\vbox to 1.1ex{}^{\rm o}(z\in P)}}
\sum_{{\rm all}\,{\vphantom{{\vec K}}} P'}
{\sum_{{\vec K}\vbox to 1.1ex{}^{\rm o}({z'\in P'})}}
U({\vec K}^{\rm o}\!(z'),{\vec K}^{\rm o}\!({z})),
\label{chipto}\end{equation}
where
\begin{eqnarray}
U({\vec K}^{\rm o}\!(z'),{\vec K}^{\rm o}\!({z}))
&=&\langle\sigma_{{\vec K}\vbox to 1.1ex{}^{\rm o}(z)}
\sigma_{{\vec K}\vbox to 1.1ex{}^{\rm o}(z')}\rangle^{\rm c}\nonumber\\
&\times&{\rm cos}\,{\rm Re}\Bigl\{q^*\sum_{n=0}^4
[K^{\rm o}_n\!(z')-K^{\rm o}_n\!({z})]\zeta^n\Bigr\},
\label{sumd}\end{eqnarray}
with $\langle\sigma\sigma'\rangle^{\rm c}$ denoting the connected pair
correlation function, subtracting the contribution from the spontaneous
magnetization.

Now we let $P=P(k_j,k_{j+1})$ and $P'=P(k_j\!+\!\ell,k_{j+1}\!+\!\ell')$.
As ${\cal M}\to\infty$, $\ell$ and $\ell'$ are kept fixed, and $k_j$ and
$k_{j+1}$ vary from $-\infty$ to $\infty$, all the $P$'s and $P'$'s are
counted once. It is also evident that the different choices of $j$
correspond to choosing one of five orientations for the parallelograms,
and they should give the same ${\bf q}$-dependent susceptibility.

In (\ref{zinP}) we defined $z=z({\myb\epsilon})$ for $z\in P$. Similarly,
for $z'\in P'$, we let
\begin{equation}
z'=\frac{\,{\rm i}\,[\zeta^j(k_{j+1}\!+\!\ell'\!-
\!\gamma_{j+1}\!-\!\epsilon'_{j+1})-
\zeta^{j+1}(k_j\!+\!\ell\!-\!\gamma_j\!-\!\epsilon'_j)]}{\sin(2\pi/5)}
\equiv z'({\myb\epsilon'}),
\label{parallelp}\end{equation}
so that $z'\leftrightarrow{\myb\epsilon'}=(\epsilon'_j,\epsilon'_{j+1})$
and $0<\epsilon'_j,\epsilon'_{j+1}<1$. The corresponding integer vectors
${\vec K}(z')={\vec K}(z'({\myb\epsilon'}))
\equiv{\vec K}'({\myb\epsilon'})$ also have two fixed components
$K_j'({\myb\epsilon'})=k_j\!+\!\ell\equiv k'_j$ and
$K_{j+1}'({\myb\epsilon'})=k_{j+1}\!+\!\ell'\equiv k'_{j+1}$. Now,
following (\ref{alpha}), we let
\begin{eqnarray}
&&\alpha'\equiv\hat\alpha(k'_{j+1})=
p^{-1}(k_{j+1}\!+\!\ell'\!-\!\gamma_{j+1})+\gamma_j+\gamma_{j+2}
=\alpha+p^{-1}\ell',\nonumber\\
&&\beta'\equiv\hat\beta(k'_j)=
p^{-1}(k_j\!+\!\ell\!-\!\gamma_j)+\gamma_{j+4}+\gamma_{j+1}
=\beta+p^{-1}\ell. 
\label{alphap}\end{eqnarray}
According to (\ref{kj12}) and (\ref{kj3}) the reference integer vector
$\vec k'$ for $P'$ is chosen to have components,
\begin{eqnarray}
&&k'_{j+2}=\lceil\alpha'\rceil-k_j-\ell
=k_{j+2}+\delta_{j+2}
+\lfloor p^{-1}\ell'\rfloor-\ell,\nonumber\\
&&k'_{j+3}=-\lfloor\alpha'\rfloor-\lfloor\beta'\rfloor
=k_{j+3}-\delta_{j+2}-\lfloor p^{-1}\ell'\rfloor
-\delta_{j+4}-\lfloor p^{-1}\ell\rfloor,\nonumber\\
&&k'_{j+4}=\lceil\beta'\rceil-k_{j+1}-\ell'
=k_{j+4}+\delta_{j+4}+\lfloor p^{-1}\ell\rfloor-\ell',
\label{kjp}\end{eqnarray}
in which\footnote{From the first members of (\ref{kjp}) and (\ref{kj12})
we find that $\delta_{j+2}=
\lceil\alpha'\rceil-\lceil\alpha\rceil-\lfloor p^{-1}\ell'\rfloor=
\lceil\alpha+p^{-1}\ell'\rceil-\lceil\alpha\rceil-
\lfloor p^{-1}\ell'\rfloor=\lceil\{\alpha\}-1+\{p^{-1}\ell'\}\rceil=
\lfloor\{\alpha\}+\{p^{-1}\ell'\}\rfloor$. For the last two steps here
and the remaining steps in the derivation of (\ref{kjp}) and
(\ref{delta24}) we must make explicit use of the fact that
$\alpha$, $\alpha'$, $\beta$ and $\beta'$ are not integer for a
regular pentagrid.}
\begin{equation}
\delta_{j+2}=\lfloor\{\alpha\}+\{p^{-1}\ell'\}\rfloor,\quad
\quad\delta_{j+4}=\lfloor\{\beta\}+\{p^{-1}\ell\}\rfloor.
\label{delta24}\end{equation} 
Substituting (\ref{parallelp}) into (\ref{mesh}) we obtain for
${\rm m}=2,3,4$
\begin{equation}
K_{j+{\rm m}}'(\myb{\epsilon}')=k'_{j+{\rm m}}+\partial
K'_{j+{\rm m}}(\myb{\epsilon}'),\quad
\partial K'_{j+{\rm m}}(\myb{\epsilon}')=\lfloor
\lambda'_{j+{\rm m}}(\myb{\epsilon}')\rfloor,
\label{kzp}\end{equation}
where
\begin{equation}
\lambda'_{j+2}(\myb{\epsilon}')
=\{\alpha'\}+\epsilon'_j-p^{-1}\epsilon'_{j+1},\quad
\lambda'_{j+4}(\myb{\epsilon}')=
\{\beta'\}+\epsilon'_{j+1}-p^{-1}\epsilon'_j.
\label{dkp2}\end{equation}
and
\begin{equation}
\lambda'_{j+3}(\myb{\epsilon}')=
p^{-1}(\epsilon'_j+\epsilon'_{j+1})-\{\alpha'\}
-\{\beta'\}+1.
\label{dkp3}\end{equation}
Comparing (\ref{kzp}) to (\ref{dkp3}) with (\ref{kz}) to (\ref{dk3}), we
find that the dependence of $\lambda'_{j+{\rm m}}(\myb{\epsilon}')$ on
$\{\alpha'\}$ and $\{\beta'\}$ is the same as the dependence of
$\lambda_{j+{\rm m}}(\myb{\epsilon})$ on $\{\alpha\}$ and $\{\beta\}$.
Consequently, the configurations of $P'$ depend on $\{\alpha'\}$ and
$\{\beta'\}$ in the same way as the configurations of $P$ on $\{\alpha\}$
and $\{\beta\}$. Therefore, the position of $\{\alpha'\}$ and $\{\beta'\}$
in the unit square in Fig.~\ref{fig7}$\,$(a) with $\{\alpha'\}$ and
$\{\beta'\}$ along the horizontal and vertical axes uniquely determines the
configuration of $P'$. The difference vectors
$\partial {\vec K}'(\myb{\epsilon}')$ for the sites in $P'$, which is in
one of the 24 configurations, are again given in Tables \ref{conftab}
and \ref{conftabb}.

The above results are valid for all spin configurations in $P$ and $P'$,
but we shall consider only the odd spins at first. By examining Tables
\ref{conftab} and \ref{conftabb}, we can find that the odd spin
configurations of the parallelograms are simpler, because several
connected regions---see C(2) to C(5), or C(9) to C(11) in
Fig.~\ref{fig7}$\,$(a) as examples---have the same odd spin
configurations. In fact, there are only eight distinct odd spin
configurations. In Fig.~\ref{fig7}$\,$(b), the regions for these 8 odd
spin configurations are shown in the unit square whose axes are the
$\{\alpha\}$ and $\{\beta\}$ directions.

Listed in Table \ref{oddtab} are the number of odd sites ${\hat N}(m)$ in
$P$, the region of validity $R(m)$, and the area ${\hat A}(m)$ of the
$m$th odd configuration for all $m=1,\cdots,8$. In the $m$th odd
configuration, the difference vectors of the ${\hat N}(m)$ spins are
denoted by $\partial {\vec K}^{[m,n]}$ for $n=1,\cdots,{\hat N}(m)$. They
are equal to the difference vectors $\partial{\vec K}^{\rm o}$ of the odd
sites in some configuration C($l$), listed in Tables \ref{conftab}
and \ref{conftabb} for $l=1,\cdots,24$. In the last column of Table
\ref{oddtab} it is indicated which C($l$)'s correspond to a given $m$. 
\def\bigstrut{\raisebox{1em}{\vbox to 0.75em{}}
\raisebox{-1em}{\vtop to 0.3em{}}}
\begin{sidewaystable}
\caption{The eight regions for the odd sublattice\hspace*{4em}}
\vspace{0.5em}
\renewcommand{\arraystretch}{1.2}
\begin{tabular}{|p{0.14in}|c|c|p{3.66in}|p{1.55in}|}
\hline
$m$ &${\hat N}(m)$&${\hat A}(m)$&\multicolumn{1}{c|}{$R(m)$}
&\hspace*{-0.7em}
$\begin{array}{c}\partial{\vec K}^{[m,n]}=\partial{\vec K}^{\rm o}\\
1\leqslant n\leqslant{\hat N}(m)\end{array}$
\cr
\hline\bigstrut
1&4&${\textstyle\frac{1}{2}}p^{-4}$&$ p^{-1}<\{\alpha\}<1\;\&\;
p-\{\alpha\}<\{\beta\}<1$&
$\partial{\vec K}^{\rm o}$ in C(1)\cr
\hline\bigstrut
2&3&${\textstyle\frac{1}{2}}p^{-1}$&\hspace*{-0.7em}
$\begin{array}{l}0<\{\alpha\}\leqslant
p^{-1}\;\&\; 1-p^{-1}\{\alpha\}<\{\beta\}<1;\\
p^{-1}\leqslant\{\alpha\}<1\;\&\;
p(1-\{\alpha\})<\{\beta\}<p-\{\alpha\}\end{array}$&\hspace*{-0.7em}
$\begin{array}{l}\hbox{$\partial{\vec K}^{\rm o}$ in C(2) or}\\
\hbox{C($l$), $l$=3,4,5,9,10,11}\end{array}$\cr
\hline\bigstrut
3&4&${\textstyle\frac{1}{2}}p^{-3}$&\hspace*{-0.7em}
$\begin{array}{l}0<\{\alpha\}\leqslant p^{-1}\;\&\;
1-\{\alpha\}<\{\beta\}<1-p^{-1}\{\alpha\};\\
p^{-1}\leqslant\{\alpha\}<1\;\&\;
1-\{\alpha\}<\{\beta\}<p(1-\{\alpha\})\end{array}$&\hspace*{-0.7em}
$\begin{array}{l}\hbox{$\partial{\vec K}^{\rm o}$ in C(6),}\\
\hbox{C(12) or C(15)}\end{array}$\cr
\hline\bigstrut
4&3&${\textstyle\frac{1}{2}}p^{-3}$&\hspace*{-0.7em}
$\begin{array}{l}0<\{\alpha\}\leqslant p^{-2}\;\&\;
1-p\{\alpha\}<\{\beta\}<1-\{\alpha\};\\
p^{-2}\leqslant\{\alpha\}<1\;\&\;
p^{-1}(1-\{\alpha\})<\{\beta\}<1-\{\alpha\}\end{array}$&\hspace*{-0.7em}
$\begin{array}{l}\hbox{$\partial{\vec K}^{\rm o}$ in C(7),}\\
\hbox{C(13) or C(16)}\end{array}$\cr
\hline\bigstrut
5&5&${\textstyle\frac{1}{2}}p^{-4}$&$0<\{\beta\}<p^{-2}\;\&\;
p^{-1}(1-\{\beta\})<\{\alpha\}<1-p\{\beta\}$&
$\partial{\vec K}^{\rm o}$ in C(8) or C(17)\cr
\hline\bigstrut
6&5&${\textstyle\frac{1}{2}}p^{-4}$&$0<\{\alpha\}<p^{-2}\;\&\;
p^{-1}(1-\{\alpha\})<\{\beta\}<1-p\{\alpha\}$&
$\partial{\vec K}^{\rm o}$ in C(14) or C(18)\cr
\hline\bigstrut
7&7&${\textstyle\frac{1}{2}}p^{-5}$&\hspace*{-0.7em}
$\begin{array}{l}0<\{\alpha\}\leqslant p^{-2}\;\&\;
p^{-1}-\{\alpha\}<\{\beta\}<p^{-1}(1-\{\alpha\});\hspace*{-3pt}\\
p^{-2}\leqslant\{\alpha\}<p^{-1}\;\&\;
p^{-1}-\{\alpha\}<\{\beta\}<1-p\{\alpha\}\end{array}$&
$\partial{\vec K}^{\rm o}$ in C(19)\cr
\hline\bigstrut
8&5&${\textstyle\frac{1}{2}}p^{-2}$&$0<\{\alpha\}<p^{-1}\;\&\;
0<\{\beta\}<p^{-1}-\{\alpha\}$&
$\partial{\vec K}^{\rm o}$ in C(20) to C(24)\cr
\hline
\end{tabular}
\label{oddtab}
\end{sidewaystable}

Let the distances $\ell$ and $\ell'$ between the two parallelograms $P$ and
$P'$ be fixed, but $k_j$ and $k_{j+1}$ vary from $-\infty$ to
$\infty$. Then $\{\alpha'\}$ and $\{\beta'\}$ given by (\ref{alphap}) are
also everywhere dense and uniformly distributed in the interval $(0,1)$.
The area ${\hat A}(m)$ is again the probability or frequency of the $m$th
configuration. From (\ref{alphap}) we find
\begin{eqnarray}
&\{\alpha'\}=&\left\{
\begin{array}{lcr}\{\alpha\}+a&\hbox{for}&\{\alpha\}+a<1\\
\{\alpha\}+a-1&\hbox{for}&\{\alpha\}+a\geqslant1\upstrut\end{array}
\right\},\quad a=\{p^{-1}\ell'\},\nonumber\\
&\{\beta'\}=&\left\{
\begin{array}{lcr}\{\beta\}+b&\hbox{for}&\{\beta\}+b<1 \\
\{\beta\}+b-1&\hbox{for}&\{\beta\}+b\geqslant1\upstrut\end{array}
\right\},\quad b=\{p^{-1}\ell\}.
\label{betap}\end{eqnarray}
\begin{figure}[tbh]
~\vskip0.01in\hskip0.1in
\epsfxsize=0.45\hsize\epsfclipon
\epsfbox{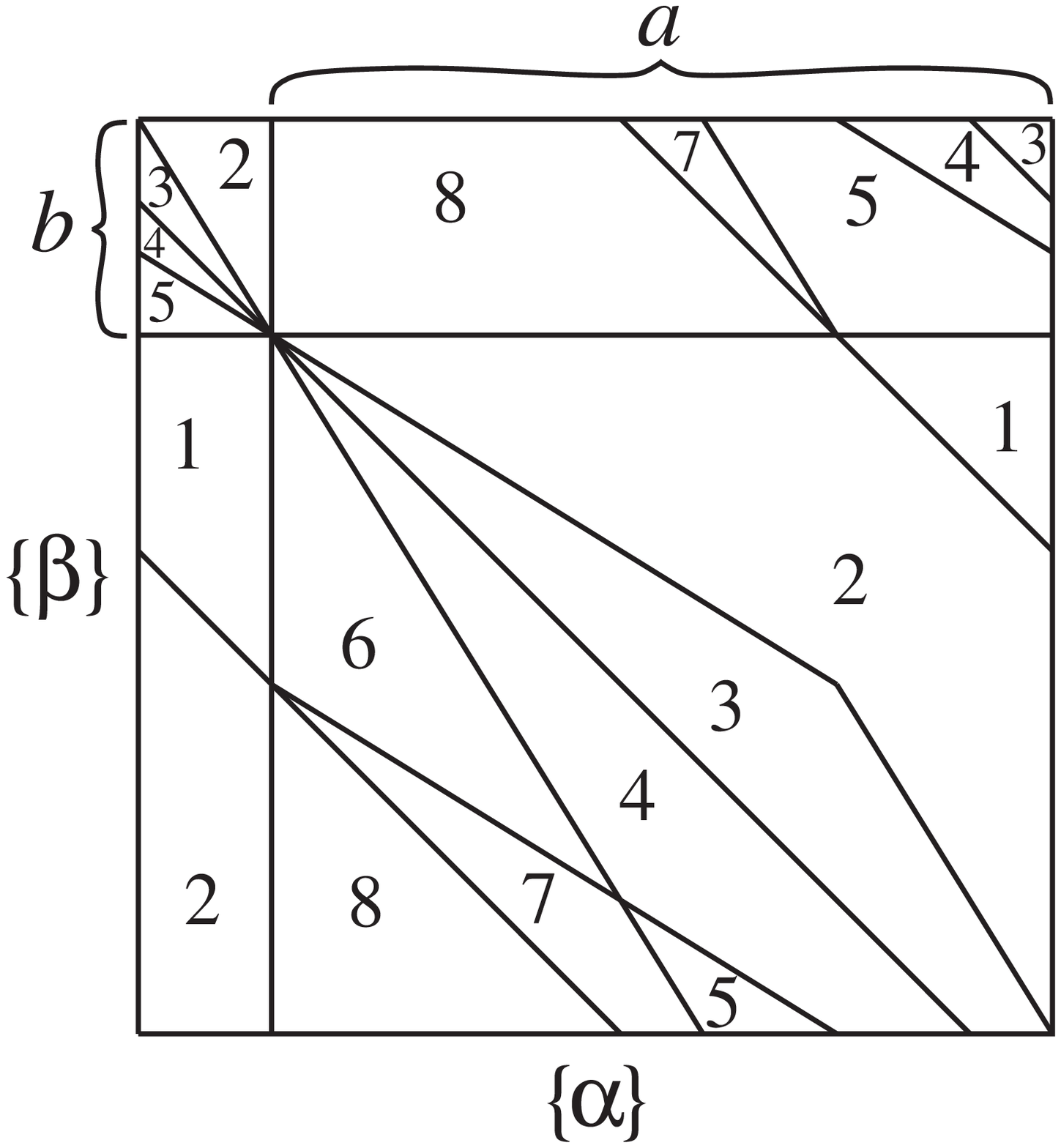}\hfil
\epsfxsize=0.45\hsize
\epsfbox{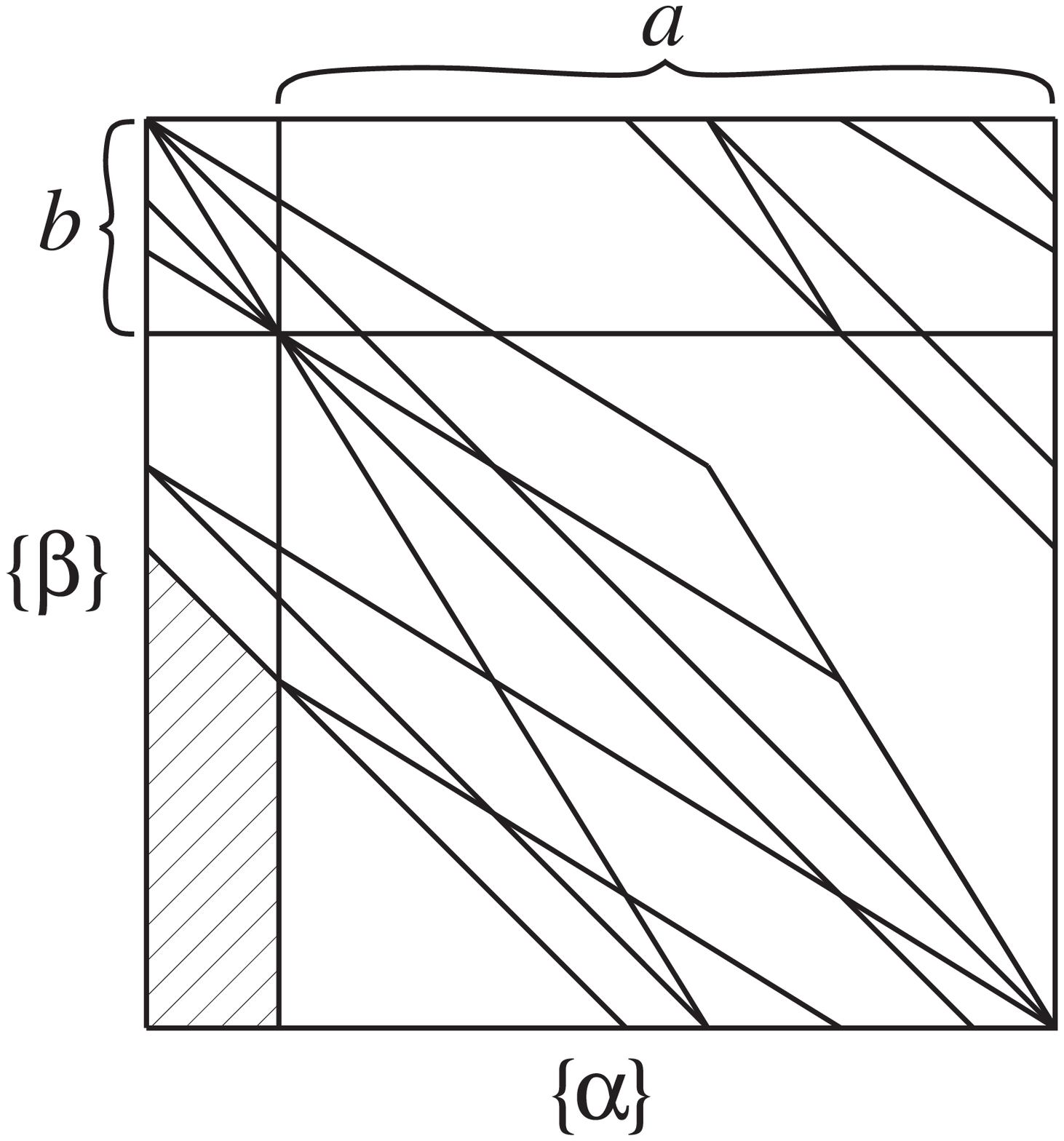}
\vskip6pt
\hbox to\hsize{\hspace*{12pt}\footnotesize
\hfil (a) \hfil\hfil (b)\hfil\hspace*{-2pt}}
\vskip-0.2in
\hskip0pt\caption{(a) The eight odd spin configurations in
$P(k_j\!+2,k_{j+1}\!+\!3)$ plotted with $\{\alpha\}$ and $\{\beta\}$ as the
horizontal and vertical axes. (b) Overlapping the plot in part (a) with
Fig.~7$\,$(b) we can determine the joint probability $A_{m,m'}(2,3)$ for
$P(k_j,k_{j+1})$ to be in the $m$th configuration and
$P(k_j\!+\!2,k_{j+1}\!+\!3)$ in $m'$th configuration, with
$m,m'=1,\cdots,8$, geometrically. The area of the shaded region represents
the probability for $m=8$ and $m'=2$.}
\label{fig8}
\end{figure} 
The plot of the eight regions for the odd configurations of
$P=P(k_j,k_{j+1})$ has been given in Fig.~\ref{fig7}$\,$(b). The plot for
$P'=P(k_j\!+\!\ell,k_{j+1}\!+\!\ell')$ is the same, only now with
$\{\alpha'\}$ and $\{\beta'\}$ along the axes. However, if the eight
regions are replotted with $\{\alpha\}$ and $\{\beta\}$ along the
horizontal and vertical axes, then we obtain unit squares as shown in
Fig.~\ref{fig8}$\,$(a) [for the special case of $\ell=2$ and $\ell'=3$].
Because of the relation (\ref{betap}), we find that Fig.~\ref{fig8}$\,$(a)
can be obtained from Fig.~\ref{fig7}$\,$(b) by cutting a horizontal slice
with width $b$ from the bottom of Fig.~\ref{fig7}$\,$(b), and pasting it
on the top; then cutting a vertical slice of width $a$ from the left and
pasting it to the right. After the cutting and pasting, the connected
region $R(m)$ for $P'$ in Fig.~\ref{fig7}$\,$(b) becomes $R'(m)$ in
Fig.~\ref{fig8}$\,$(a), which---for $\ell\ne 0$ or $\ell'\ne 0$---may
consist of disjointed pieces pasted in up to four different sections of
the unit square. The probability for $P'$ to be in the $m$th configuration
is still the area ${\hat A}(m)={\rm area}(R(m))={\rm area}(R'(m))$, which
in the latter case, could be a sum of areas of disjointed pieces.

For fixed $\ell$ and $\ell'$ [chosen to be $\ell=2$ and $\ell'=3$ in
Fig.~\ref{fig8}$\,$(a)], the position of $\alpha=\hat\alpha(k_{j+1})$ and
$\beta=\hat\beta(k_{j})$ in the unit square shown in Fig.~\ref{fig7}$\,$(b)
uniquely determines the configuration of $P(k_j,k_{j+1})$, while the
position of $\alpha'=\hat\alpha(k_{j+1}\!+\!\ell')$ and
$\beta'=\hat\beta(k_{j}\!+\!\ell)$ in Fig.~\ref{fig8}$\,$(a) determines the
configuration of $P(k_j\!+\!\ell,k_{j+1}\!+\!\ell')$. As $k_j$ and
$k_{j+1}$ run over all the values from $-\infty$ to $\infty$, we find from
Kronecker's theorem \cite{HW} that every point in either
Fig.~\ref{fig7}$\,$(b) or Fig.~\ref{fig8}$\,$(a) is equally probable.
However, the positions of $(\alpha,\beta)$ and $(\alpha',\beta')$ are
completely correlated by the shift $(a,b)$ in (\ref{betap}), which is
fixed as long as $\ell$ and $\ell'$ are unchanged. Thus, the joint
probability for $P(k_j,k_{j+1})$ to be in the $m$th configuration and
$P(k_j\!+\!\ell,k_{j+1}\!+\!\ell')$ to be in the $m'$th configuration is
the area of the intersection of the two regions $R(m)$ and $R'(m')$, and
is denoted by
\begin{equation}
A_{m,m'}(\ell,\ell')={\rm area}(R(m)\cap R'(m')).
\label{overlap}\end{equation}
By superimposing Fig.~\ref{fig7}$\,$(b) on top of Fig.~\ref{fig8}$\,$(a),
we obtain Fig.~\ref{fig8}$\,$(b). This figure gives the intersections of
all the regions of Fig.~\ref{fig7}$\,$(b) with all the regions of
Fig.~\ref{fig8}$\,$(a), and the joint probabilities can be read off as the
areas of these intersections.

For different values of $\ell$ and $\ell'$, we get different values of the
width $a$ given in (\ref{betap}) of the vertical slice in
Fig.~\ref{fig8}$\,$(a) and also of the width $b$ of the horizontal slice.
After cutting and pasting the difference slices, the resulting figures are
very different, so are the superimposed figures. Thus, the area of
intersection $A_{m,m'}(\ell,\ell')$ depends on the choice of $\ell$ and
$\ell'$. However, just as ${\hat A}(m)$ and ${\hat A}(m')$ do not depend
on the shifts $\gamma_j$, $A_{m,m'}(\ell,\ell')$ also does not depend on
these shifts. Moreover, it is easily seen that this joint probability
$A_{m,m'}(\ell,\ell')$ is not only the same for all the different regular
pentagrids, but also the same for the different orientations of the
parallelograms (i.e.\ different choices of $j=0,\cdots,4$).

We let ${\vec K^{[m,n]}}$ denote the $n$th integer vector in the $m$th odd
configuration for odd spins inside $P$, where $n=1,\cdots,N(m)$ and
$m=1,\cdots,8$. Similarly ${\vec K^{[m',n']}}$ denotes the $n'$th integer
vector of the $m'$th odd configuration of odd spins inside $P'$, with
$n'=1,\cdots,N(m')$ and $m=1,\cdots,8$. From Tables \ref{conftab},
\ref{conftabb} and \ref{oddtab}, the possible difference vectors
$\partial\vec K^{[m,n]}$ and $\partial\vec K^{[m',n']}$ for two spins in
$P$ and $P'$ may be found. Adding, as in (\ref{kz}) and (\ref{kzp}), these
to their corresponding reference integer vectors, with three of their
components given in (\ref{kj12}), (\ref{kj3}) and (\ref{kjp}), we obtain
the two integer vectors $\vec K^{\rm o}(z)={\vec K^{[m,n]}}$ for the spin
in $P$ and $\vec K^{\rm o}(z')={\vec K^{[m',n']}}$ for the spin in $P'$.
In (\ref{chipt}) and (\ref{sumd}) we only need their difference
\begin{equation} 
{\vec K^{\rm o}}(z')-{\vec K^{\rm o}}(z)=
{\vec K^{[m',n']}}-{\vec K^{[m,n]}}
\equiv(\ell_0,\ell_1,\cdots,\ell_4).
\label{dfell}\end{equation}
We find from (\ref{kz}), (\ref{kjp}) and (\ref{kzp}) that
\begin{eqnarray}
&&\ell_j=\ell, \qquad \ell_{j+1}=\ell',\nonumber\\
&&\ell_{j+2}=\delta_{j+2}+\lfloor p^{-1}\ell'\rfloor-\ell
+\partial K^{[m'\!,n']}_{j+2}-\partial K^{[m,n]}_{j+2}
\equiv\ell'',\nonumber\\
&&\ell_{j+3}=-\delta_{j+2}-\delta_{j+4}
-\lfloor p^{-1}\ell'\rfloor-\lfloor p^{-1}\ell\rfloor
+\partial K^{[m'\!,n']}_{j+3}-\partial K^{[m,n]}_{j+3}
\equiv\ell''',\nonumber\\
&&\ell_{j+4}=\delta_{j+4}+\lfloor p^{-1}\ell\rfloor-\ell'
+\partial K^{[m'\!,n']}_{j+4}-\partial K^{[m,n]}_{j+4}\equiv\ell''''.
\label{ell234}\end{eqnarray}
It is easy to see from (\ref{delta24})
and (\ref{betap}) that
\begin{eqnarray}
&&\delta_{j+2}=\cases{\begin{array}{lcr}
0&\hbox{if}&0\leqslant\{\alpha\}<1-a,\\
1&\hbox{if}&1-a\leqslant\{\alpha\}<1,\upstrut
\end{array}}\,\nonumber\\
&&\delta_{j+4}\!=\!\cases{\begin{array}{lcr}
0&\hbox{if}&0\leqslant\{\beta\}<1-b,\\
1&\hbox{if}&1-b\leqslant\{\beta\}<1,\upstrut
\end{array}}\end{eqnarray}
which shows that in the four sectors of the unit square shown in
Fig.~\ref{fig8}$\,$(b), the $\delta_{i}$ pairs are different. As $a$ and
$b$ are functions of $\ell$ and $\ell'$ only, and the entries in Tables
\ref{conftab}, \ref{conftabb} and \ref{oddtab} are also independent of
$j$, the results in (\ref{ell234}) are easily seen to be functions of
$\ell$ and $\ell'$, and of $\{\alpha\}$ and
$\{\beta\}$, but they are {\em independent} of $j$. Therefore, we use the
primed variables $\ell,\cdots,\ell''''$ to denote these $j$-independent
values of $\ell_j,\cdots,\ell_{j+4}$, i.e.\
\begin{equation}
\tilde{\myb\ell}\equiv
[\ell,\ell',\cdots,\ell'''']=[\ell_j,\ell_{j+1},\cdots,\ell_{j+4}].
\label{sumd3}\end{equation}
For different choices of $j$, $[\ell_0,\ell_1,\ell_2,\ell_3,\ell_4]$ is
just a cyclic permutation of $\tilde{\myb\ell}$.

If we let the distance vector $(\ell,\ell')$ between the two
parallelograms $P=P(k_j,k_{j+1})$ and
$P'=P(k_j\!+\!\ell,k_{j+1}\!+\!\ell')$ in (\ref{chipto}) be fixed, while
letting both $k_j$ and $k_{j+1}$ vary from $-\infty$ to $\infty$ (which is
equivalent to ${\cal M}\to\infty$), then the parallelograms $P$ and $P'$
each are in one of eight different odd configurations. Since the
joint probability for $P$ being in the $m$th configuration and $P'$ in the
$m'$th configuration is
$A_{m,m'}(\ell,\ell')$ given in (\ref{overlap}), the double sum in
(\ref{chipto}) can be rewritten as
\begin{eqnarray}
&&{\hat\chi}^{\rm o}({\bf q})=
\lim_{{\cal M}\to\infty}{\frac{1}{{\cal N}{\cal M}^2}}
\sum_{\ell,\ell'}\sum_{k_j, k_{j+1}}\,
\sum_{{\vec K}\vbox to 1.1ex{}^{\rm o}({\myb{\epsilon}})}
{\sum_{{\vec K}{\vbox to 1.1ex{}^{\rm o}}'({\myb{\epsilon}'})}}
U({\vec K^{\rm o}}({\myb{\epsilon}}),{\vec K^{\rm o}}\vphantom{K}'
({\myb{\epsilon}'}))\nonumber\\
&&\quad={\frac{1}{\cal N}}\sum_{\ell,\ell'}\sum_{m=1}^{8}\sum_{m'=1}^{8}
A_{m,m'}(\ell,\ell')
\sum_{n=1}^{N(m)}\sum_{n'=1}^{N(m')}
U({\vec K^{[m,n]}},{\vec K^{[m',n']}}).\hspace*{3em}
\label{dsum}\end{eqnarray}
with ${\cal N}=5p$, cf.\ (\ref{spsum}). Using (\ref{cor}) and
(\ref{dfell}), we find (\ref{sumd}) becomes
\begin{equation}
U({\vec K^{[m,n]}},{\vec K^{[m',n']}})={\rm cos}
\bigl[{\rm Re}(q^*{\textstyle{\sum_{k=0}^4}}\ell_k\zeta^k)\bigr]\,
\langle\sigma\sigma'\rangle_{[\ell_0,\ell_1,\cdots,\ell_4]}^{\rm c},
\label{sumd2}\end{equation}
which is different for different $j$ in view of (\ref{sumd3}).
Since the correlation functions have the cyclic property shown in
(\ref{cyclic}), and ${\hat\chi}^{\rm o}({\bf q})$ can be evaluated by
choosing parallelograms $P$ and $P'$ in (\ref{chipto}) oriented in any one
of the five directions (any choice of $j$), we can rewrite
${\hat\chi}^{\rm o}(q_x,q_y)$ in a more symmetric way by expressing it as
the sum over the five different orientations $j$, and then dividing the
result by 5. This means,
\begin{equation}
{\hat\chi}^{\rm o}({\bf q})=
\sum_{\ell=-\infty}^{\infty}\sum_{\ell'=-\infty}^{\infty}
{\hat\chi}^{\rm o}({\bf q})_{\ell,\ell'}
\label{ochi2}\end{equation}
where (\ref{dsum}) to (\ref{sumd3}) and (\ref{cyclic}) are used to find
\begin{eqnarray}
&&{\hat\chi}^{\rm o}({\bf q})_{\ell,\ell'}={\frac 1{\cal N}}
\sum_{m=1}^{8}\sum_{m'=1}^{8} A_{m,m'}(\ell,\ell')
\sum_{n=1}^{N(m)}\sum_{n'=1}^{N(m')}c({\bf q},\tilde{\myb\ell})\,\langle
\sigma\sigma'\rangle_{[\ell,\ell',\ell'',\ell''',\ell'''']},\nonumber\\
&&c({\bf q},\tilde{\myb\ell})={\frac 1 5}\sum_{j=1}^{5}
\cos {\rm Re}\bigl[q^*\zeta^{j}
(\ell+\ell'\zeta+\ell''\zeta^2+\ell'''\zeta^3+\ell''''\zeta^4)\bigr].\quad
\label{ochiell}\end{eqnarray}
It satisfies the following identities,
\begin{equation}
{\hat\chi}^{\rm o}({\bf q})_{-\ell,-\ell'}=
{\hat\chi}^{\rm o}({\bf q})_{\ell,\ell'},\quad
{\hat\chi}^{\rm o}({\bf q^*})_{\ell',\ell}=
{\hat\chi}^{\rm o}({\bf q})_{\ell,\ell'},
\label{qsymm}\end{equation}
in which ${\bf q^*}\leftrightarrow(q_x,-q_y)$. The former identity in
(\ref{qsymm}) is easily seen as a consequence of the reflection symmetry
in the correlation function
$\langle\sigma\sigma'\rangle=\langle\sigma'\sigma\rangle$; the latter one
is due to five-fold rotation and reflection symmetry.\footnote{In
particular, one can start with the reflection symmetry about the direction
of the $k_{j+3}$ grid-line and its action on the parallelograms
$P(k_j,k_{j+1})$ and $P(k'_j,k'_{j+1})$. One arrives at
$\langle\sigma\sigma'\rangle_{[\ell,\ell',\ell'',\ell''',\ell'''']}=
\langle\sigma\sigma'\rangle_{[\ell',\ell,\ell'''',\ell''',\ell'']}$.
Replacing $j\rightarrow1-j$ in (\ref{ochiell}) then completes the
proof of the second identity in (\ref{qsymm}).}

In the actual calculation, because $\delta_2$ and $\delta_4$ generally
differ in four sectors of the unit square, the contributions to the
susceptibility from these different sectors are evaluated separately.

\subsection{Results}\label{sect52}

To evaluate the wavevector-dependent susceptibility (\ref{ochi2}),
(\ref{ochiell}), we can compute the $A_{m,m'}(\ell,\ell')$ as the
overlap area (\ref{overlap}) and the $\ell$'s from (\ref{ell234}).
In Table \ref{corrtab} and Eq.\ (\ref{corems}) in Section \ref{sect3} we
have expressed the pair-correlation function
$\langle\sigma\sigma'\rangle_{[\ell_0,\cdots,\ell_4]}$ in terms of
Baxter's universal functions $g$ \cite{BaxZI}, which can be evaluated
using methods in our earlier work \cite{AJPq,APmc1,APmc2}.

Near the critical point $k=1$ the leading asymptotic behavior of
the pair-correlation function is the same Painlev\'e III or V scaling
function \cite{AJPq,APmc2} as in the uniform rectangular lattice
\cite{WMTB76}. Therefore, the scaling behavior of the central peak of
$\chi({\bf q})$ in our Penrose Ising model is also known
\cite{APmc2,APsus1,APsus2} to be the same as for the regular Ising model.

More interesting is the incommensurate behavior of $\chi({\bf q})$ as a
function of wavevector ${\bf q}$ and how it changes with temperature or
$k$. At the critical point we expect $\chi({\bf q})$ to be a function
that has everywhere dense $7/4$-th power divergencies, but is locally
integrable. It is nontrivial to show this directly, but this conclusion
seems to impose itself as one approaches the critical point from either
side.

Since the correlation functions decay exponentially away from the critical
point, we find that the ${\hat\chi}^{\rm o}({\bf q})_{\ell,\ell'}$ are
rapidly decreasing functions of $\ell$ and $\ell'$. Putting terms of about
the same order of magnitude together, we find
\begin{equation}
{\hat\chi}^{\rm o}({\bf q})={\hat\chi}^{\rm o}({\bf q})_{0,0}+
\sum_{\ell=1}^{\infty}{\cal S}_{\ell},\quad
{\cal S}_{\ell}=2\!\sum_{n=-\ell+1}^{\ell}[{\hat\chi}^{\rm o}({\bf q})_{\ell,n}
+{\hat\chi}^{\rm o}({\bf q^*})_{\ell,n-1}],
\label{sl}\end{equation}
using both identities in (\ref{qsymm}). We shall give density plots of
several cases next, displaying the temperature dependence more clearly.
However, it must be said that we can calculate
${\hat\chi}^{\rm o}({\bf q})$ to very high precision in the cases shown,
which fact is not clear from looking at these density plots.

We shall give plots both above the critical temperature $T_{\rm c}$,
($k_>\!\equiv\!k\!<\!1$), and below $T_{\rm c}$,
($k_<\!\equiv\!1/k\!<\!1$). The value of modulus $k$ corresponds to the
row correlation length\cite{MWbk}
\begin{equation}
\xi=1/|\hbox{arsinh}(1/\sqrt{k})-\hbox{arsinh}(\sqrt{k})|
\label{onscorr}\end{equation}
of the symmetric square-lattice Ising model for $T>T_{\rm c}$.
For $T<T_{\rm c}$ the true value of this row correlation length is
$\xi/2$ with $\xi$ again given by (\ref{onscorr}).\cite{MWbk,WMTB76} 

\begin{figure}[tbh]
~\vskip0in\hskip0in\epsfclipon
\epsfxsize=0.475\hsize\epsfbox{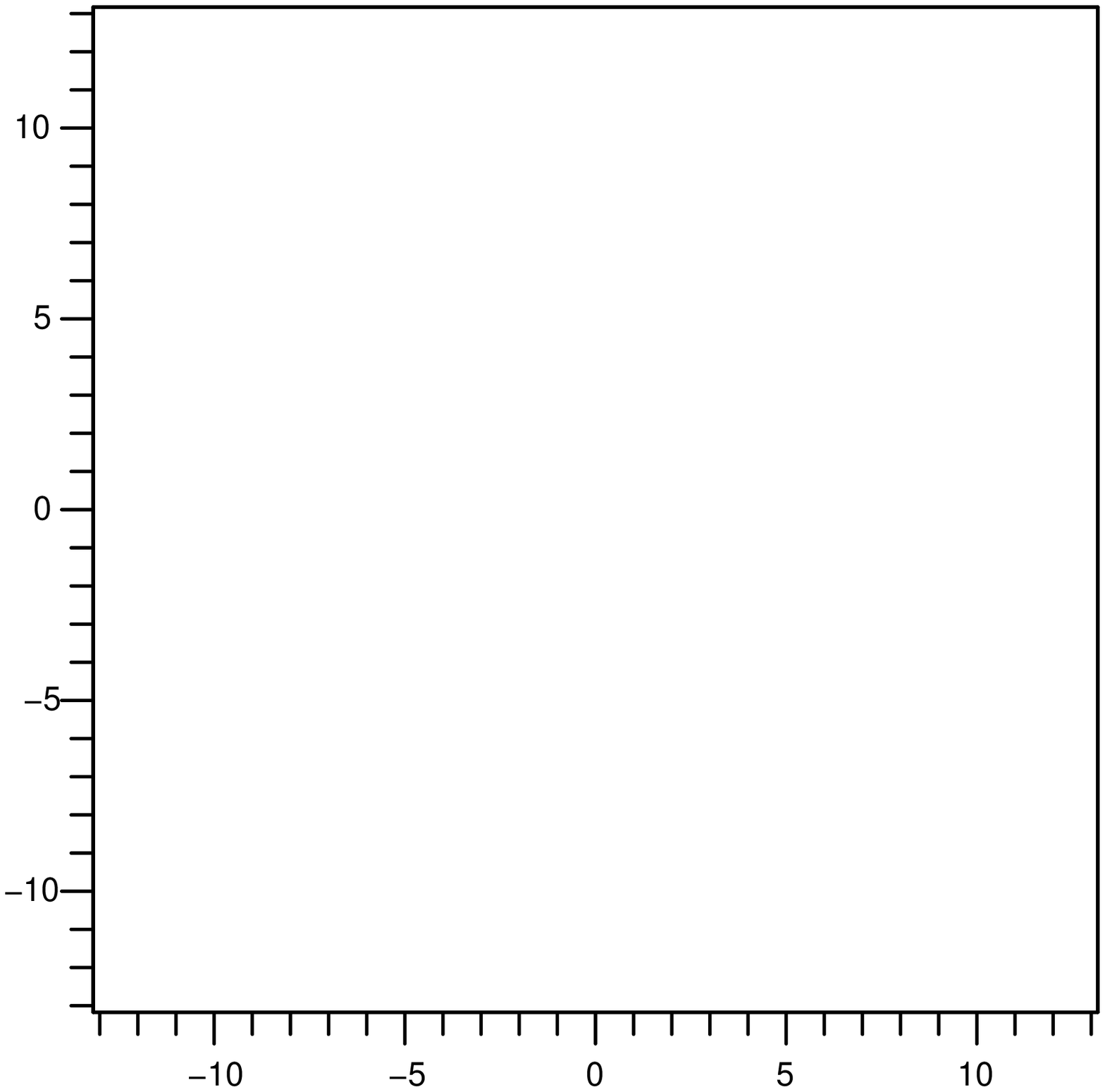}\hspace*{-0.4225\hsize}%
\epsfxsize=0.370\hsize\raisebox{0.0517\hsize}{\epsfbox{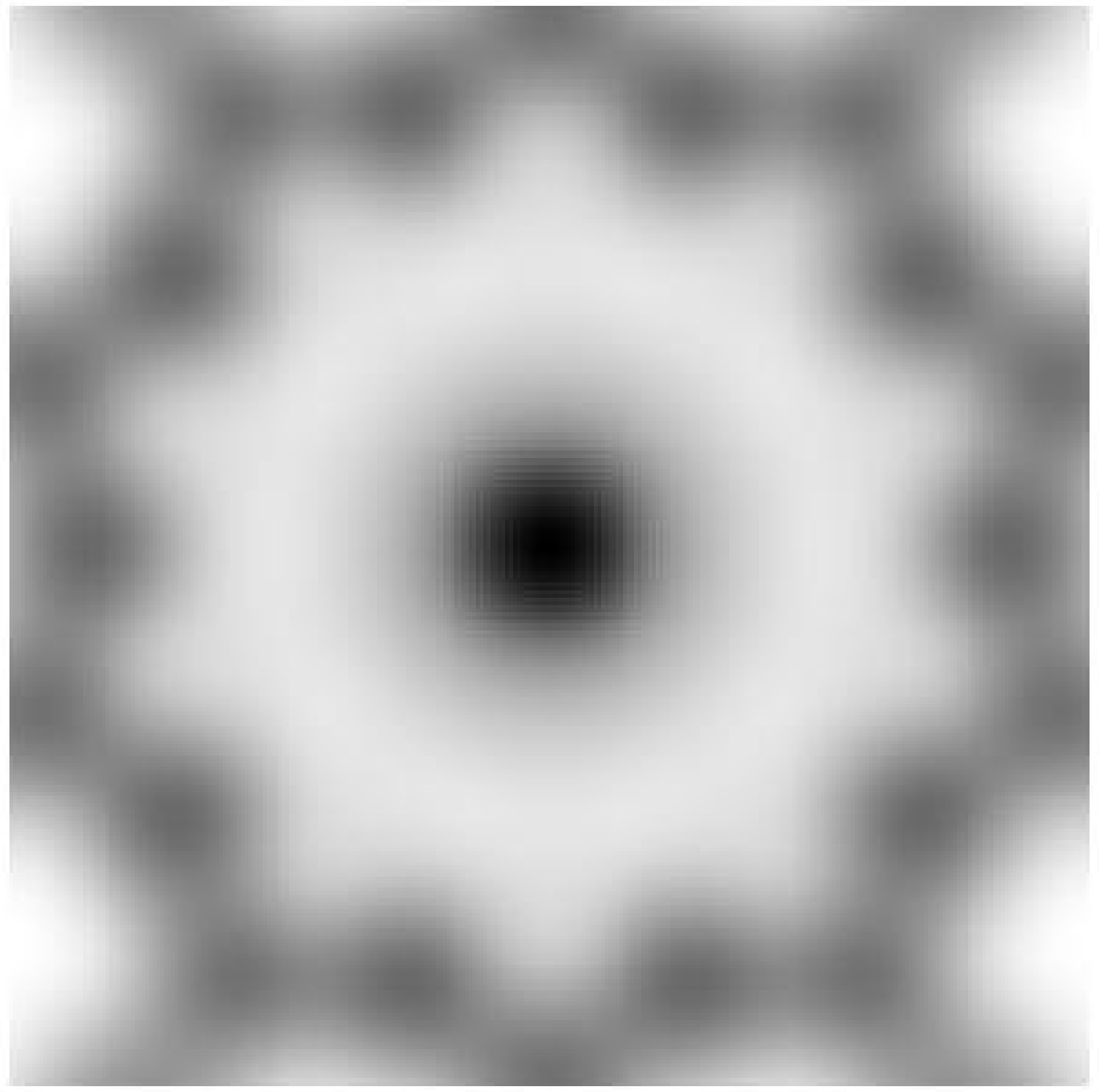}}\hfil
\epsfxsize=0.475\hsize\epsfbox{fig910z.eps}\hspace*{-0.4225\hsize}%
\epsfxsize=0.370\hsize\raisebox{0.0517\hsize}{\epsfbox{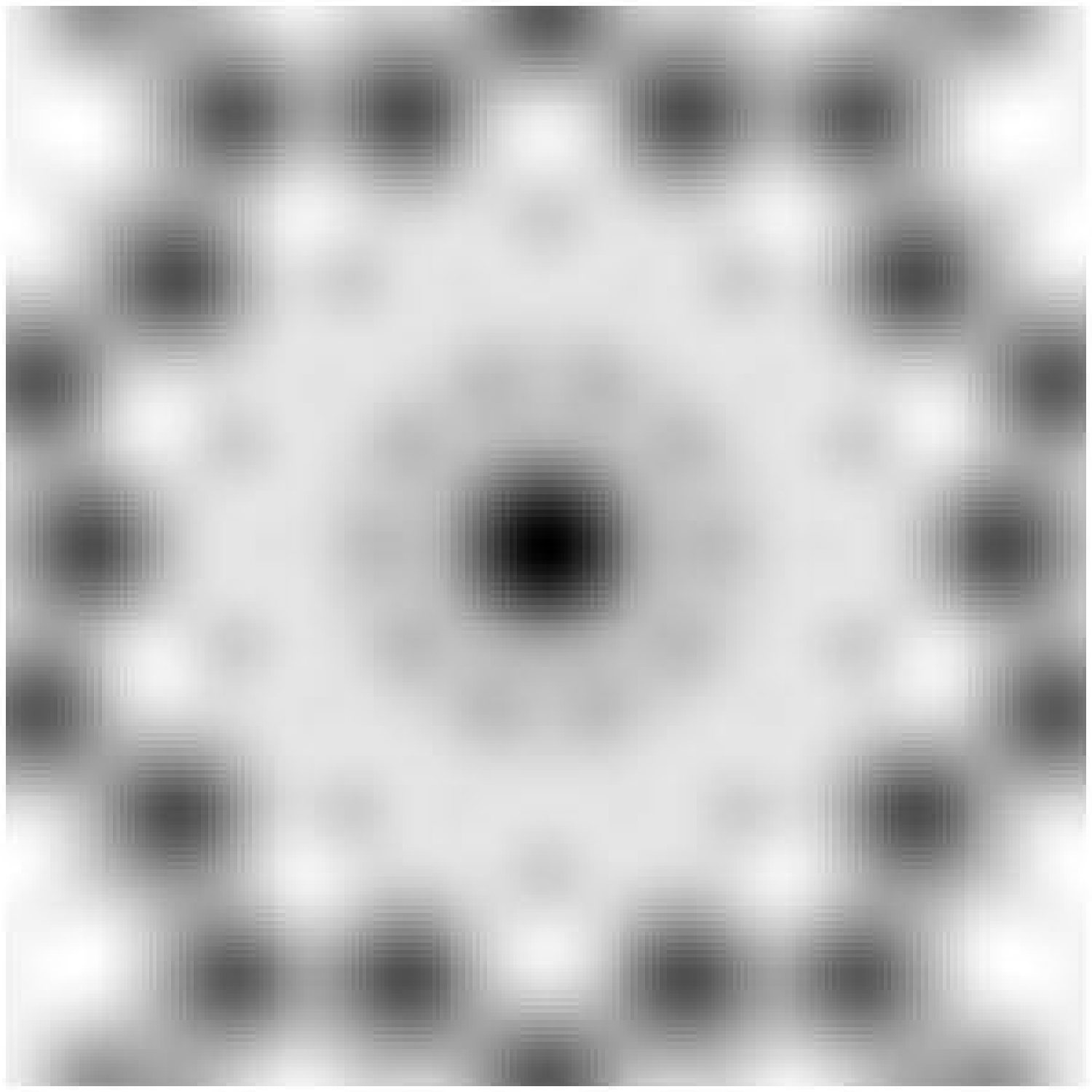}}
\vskip6pt
\hbox to\hsize{\hspace*{12pt}\footnotesize
\hfil (a) \hfil\hfil (b)\hfil\hspace*{-2pt}}
\vskip-0.2in
\hskip0pt\caption{Density plot of the ${\bf q}$-dependent susceptibility
showing $1/\chi(q_x,q_y)$ for the $Z$-invariant Ising model on a Pentagrid
Lattice (Penrose Tiles) at very low temperature: (a) $\xi\approx 0.5$, or
$k_<=.04847302$; (b) $\xi\approx 1$, or $k_<=.2363562$, both for
$T<T_{\rm c}$.}
\label{fig9}
\end{figure}

At very low temperature, we only need to consider ${\cal S}_{\ell}$ for
very small $\ell$. For $\ell,\ell'\leqslant2$ the joint probabilities
$A_{m,m'}(\ell,\ell')$ can be easily evaluated by hand as it only involves
the calculation of areas of triangles and rectangles. For $k_<=.04847302$,
which corresponds to $\xi\approx1/2$, we find ${\cal S}_{\ell}<10^{-10}$
for $\ell>2$. The density plot for $1/{\hat\chi}^{\rm o}({\bf q})$ is shown
in Fig.~\ref{fig9}$\,$(a) for $-4\pi\leqslant q_x,q_y\leqslant 4\pi$ where
${\bf q}=(q_x,q_y)$. We find ten-fold symmetry, corresponding to the
five-fold symmetry of the Penrose tiling. 

For $k_<=.2363562$ ($\xi\approx1$), we find it necessary to consider all
${\cal S}_{\ell}$ for $\ell\leqslant4$. As the temperature increases,
larger and larger $\ell$'s are needed. To evaluate the joint probability
by hand is no longer feasible. To symbolically program the calculation for
any values of $\ell$ and $\ell'$ is highly nontrivial, as there are
many different situations to take into account. It took us several months
to sort out all cases, programming the calculation using Maple.
\begin{figure}[ptbh]
~\vskip0in\hskip0in\epsfclipon
\epsfxsize=0.475\hsize\epsfbox{fig910z.eps}\hspace*{-0.4225\hsize}%
\epsfxsize=0.370\hsize\raisebox{0.0517\hsize}{\epsfbox{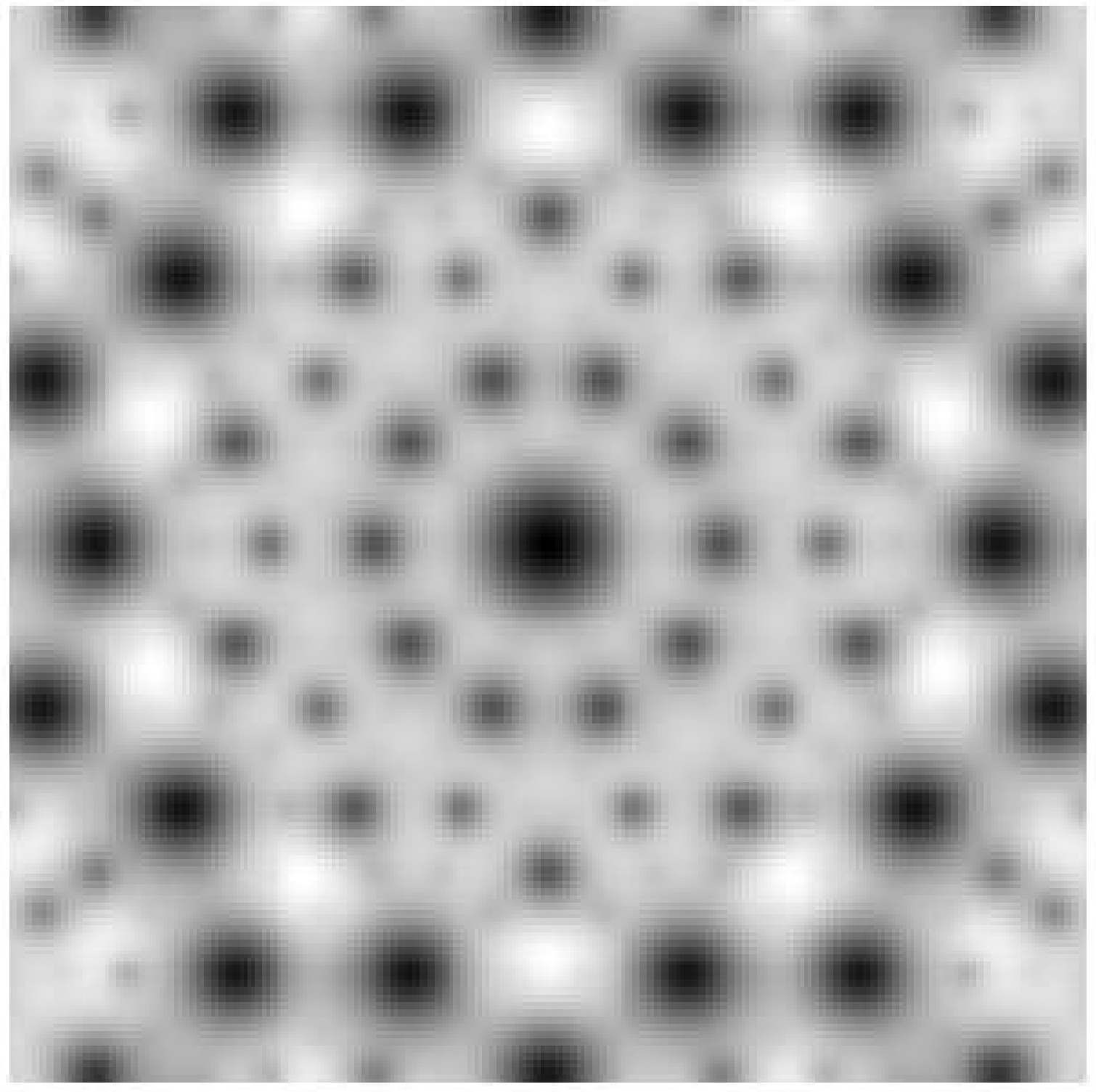}}\hfil
\epsfxsize=0.475\hsize\epsfbox{fig910z.eps}\hspace*{-0.4225\hsize}%
\epsfxsize=0.370\hsize\raisebox{0.0517\hsize}{\epsfbox{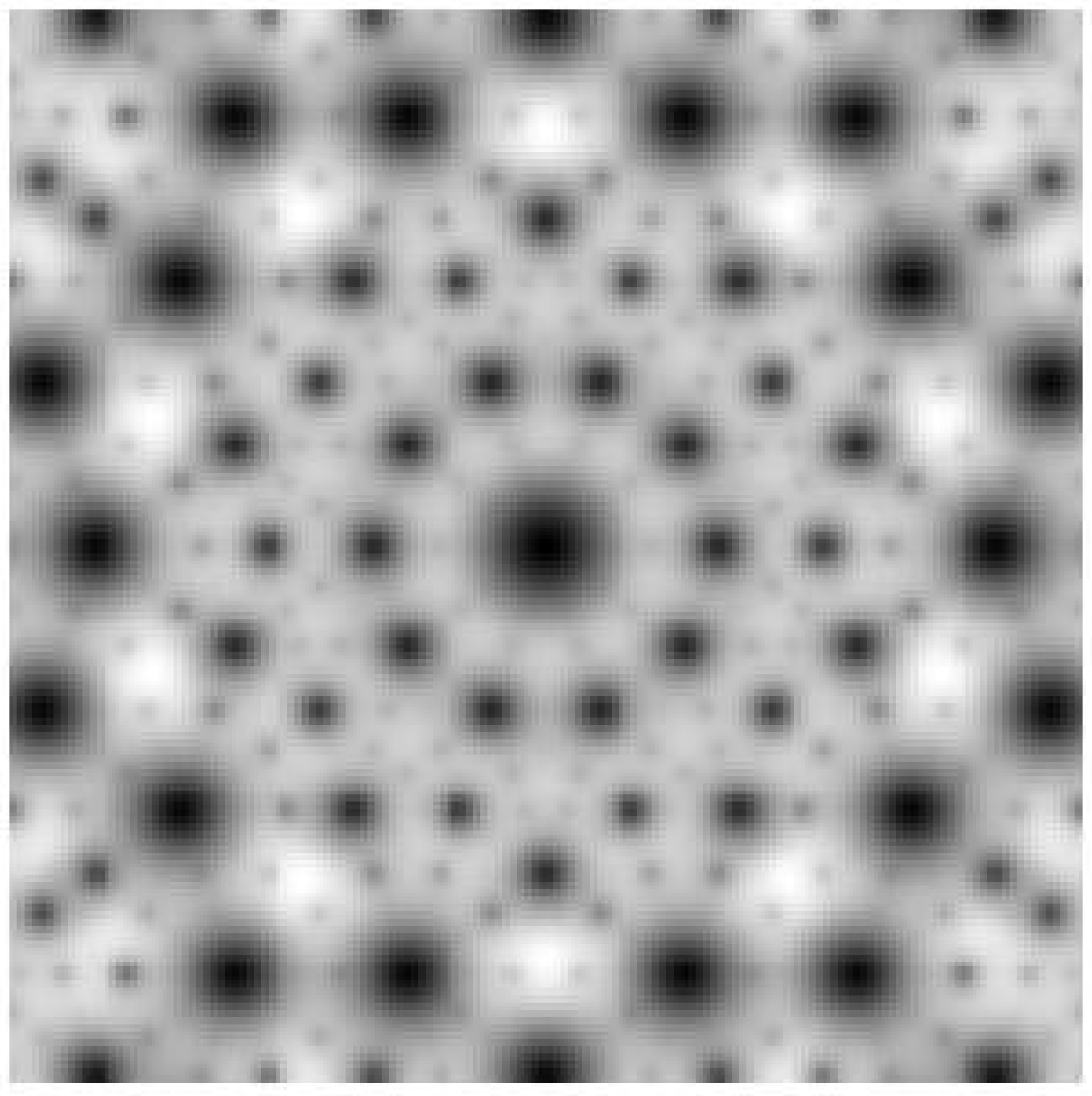}}
\vskip6pt
\hbox to\hsize{\hspace*{12pt}\footnotesize
\hfil (a) \hfil\hfil (b)\hfil\hspace*{-2pt}}
\vskip0.1in
\epsfxsize=0.475\hsize\epsfbox{fig910z.eps}\hspace*{-0.4225\hsize}%
\epsfxsize=0.370\hsize\raisebox{0.0517\hsize}{\epsfbox{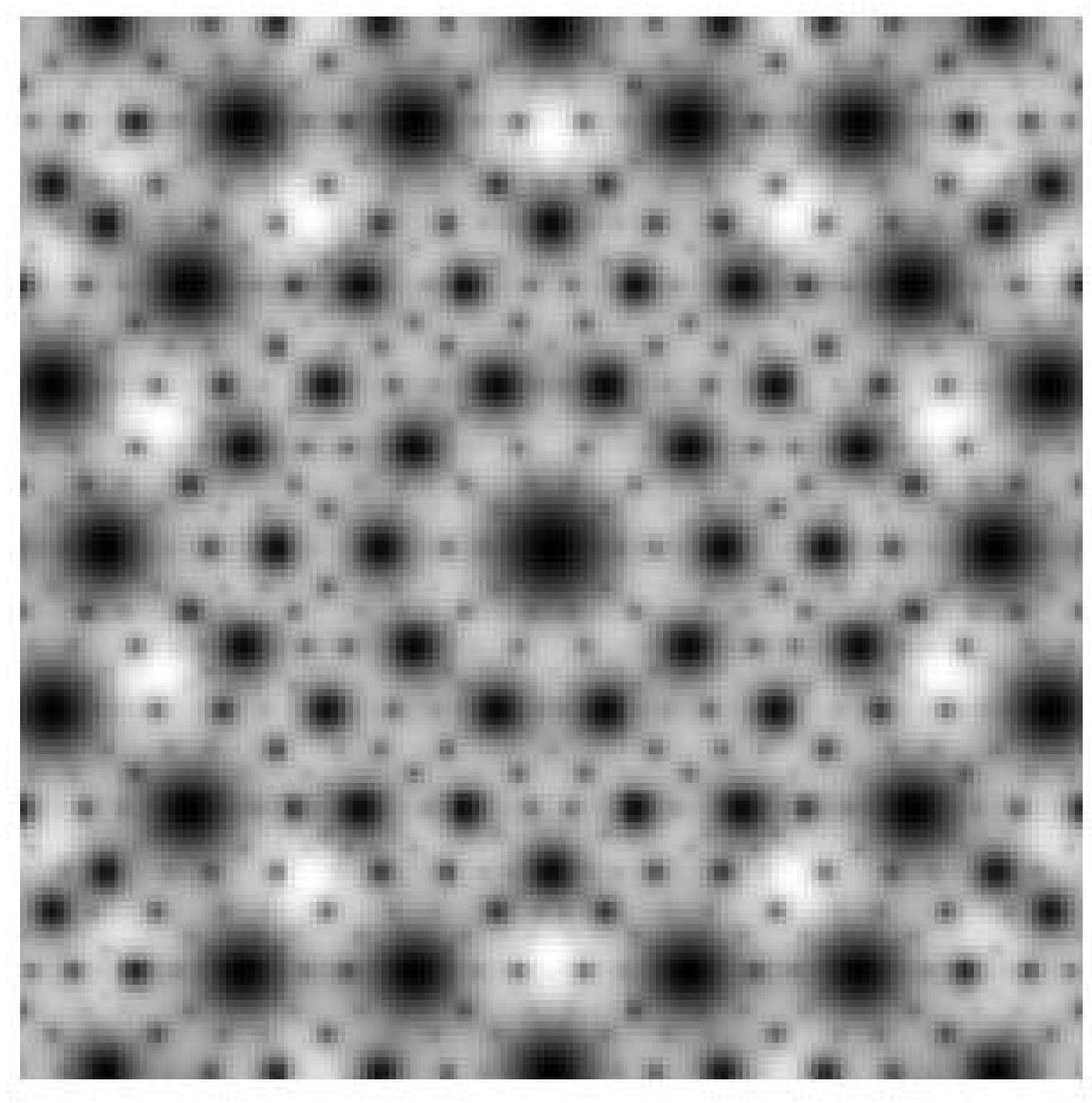}}\hfil
\epsfxsize=0.475\hsize\epsfbox{fig910z.eps}\hspace*{-0.4225\hsize}%
\epsfxsize=0.370\hsize\raisebox{0.0517\hsize}{\epsfbox{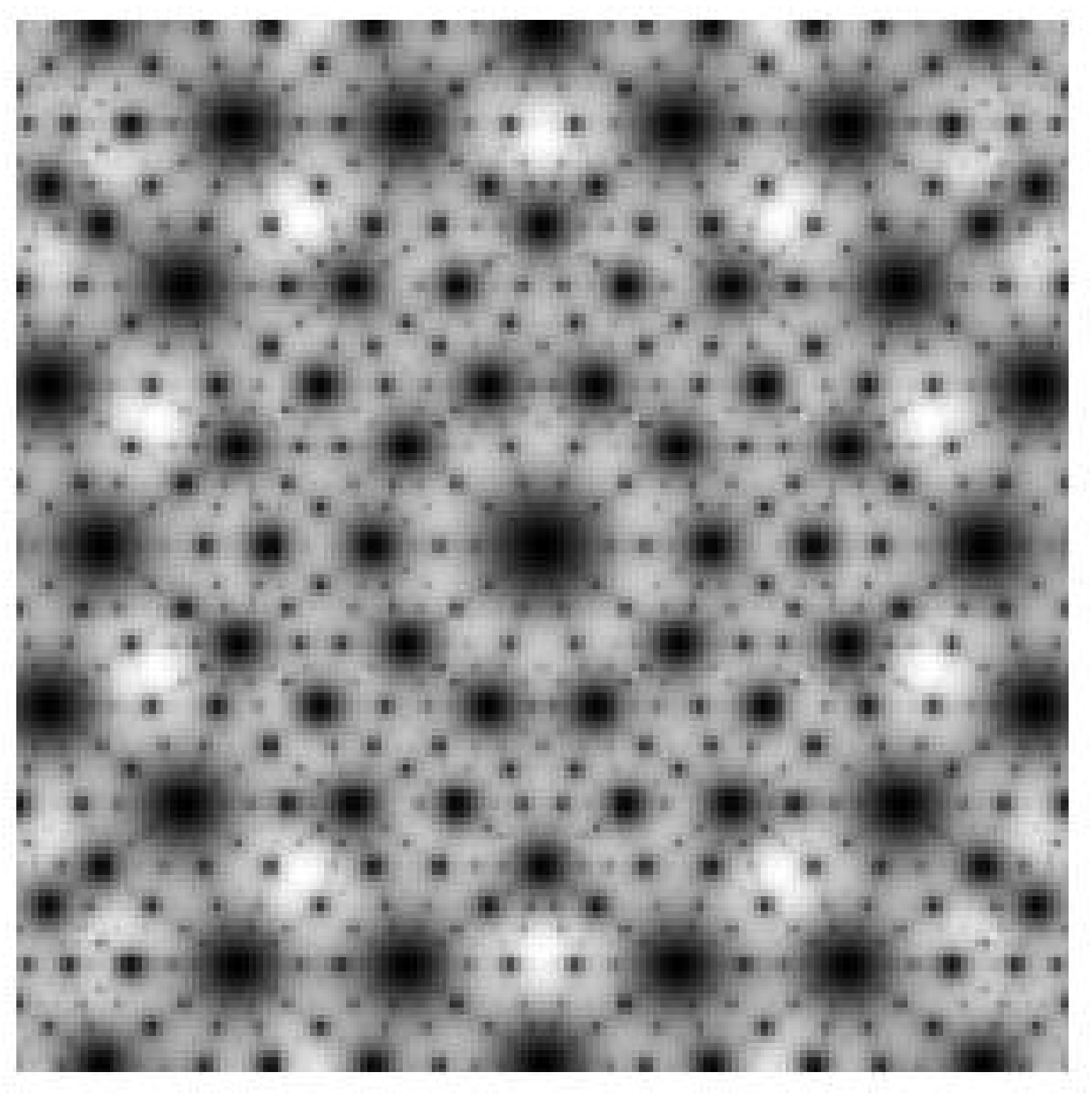}}
\vskip6pt
\hbox to\hsize{\hspace*{12pt}\footnotesize
\hfil (c) \hfil\hfil (d)\hfil\hspace*{-2pt}}
\vskip-0.2in
\hskip0pt\caption{Density plots of the ${\bf q}$-dependent susceptibility,
plotting $1/\chi({\bf q})$ versus $q_x$ and $q_y$: (a) $k_<=.7018662$ or
$\xi\approx 4$; $\,$(b) $k_<=.8379187$ or $\xi\approx 8$, both for
$T<T_{\rm c}$. Two corresponding plots for the dual models, with
$T>T_{\rm c}$ and identical values of $k_>$ and $\xi$, are given
in (c) and (d).}
\label{fig10}
\end{figure} 

\begin{figure}[tbph]
~\vskip-1in\hskip0in\epsfclipon
\epsfxsize=1.0\hsize\epsfbox{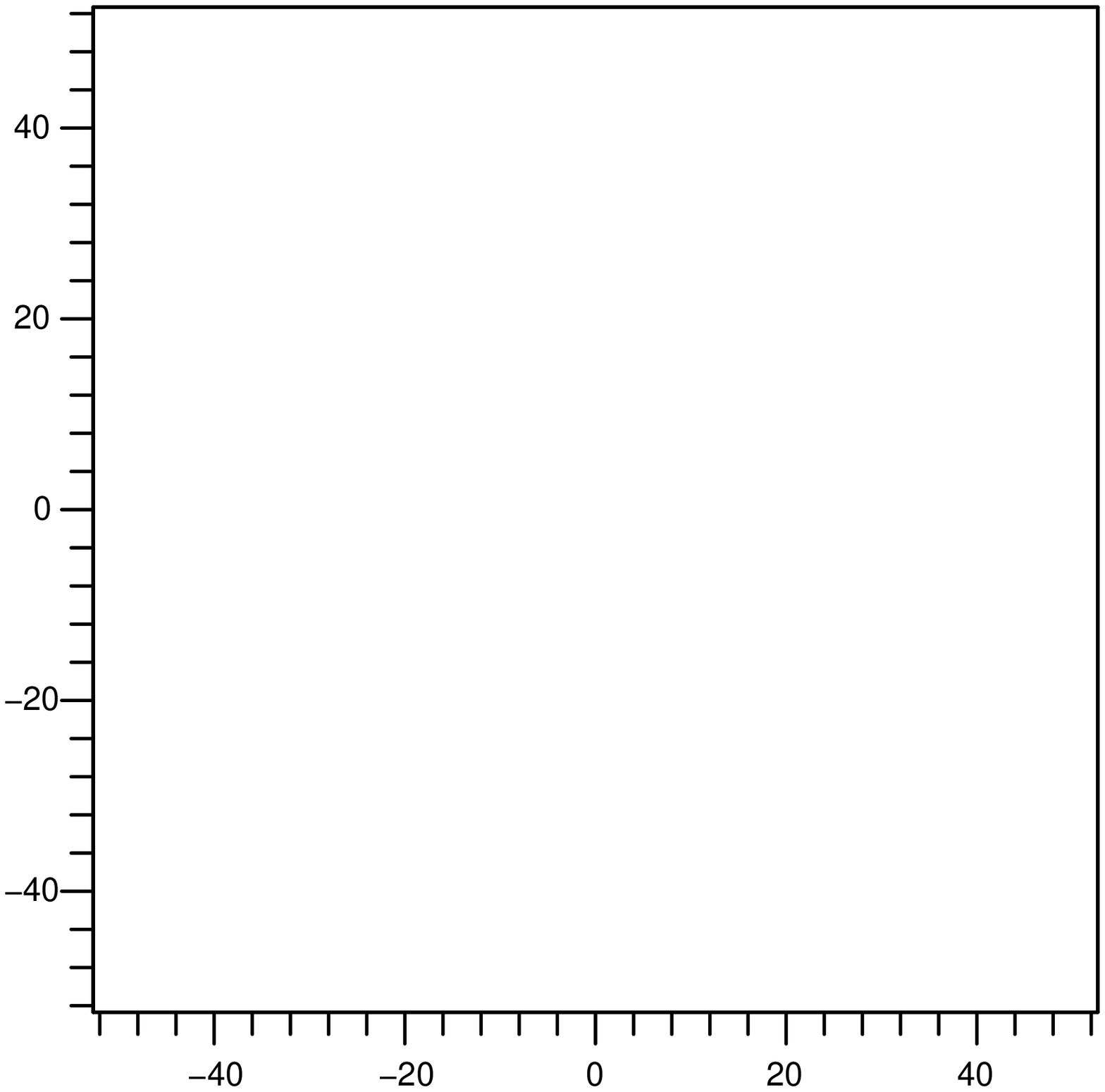}\hspace*{-0.8895\hsize}%
\epsfxsize=0.779\hsize\raisebox{0.1088\hsize}{\epsfbox{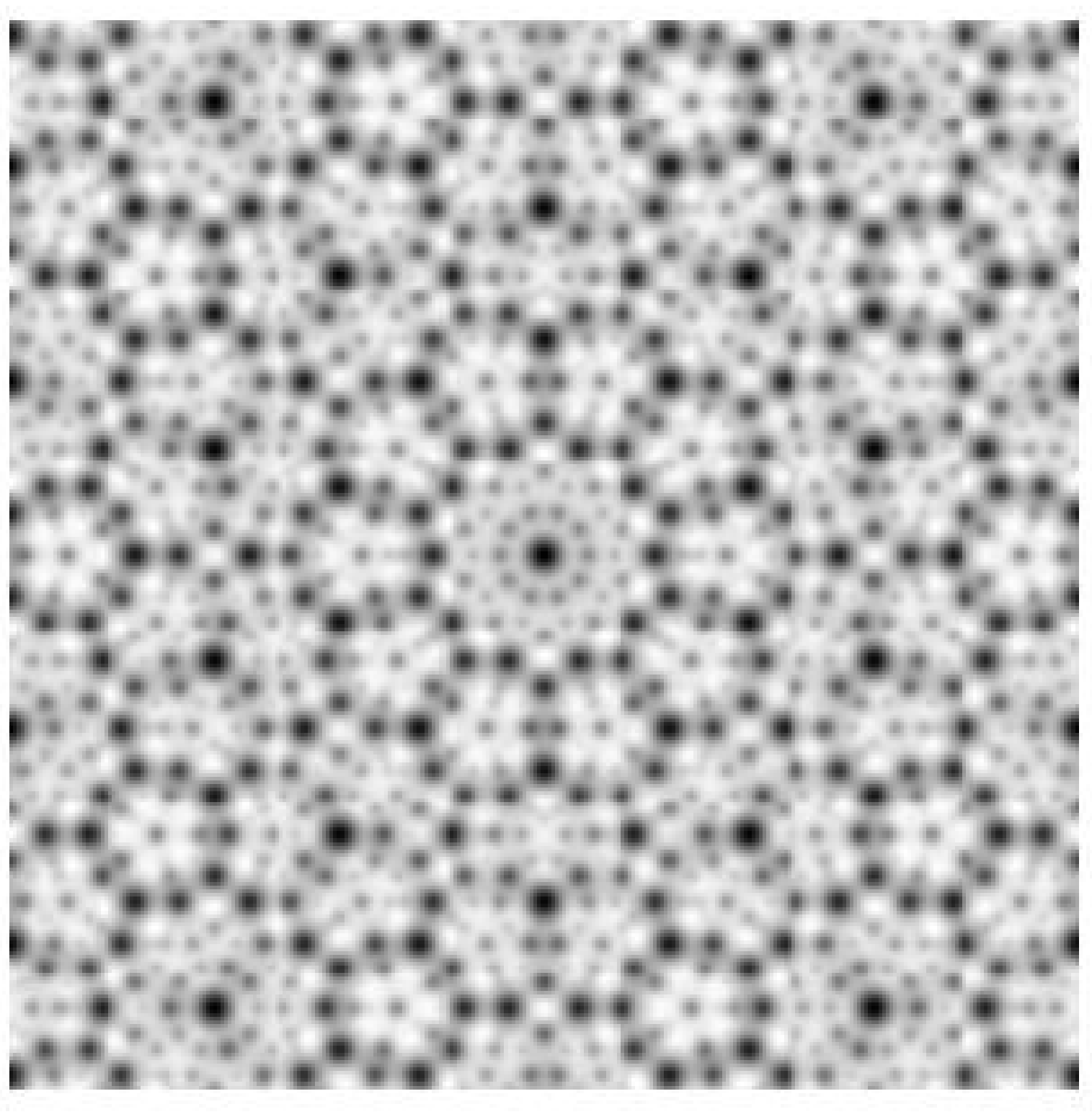}}
\vskip0.2in
\hskip0pt\caption{Density plot of the ${\bf q}$-dependent susceptibility
showing $1/\chi({\bf q})$ at $k_<=.4912758$ or $\xi\approx 2$ and
$T<T_{\rm c}$.}
\label{fig11}
\end{figure}

A density plot for $k_<=.2363562$ is shown in Fig.~\ref{fig9}$\,$(b).
Plots of $1/{\hat\chi}^{\rm o}({\bf q})$ for $\xi\approx 4$ and
$\xi\approx 8$ are shown in Fig.~\ref{fig10}$\,$(a) and
Fig.~\ref{fig10}$\,$(b), together with corresponding plots for the
dual cases with $T>T_{\rm c}$ in Fig.~\ref{fig10}$\,$(c) and
Fig.~\ref{fig10}$\,$(d). We can see clearly that the number of visible
peaks increases as $T\to T_{\rm c}$ and that this effect is more
pronounced as $T_{\rm c}$ is approached from above. In Fig.~\ref{fig11},
a density plot for $1/{\hat\chi}^{\rm o}({\bf q})$ at $\xi=2$ is given for
$-16\pi\leqslant q_x,q_y\leqslant 16\pi$, and we can already see some
evidence for the quasiperiodic pattern of the ${\bf q}$-dependent
susceptibility in the full ($q_x,q_y$)-plane.

\subsection{The q-dependent susceptibility
$\myb{\chi^{\rm e}({\bf q})}$}\label{sect53}

\begin{figure}[tbh]
~\vskip0.01in\hskip0.1in
\epsfxsize=0.45\hsize\epsfclipon
\epsfbox{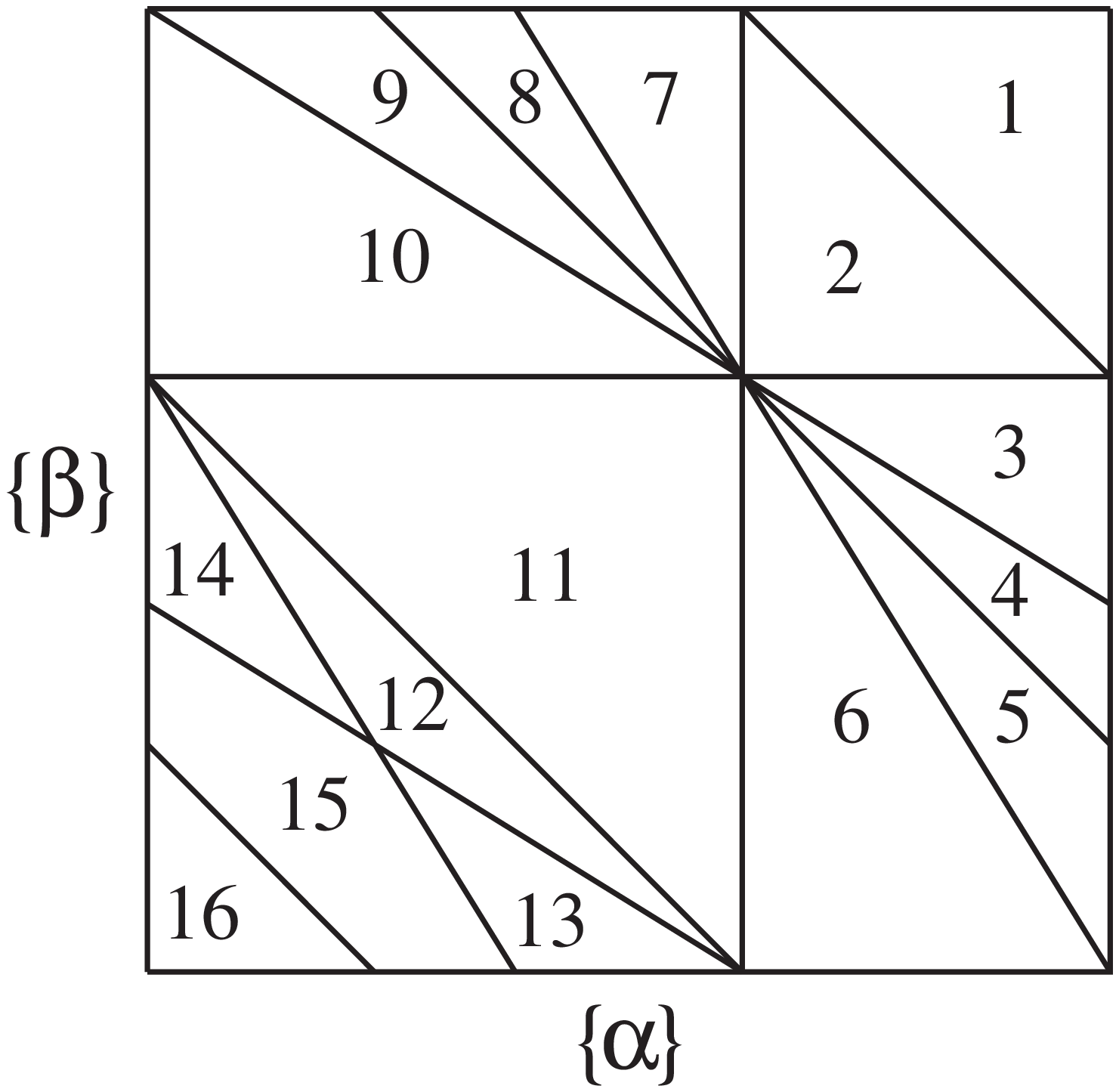}\hfil
\epsfxsize=0.45\hsize
\epsfbox{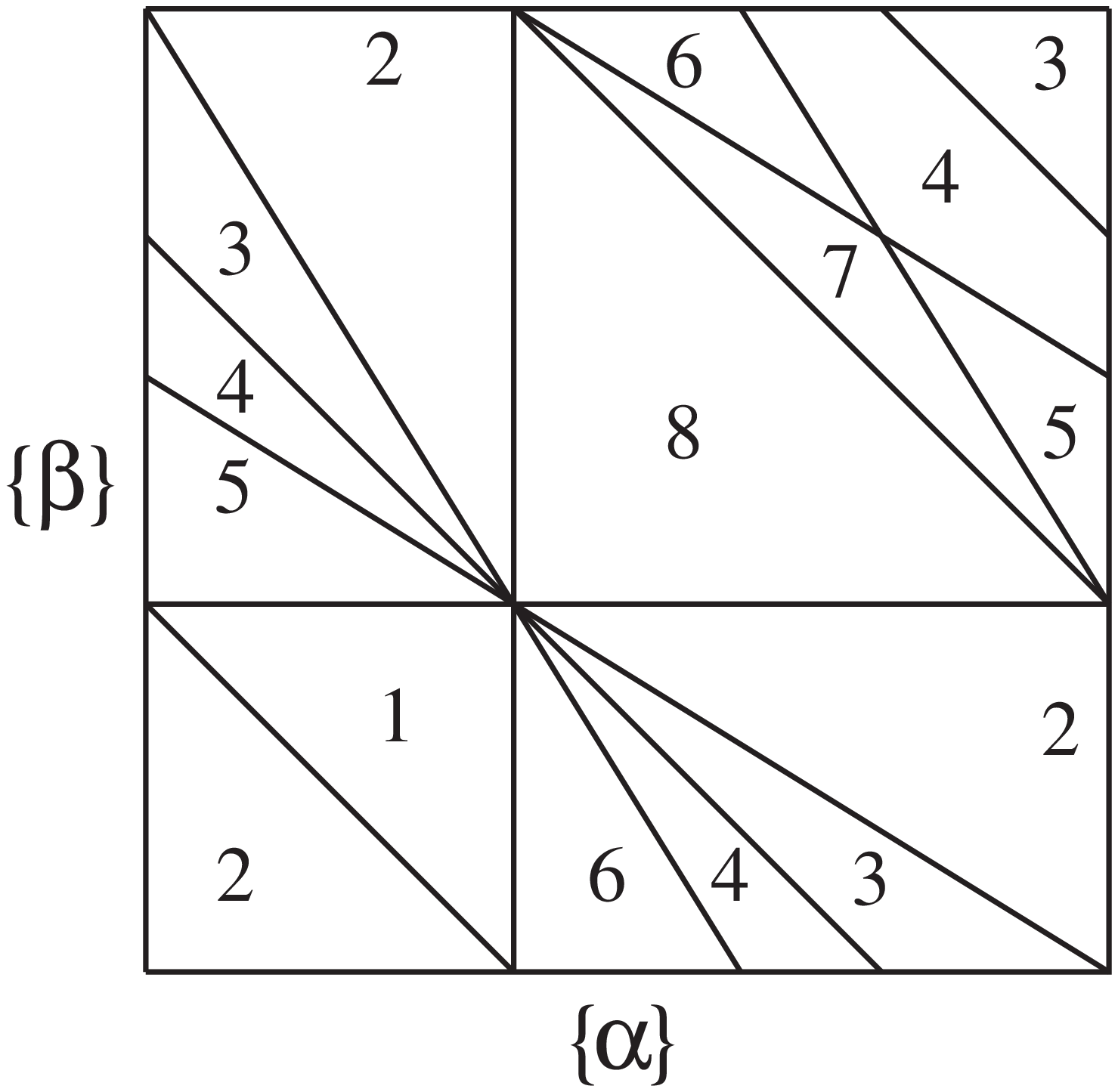}
\vskip6pt
\hbox to\hsize{\hspace*{12pt}\footnotesize
\hfil (a) \hfil\hfil (b)\hfil\hspace*{-2pt}}
\vskip-0.2in
\hskip0pt\caption{(a) The sixteen even spin configurations of
$P(k_j,k_{j+1})$ plotted with
$\{\alpha\}$ and $\{\beta\}$ as the horizontal and vertical axis. (b) The 
eight odd spin configurations of
$P(k_j\!+\!1,k_{j+1}\!+\!1)$ plotted with
$\{\alpha\}$ and $\{\beta\}$.}
\label{fig12}
\end{figure}

Finally, we show that the ${\bf q}$-dependent susceptibility 
${\hat\chi}^{\rm e}({\bf q})$ of the even sublattice is identical to
${\hat\chi}^{\rm o}({\bf q})$ of the odd sublattice. From Tables
\ref{conftab} and \ref{conftabb}, we find there are 16 different even
configurations for $P(k_j,k_{j+1})$. The corresponding 16 regions are
plotted in the unit square with $\{\alpha\}$ and $\{\beta\}$ along the
axes in Fig.~\ref{fig12}$\,$(a). The regions for the 8 odd spin
configurations of $P(k_j\!+\!1,k_{j+1}\!+\!1)$ are also plotted with
$\{\alpha\}$ and $\{\beta\}$ as axes in Fig.~\ref{fig12}$\,$(b). It is easy
to see that after inverting one of the squares, the two figures are
identical up to labeling.

Furthermore, looking at Fig.~\ref{fig12}$\,$(b), we see that the five
disjoint regions $2,\cdots,6$ become connected, if we impose periodic
boundary conditions on the square. This ``wrapping on a torus" is
consistent with moving the slices as discussed below Eq.\ (\ref{betap}).
We next compare how the even regions in Fig.~\ref{fig12}$\,$(a) relate
under the same periodic boundary conditions, to see if this is somehow
true here also. As first examples we look at the two even configurations
$\mathrm{e}(7)$ and $\mathrm{e}(13)$ and find that they have the same
number of sites, and their difference vectors are related by
\begin{equation}
\delta{\vec K^{\mathrm{e},7}}=(0,-1,1)+\delta{\vec K^{\mathrm{e},13}}.
\label{eveneven1}\end{equation}
For the other regions, we similarly find equal numbers of sites and
\begin{eqnarray}
&&\delta{\vec K^{\mathrm{e},10}}=(1,-1,0)+\delta{\vec
K^{\mathrm{e},1}},\quad 
\delta{\vec K^{\mathrm{e},3+m}}=(1,-1,0)+\delta{\vec K^{\mathrm{e},14+m}},
\hspace*{2em}\nonumber\\
&&\delta{\vec K^{\mathrm{e},6}}=(0,-1,1)+\delta{\vec
K^{\mathrm{e},1}},\quad 
\delta{\vec K^{\mathrm{e},8+n}}=(0,-1,1)+\delta{\vec K^{\mathrm{e},15+n}},
\label{eveneven2}\end{eqnarray}
where $m=0,1,2$ and $n=0,1$.

The difference vectors of the odd configurations in Fig.~\ref{fig12}$\,$(b)
are also related to those of the even configurations in
Fig.~\ref{fig12}$\,$(a), i.e.
\begin{eqnarray}
&&\delta{\vec K^{\mathrm{e},1}}=(1,-1,1)-\delta{\vec
K^{\mathrm{o},2}},\quad 
\delta{\vec K^{\mathrm{e},2}}=(1,-1,1)-\delta{\vec K^{\mathrm{o},1}},
\hspace*{2em}\nonumber\\
&&\delta{\vec K^{\mathrm{e},11+n}}=(0,1,0)-\delta{\vec
K^{\mathrm{o},8-n}},\quad n=0,\cdots,5.
\label{evenodd}\end{eqnarray}
If the dependences of difference vectors on the values of
$\{\alpha\}$ and $\{\beta\}$ are included in the equations, we may relate
the even spins in $P(k_j,k_{j+1})$ with the odd spins in
$P(k_j\!+\!1,k_{j+1}\!+\!1)$ by
\begin{equation}
\delta{\vec K^{\rm e}}[\{\alpha\},\{\beta\}]=(1,-1,1)-(\delta_2,
-\delta_2\!-\!\delta_4,
\delta_4)-\delta{\vec K^{\rm o}}[\{-\alpha\},\{-\beta\}],
\label{devenodd}\end{equation}
where $1-\{x\}=\{-x\}$, for $x$ not an integer, and
\begin{equation}
\delta_{2}=\lfloor{1-\{\alpha\}+p^{-1}}\rfloor,\quad
\delta_{4}=\lfloor{1-\{\beta\}+p^{-1}}\rfloor.
\end{equation}
Consider the pentagrid with $\gamma_j\to -\gamma_j$
and denote its parallelograms by ${\bar P}(k_j,k_{j+1})$ such that
\begin{equation}
\bar\alpha(-k_{j+1})=-\alpha(k_{j+1}),\quad
\bar\beta(-k_{j})=-\beta(k_{j}).
\label{balpha}\end{equation}
Then it is easy to show that
\begin{equation}
\delta_{2}+\lfloor{-\alpha(k_{j+1})}\rfloor
=\lfloor{\bar\alpha(1\!-\!k_{j+1})}\rfloor,\quad
\delta_{4}+\lfloor{-\beta(k_j)}\rfloor
=\lfloor{\bar\beta(1\!-\!k_j)}\rfloor.
\end{equation} 
Now, the integer vectors of even spins in $P(k_j,k_{j+1})$ given by
(\ref{kz}), (\ref{kj12}) and (\ref{kj3}) can be shown to relate to the
integer vectors of the odd spins in ${\bar P}(1\!-\!k_j,1\!-\!k_{j+1})$ by
\begin{equation}
{\vec K^{\rm e}}({\myb{\epsilon'}})
=(1,1,1,1,1)-{\vec K^{\rm o}}({\myb{\epsilon}}).
\label{ievenodd}\end{equation}
This equation is consistent with the fact that the index of the odd spins
is either 1 or 3, while the index of the even spins is either 2 or 4. Since
the even spins in the original pentagrid are related to the odd spins in a
different pentagrid, and the joint probabilities are independent of
shifts, we have shown that the susceptibility of the even sublattice is
identically the same as the one of the odd sublattice.

\section{Conclusions and Final Remarks}\label{sect6}

In this paper we have presented a systematic way of evaluating the
averaged pair-correlation function of a $Z$-invariant ferromagnetic Ising
model with spins on half the sites of a Penrose tiling and Ising
interactions across the diagonals of the rhombuses.

Next, we have found that the ${\bf q}$-dependent susceptibility of this
model is a superposition of incommensurate everywhere-dense peaks, though
not many peaks are visible at temperatures very far away from $T_{\rm c}$.
For $T<T_{\rm c}$ these peaks add a diffuse background to the Bragg
peaks due to the spontaneous magnetization.
Since the $S_{\ell}$ in (\ref{sl}) consists of $4\ell$ terms of the same
order of magnitude, we compare their contributions. We find that the
number of peaks of $S_{\ell}$ increases as $\ell$ increases, but that the
numbers at fixed $\ell$ are almost independent of temperature, even though
the magnitudes of the peaks change as the temperature varies.

As $T\to T_{\rm c}$, the correlations decay more and more slowly, so the
$S_{\ell}$'s increase, and more and more of these $S_{\ell}$'s are to be
included in the numerical evaluation of the ${\bf q}$-dependent
susceptibility, which accounts for the ever-increasing number of peaks.
This is unlike the behavior of the Fibonacci Ising models, considered
earlier,\cite{AJPq,APmc1} where the ferromagnetic aperiodic Fibonacci
lattice behaves almost like the regular Ising model.

Moreover, the ${\bf q}$-dependent susceptibility is not a periodic function
of $q_x$ or $q_y$. This behavior is different from that of aperiodic
models defined on regular lattices.\cite{AJPq} This is because, when the
lattice is aperiodic, we cannot separate the average of the correlation
functions from the exponential (or cosine) terms, which contain the
information about the lattice structure, as can be seen from (\ref{ochi2})
and (\ref{ochiell}).

At $T_{\rm c}$, the ${\bf q}$-dependent susceptibility has everywhere-dense
divergences with the Ising exponent $7/4$, but is still locally integrable.
Away from $T_{\rm c}$ the $\chi({\bf q})$ is a continuous function.
The Bragg peaks below $T_{\rm c}$ form a set of everywhere-dense Dirac
delta functions of various strengths, but their sum is also locally
integrable. These are strange objects and de Bruijn
initiated their mathematical study\cite{Bruijn3}.

One of the main results of the current paper is that it provides a new
method for doing calculations of probabilities on Penrose tilings. In 
Section \ref{sect4}, the calculation of the joint probability of the
configurations of two parallelograms on the pentagrid is reduced to linear
programming.

Penrose tilings may be obtained by projecting certain subsets of the
$\mbox{\mymsbm Z}^5$ lattice into the plane\cite{Bruijn1}. The
frequencies of the different types of vertices are then given as areas in
the orthogonal spaces\cite{Bruijn1}. This method, which is known as the
cut-and-project method, has been applied and generalized by many authors
\cite{Mackay2,CN,GRh,KGR,DK,KKL,LSt0,Jar,Elser,RGS,RRG,BKSZ,BJKS}.
The equivalence of the projection method and a generalized grid method
has been demonstrated\cite{CN,GRh,KGR}. The positions of the Bragg peaks
have been worked out \cite{Mackay2,DK,KKL,LSt0,Jar,Elser}, together with
the values of probabilities of local configurations. The Penrose tilings
can even be obtained from projections in a four-dimensional root
lattice\cite{BKSZ,BJKS}.

It would be interesting to obtain results for joint probabilities similar
to ours also by cut-and-project methods. This would generalize the
windowing method of Baake and Grimm\cite{BG}. We have not pursued this
here, as our model is defined in terms of rapidity lines on the pentagrid.

Another possible generalization of our work is to consider Penrose
tilings that are periodic in either one or both directions. Finite
approximants to the aperiodic Penrose tiling have been constructed
through periodic pentagrids or projection
methods.\cite{TFUT,TFUT2,TFUT3,RGS}

The mathematician Robinson has brought to our attention an exercise in the
book by Gr\"unbaum and Shephard \cite{GrSh} where a tiling with Penrose
rhombuses can be cut into patches, and then converted into an aperiodic
set of 24 Wang's tiles. We have found that each patch in the Penrose
tiling described in the exercise is in fact the image of a parallelogram
under the mapping (\ref{penrose}). Thus these 24 configurations of the
parallelogram can be easily converted into Wang's tiles.

Finally, in our previous studies \cite{AJPq} we have examined the
$q$-dependent susceptibility $\chi({\bf q})$ of some quasiperiodic Ising
models on the square lattice defined in terms of Fibonacci sequences. It
may be of interest to study models based on other sequences such as the
aperiodic sequences studied by de Bruijn\cite{Bruijn2} or Tracy\cite{Tr2}.
We may ask what effect this has on the mixed interaction cases and also
if it makes any difference for purely ferromagnetic models.

\section*{\sffamily\bfseries\normalsize ACKNOWLEDGMENTS}
\par\hspace*{\parindent}%
An earlier version of this work has been presented to Professor
M.~E.\ Fisher at a conference in honor of his seventieth birthday. We are
most thankful to Dr.\ M.\ Widom for his interest in using exactly
solvable models to study quasicrystals, which led us to start this
work. We also thank Dr.\ M.\ Baake, Dr.\ U.\ Grimm, and Dr.\ E.~A.\
Robinson for providing us with many useful references. This work has
been supported by NSF Grant PHY 01-00041.

\catcode`\@=11
\def\footnotesize{\@setsize\footnotesize{12pt}\xpt\@xpt
\abovedisplayskip 10pt plus2pt minus5pt\belowdisplayskip \abovedisplayskip
\abovedisplayshortskip
\z@ plus3pt\belowdisplayshortskip 6pt plus3pt minus3pt
\def\@listi{\leftmargin\leftmargini
\topsep 6pt plus 2pt minus 2pt\parsep 0pt
plus 0pt minus 0pt
\itemsep \parsep}}
\catcode`\@=12
\footnotesize

\end{document}